\documentclass[aps,pre,twocolumn]{revtex4}
\usepackage{amssymb,macros2erev4,graphicx,amsmath}
\usepackage{times}
\newcommand{\be}{\begin{equation}}
\newcommand{\ee}{\end{equation}}
\newcommand{\bea}{\begin{eqnarray}}
\newcommand{\eea}{\end{eqnarray}}
\newcommand{\half}{\frac{1}{2}}
\newcommand{\ts}{\hskip0.1ex\raisebox{-1ex}[0ex][0.8ex]{\rule{0.1ex}{2.75ex}\hskip0.2ex}} 

\newcommand{\fig}[2]{\includegraphics[width=#1]{./figures/#2}}
\newcommand{\Fig}[1]{\includegraphics[width=\columnwidth]{./figures/#1}}
\newlength{\bilderlength}
\newcommand{\bilderscale}{0.35}
\newcommand{\storebilderscale}{\bilderscale}
\newcommand{\bilderskip}{\hspace*{0.8ex}}
\newcommand{\textdiagram}[1]{%
\renewcommand{\bilderscale}{0.25}%
\diagram{#1}\renewcommand{\bilderscale}{\storebilderscale}}
\newcommand{\diagram}[1]{%
\settowidth{\bilderlength}{\bilderskip%
\includegraphics[scale=\bilderscale]{./figures/#1}\bilderskip}%
\parbox{\bilderlength}{\bilderskip%
\includegraphics[scale=\bilderscale]{./figures/#1}\bilderskip}}

\newcommand{\sgn}{{\mathrm{sgn}}}
\newcommand{\rme}{{\mathrm{e}}}
\newcommand{\rmd}{{\mathrm{d}}}
\newcommand{\nn}{\nonumber}

\newcommand{\E}{\epsilon}

\begin{document}

%
\title{\sffamily\bfseries\Large Functional Renormalization Group and
the Field Theory of \\
Disordered Elastic Systems}
\author{\sffamily\bfseries\normalsize Pierre Le Doussal{$^1$}, Kay
J\"org Wiese{$^2$} and Pascal Chauve{$^3$} \vspace*{3mm}}
\affiliation{{$^1$} CNRS-Laboratoire de Physique Th{\'e}orique de
l'Ecole Normale Sup{\'e}rieure,
24 rue Lhomond, 75005 Paris, France\\
{$^2$} Kavli Institute of Theoretical Physics, University of
California at
Santa Barbara, Santa Barbara, CA 93106-4030, USA\\
{$^3$} CNRS-Laboratoire de Physique des Solides, Universit{\'e} de
Paris-Sud, B{\^a}t.\ 510, 91405 Orsay, France\medskip }
\date{\small April 26, 2003}
\begin{abstract}
We study elastic systems such as interfaces or lattices, pinned by
quenched disorder. To escape triviality as a result of ``dimensional
reduction'', we use the functional renormalization group.
Difficulties arise in the calculation of the renormalization group
functions beyond 1-loop order. Even worse, observables such as the
2-point correlation function exhibit the same problem already at
1-loop order. These difficulties are due to the non-analyticity of the
renormalized disorder correlator at zero temperature, which is
inherent to the physics beyond the Larkin length, characterized by
many metastable states.  As a result, 2-loop diagrams, which involve
derivatives of the disorder correlator at the non-analytic point, are
naively ``ambiguous''. We examine several routes out of this dilemma,
which lead to a unique renormalizable field-theory at 2-loop order. It
is also the only theory consistent with the potentiality of the
problem. The $\beta$-function differs from previous work and the one
at depinning by novel ``anomalous terms''. For interfaces and random
bond disorder we find a roughness exponent $\zeta=0.20829804 \epsilon
+ 0.006858 \epsilon^2$, $\epsilon=4-d$. For random field disorder we
find $\zeta=\epsilon/3$ and compute universal amplitudes to order
$O(\epsilon^2)$. For periodic systems we evaluate the universal
amplitude of the 2-point function.  We also clarify the dependence of
universal amplitudes on the boundary conditions at large scale. All
predictions are in good agreement with numerical and exact results,
and an improvement over one loop. Finally we calculate higher
correlation functions, which turn out to be equivalent to those at
depinning to leading order in $\epsilon$.

\end{abstract}
\maketitle

\section{Introduction} Elastic objects pinned by quenched disorder are
central to the physics of disordered systems. In the last decades a
considerable amount of research has been devoted to them. From the
theory side they are among the simplest, but still quite non-trivial,
models of glasses with complex energy landscape and many metastable
states. They are related to a remarkably broad set of problems, from
subsequences of random permutations in mathematics
\cite{Johansson2000,Johansson1999,BaikDeiftJohansson1999}, random
matrices \cite{PraehoferSpohn2000a,PraehoferSpohn2000} to growth
models
\cite{KPZ,FreyTaeuber1994,Laessig1995,FreyTaeuberHwa1996,Wiese1997c,Wiese1998a,MarinariPagnaniParisi2000,PraehoferSpohn1997,Krug1997}
and Burgers turbulence in physics
\cite{Mezard1997,MedinaHwaKardarZhang1989}, as well as directed
polymers \cite{KPZ,HwaFisher1994b} and optimization problems such as
sequence alignment in biology
\cite{BundschuhHwa2000,BundschuhHwa1999,HwaLaessig1998}.  Foremost,
they are very useful models for numerous experimental systems, each
with its specific features in a variety of situations.  Interfaces in
magnets
\cite{NattermannBookYoung,LemerleFerreChappertMathetGiamarchiLeDoussal1998}
experience either short-range disorder (random bond RB), or long range
(random field RF). Charge density waves (CDW) \cite{Gruner1988} or the
Bragg glass in superconductors
\cite{BlatterFeigelmanGeshkenbeinLarkinVinokur1994,GiamarchiBookYoung,GiamarchiLeDoussal1995,GiamarchiLeDoussal1994,NattermannScheidl2000}
are periodic objects pinned by disorder. The contact line of liquid
helium meniscus on a rough substrate is governed by long range
elasticity
\cite{PrevostRolleyGuthmann2002,PrevostThese,ErtasKardar1994b}.  All
these systems can be parameterized by a $N$-component height or
displacement field $u_x$, where $x$ denotes the $d$-dimensional
internal coordinate of the elastic object (we will use $u_q$ to denote
Fourier components). An interface in the 3D random field Ising model
has $d=2$, $N=1$, a vortex lattice $d=3$, $N=2$, a contact-line $d=1$
and $N=1$. The so-called directed polymer ($d=1$) has been much
studied \cite{KardarLH1994} as it maps onto the Kardar-Parisi-Zhang
growth model \cite{KPZ} for any $N$. The equilibrium problem is
defined by the partition function ${\cal Z} = \int {\cal D}[u]\,
\exp(-{\cal H}[u]/T)$ associated to the Hamiltonian
\begin{equation} \label{ham}
 {\cal H}[u]= \int \rmd^d x\, \frac{1}{2} (\nabla u)^2 + V(u_x,x)
\ ,
\end{equation}
which is the sum of an elastic energy which tends to suppress
fluctuations away from the perfectly ordered state $u=0$, and
a random potential which enhances them. The resulting roughness
exponent $\zeta$
\begin{equation}\label{lf5}
\overline{\left<(u(x) - u(x'))^2\right>} \sim |x-x'|^{2 \zeta}
\end{equation}
is measured in experiments for systems at equilibrium ($\zeta_{\rm
eq}$) or driven by a force $f$. Here and below $\left<\dots \right>$
denote thermal averages and $\overline {(\dots) }$
disorder ones. In some cases, long-range elasticity appears e.g.\ for
the contact line by integrating out the bulk-degrees of freedom
\cite{ErtasKardar1994b}, corresponding to $q^2 \to |q|$ in the elastic
energy. As will become clear later, the random potential can without
loss of generality be chosen Gaussian with second cumulant
\begin{equation}\label{corrstat}
\overline{V (u, x) V(u',x')} = R(u-u') \delta^d(x-x') \ .
\end{equation}
with various forms: Periodic systems
are described by a periodic function $R(u)$, random
bond disorder by a short-range function and random field disorder of
variance $\sigma$ by $R(u) \sim - \sigma |u|$ at large $u$.
Although this paper is devoted to equilibrium statics, some comparison
with dynamics will be made and it is thus useful to indicate the
equation of motion
\begin{equation}\label{eqn.motion}
\eta \partial_t u_{xt} = c \nabla_x^2 u_{xt}  + F(x, u_{xt} ) + f\ ,
\end{equation}
with friction $\eta$. The pinning force is $F(u,x) = - \partial_u V(u,x)$
of correlator $\Delta(u) = - R''(u)$ in the bare model.

Despite some significant progress, the model (\ref{ham}) has mostly
resisted analytical treatment, and one often has to rely on
numerics. Apart from the case of the directed polymer in $1+1$
dimensions ($d=1$, $N=1$), where a set of exact and rigorous results
was obtained
\cite{Kardar1987,BrunetDerrida2000,BrunetDerrida2000a,Johansson1999,PraehoferSpohn2000},
analytical methods are scarce. Two main analytical methods exist at
present, both interesting but also with severe limitations. The first
one is the replica Gaussian Variational Method (GVM)
\cite{MezardParisi1991}.  It is a mean field method, which can be
justified for $N=\infty$ and relies on spontaneous replica symmetry
breaking (RSB) \cite{ParisiDiracMedal,MezardParisiVirasoro}. Although
useful as an approximation, its validity at finite $N$ remains
unclear. Indeed, it seems now generally accepted that RSB does not
occur for low $d$ and $N$. The remaining so-called weak RSB in
excitations \cite{Mezard1990,WeigtMonasson1996,SalesYoshino2002} may
not be different from a more conventional droplet picture. Another
exactly solvable mean field limit is the directed polymer on the
Cayley tree, which also mimics $N \to \infty$ and there too it is not
fully clear how to meaningfully expand around that limit
\cite{DerridaSpohn1988,CookDerrida1989,CookDerrida1989a}.  The second
main analytical method is the Functional Renormalization Group (FRG)
which attempts a dimensional expansion around $d=4$
\cite{Fisher1985b,DSFisher1986,BalentsDSFisher1993,GiamarchiLeDoussal1995,GiamarchiLeDoussal1994}.  The hope there
is to include fluctuations, neglected in the mean field approaches.
However, until now this method has only been developed to one loop,
for  good reasons, as we discuss below. Its consistency has
never been  checked or tested in any calculation beyond one loop
(i.e.\ lowest order in $\epsilon=4-d$). Thus contrarily to pure
interacting elastic systems (such as e.g.\ polymers) {\it there is at
present no quantitative method, such as a renormalizable field theory,
which would allow to compute accurately all universal observables in
these systems}.

The central reason for these difficulties is the existence of many
metastable states (i.e.\ local extrema) in these systems.  Although
qualitative arguments show that they arise beyond the Larkin length
\cite{LarkinOvchinnikov}, these are hard to capture by conventional
field theory methods. The best illustration of that is the so called
dimensional reduction (DR) phenomenon, which renders naive
perturbation theory useless
\cite{NattermannBookYoung,EfetovLarkin1977,AharonyImryShangkeng1976,Grinstein1976,ParisiSourlas1979,Cardy1983}
in pinned elastic systems as well as in a wider class of disordered
models (e.g.\ random field spin models). Indeed it is shown that to
{\em any} order in the disorder at zero temperature $T=0$, any
physical observable is found to be {\it identical} to its (trivial)
average in a Gaussian random force (Larkin) model, e.g.\
$\zeta=(4-d)/2$ for RB disorder.  Thus perturbation theory appears
(naively) unable to help in situations where there are many metastable
states.  The two above mentioned methods (GVM and FRG) are presently
the only known ways to {\it escape dimensional reduction} and to
obtain non-trivial values for $\zeta$ (in two different limits but
consistent when they can be compared
\cite{BalentsDSFisher1993,GiamarchiLeDoussal1995,GiamarchiLeDoussal1994}). The
mean field method accounts for metastable states by RSB. This however
may go further than needed since it implies a large number of pure
states (i.e.\ low (free) energy states differing by $O(T)$ in (free)
energy).  The other method, the FRG, captures metastability through a
non-analytic action with a cusp singularity.  Both the RSB and the
cusp arise dynamically, i.e.\ spontaneously, in the limits studied.

The 1-loop FRG has had some success in describing pinned systems. It
was noted by Fisher \cite{DSFisher1986} within a Wilson scheme
analysis of the interface problem in $d=4 -\epsilon$ that the coarse
grained disorder correlator becomes {\em non-analytic} beyond the
Larkin scale $L_c$, yielding large scale results distinct from naive
perturbation theory. Within this approach an infinite set of operators
becomes relevant in $d<4$, parameterized by the second cumulant $R(u)$
of the random potential. Explicit solution of the 1-loop FRG for
$R(u)$ gives several non-trivial attractive fixed points (FP) to
${O}(\epsilon)$ proposed in \cite{DSFisher1986} to describe RB, RF
disorder and in \cite{GiamarchiLeDoussal1995,GiamarchiLeDoussal1994},
periodic systems such as CDW or vortex lattices.  All these fixed
points exhibit a ``cusp'' singularity as $R^{*\prime \prime}(u) -
R^{*\prime \prime}(0) \sim |u|$ at small $|u|$. The cusp was
interpreted in terms of shocks in the renormalized force
\cite{BalentsBouchaudMezard1996}, familiar from the study of Burgers
turbulence (for $d=1$, $N=1$). The dynamical FRG was also developed to
one loop
\cite{NattermanStepanowTangLeschhorn1992,LeschhornNattermannStepanow1996,%
NarayanDSFisher1993a} to describe the depinning transition. The mere
existence of a non-zero critical threshold force $f_c \sim
|\Delta'(0^+)|>0$ is a direct consequence of the cusp (it vanishes for
an analytic force correlator $\Delta(u)$).  Extension to non-zero
temperature $T>0$ suggested that the cusp is rounded within a thermal
boundary layer $u \sim T L^{- \theta}$.  This was interpreted to
describe thermal activation and leads to a reasonable derivation of
the celebrated creep law for activated motion
\cite{ChauveGiamarchiLeDoussal1998,ChauveGiamarchiLeDoussal2000}.

In standard critical phenomena a successful 1-loop calculation
usually quickly opens the way for higher loop computations, allowing
for accurate calculation of universal observables and comparison with
simulations and experiments, and eventually a proof of
renormalizability.  In the present context however, no such work has
appeared in the last fifteen years since the initial proposal of
\cite{DSFisher1986}, a striking sign of the high difficulties which
remain. Only recently a 2-loop calculation was performed
\cite{BucheliWagnerGeshkenbeinLarkinBlatter1998,WagnerGeshkenbeinLarkinBlatter1999}
but since this study is confined to an analytic $R(u)$ it only applies
below the Larkin length and does not consistently address the true
large scale critical behavior. In fact doubts were even raised
\cite{BalentsDSFisher1993} about the validity of the
$\epsilon$-expansion beyond  order $\epsilon$.

It is thus crucial  to construct a renormalizable field theory, which
describes  statics and depinning of disordered elastic systems, and
which  allows for a systematic expansion in $\epsilon=4-d$.  As
long as this is not achieved, the physical meaning and validity of the
1-loop approximation does not stand on solid ground and thus,
legitimately, may itself be called into question.  Indeed, despite
its successes, the 1-loop approach has  obvious
weaknesses.  One example is that the FRG flow equation for the
equilibrium statics and for depinning are  identical, while
it is clear that these are two vastly different physical phenomena,
depinning being irreversible.  Also, the detailed mechanism by which
the system escapes dimensional reduction in both cases is not really
elucidated. Finally, there exists no convincing scheme to
compute correlations, and in fact no calculation of higher than
2-point correlations has been performed.

Another motivation to investigate the FRG is that it should apply to
other disordered systems, such as random field spin models, where
dimensional reduction also occurs and progress has been slow
\cite{Fisher1985b,BrezinDeDominicis1998,Feldman2000,Feldman2002,LeDoussalWieseRandomField}.
Insight into model (\ref{ham}) will thus certainly lead to progresses
in a broader class of disordered systems.

In this paper we construct a renormalizable field theory for the
statics of disordered elastic systems beyond one loop.  The main
difficulty is the non-analytic nature of the theory (i.e.\ of the
fixed point effective action) at $T=0$. This makes it a priori quite
different from conventional field theories for pure systems. We find
that the 2-loop diagrams are naively ``ambiguous'', i.e.\ it is not
obvious how to assign a value to them. We want to emphasize that this
difficulty already exists at one loop, e.g.\ {\it even the simplest one
loop correction to the two point function is naively
``ambiguous''}. Thus it is not a mere curiosity but a fundamental
problem with the theory, ``swept under the rug'' in all previous
studies, but which becomes unavoidable to confront at 2-loop order. It
originates from the metastability inherent in the problem.  For the
related theory of the depinning transition, we have shown in companion
papers \cite{ChauveLeDoussalWiese2000a,LeDoussalWieseChauve2002} how
to surmount this problem and we constructed a 2-loop renormalizable
field theory {\it from first principles}. There, all ambiguities are
naturally lifted using the known exact property that the manifold only
moves forward in the slowly moving steady state. Unfortunately in the
statics there is no such helpful property and the ambiguity problem is
even more arduous.  Here we examine the possible ways of curing these
difficulties.  We find that the natural physical requirements, i.e.\
that the theory should be (i) {\it renormalizable} (i.e.\ that a
universal continuum limit exists independent of short-scale details),
(ii) that the renormalized force should remain {\it potential}, and
(iii) that no stronger singularity than the cusp in $R''(u)$ should
appear to two loop (i.e.\ {\it no ``supercusp''}), are rather
restrictive and constrain possible choices. We then propose a theory
which satisfies all these physical requirements and is consistent to
two loops.  The resulting $\beta$-function differs from the one
derived in previous studies
\cite{BucheliWagnerGeshkenbeinLarkinBlatter1998,WagnerGeshkenbeinLarkinBlatter1999}
by novel static ``anomalous terms''. These are different from the
dynamical ``anomalous terms'' obtained in
\cite{ChauveLeDoussalWiese2000a,LeDoussalWieseChauve2002,LeDoussalWiese2002a}
showing that indeed depinning and statics differ at two loop,
fulfilling another physical requirement.

We then study the fixed points describing several universality
classes, i.e.\ the interface with RB and RF disorder, the random
periodic problem, and the case of LR elasticity. We obtain the
$O(\epsilon^2)$ corrections to several universal quantities. The
prediction for the roughness exponent $\zeta$ for random bond disorder
has the correct sign and order of magnitude to notably improve the
precision as compared to numerics in $d=3,2$ and to match the exact result
$\zeta=2/3$ in $d=1$.  For random field disorder we find
$\zeta=\epsilon/3$ which, for equilibrium is likely to hold to all
orders. By contrast, non-trivial corrections of order $O(\epsilon^2)$
were found for depinning
\cite{ChauveLeDoussalWiese2000a,LeDoussalWieseChauve2002}.  The
amplitude, which in that case is a universal function of the random
field strength is computed and it is found that the 2-loop result
also improves the agreement as compared to the exact result known
\cite{LeDoussalMonthus2003} for $d=0$.  For the periodic CDW case we
compare with the numerical simulations in $d=3$ and obtain reasonable
agreement. Some of the results of this paper were briefly described in
a short version \cite{ChauveLeDoussalWiese2000a} and agree with a
companion study using exact RG \cite{ChauveLeDoussal2001,scheidl2}).

Since the physical results also seem to favor this theory we then look
for better methods to justify the various assumptions.  We found
several methods which allow to lift ambiguities and all yield
consistent answers.  A detailed discussion of these methods is given.
In particular we find that correlation functions can be unambiguously
defined in the limit of a small background field which splits apart
quasi-degenerate states when they occur. This is very similar to what
was found in a related study where we obtained the exact solution of
the FRG in the large $N$ limit \cite{LeDoussalWiese2001}. Finally, the
methods introduced here will be used and developed further to obtain a
renormalizable theory to three loops, and compute its $\beta$-function
in \cite{LeDoussalWiesePREPb}.  Let us mention that a first principles
method which avoids ambiguities is to study the system at $T>0$.
However, this turns out to be highly involved. It is attempted via
exact RG in \cite{ChauveLeDoussal2001} and studied more recently in
\cite{BalentsLeDoussal2002a,BalentsLeDoussal2002b} where a field
theory of thermal droplet excitation was constructed. A short account
of our work has appeared in \cite{ChauveLeDoussalWiese2000a}, and a
short pedagogical introduction is given in \cite{Wiese2002}.

The outline of this paper is as follows. In Section \ref{sec:model} we
explain in a detailed and pedagogical way the perturbation theory and
the power counting. In Section \ref{sec:renprog} we compute the 1-loop
(Section \ref{sec:oneloop}) and 2-loop (Section \ref{sec:twoloop})
corrections to the disorder.  The calculation of the repeated 1-loop
counter-term is given in Section \ref{sec:rgdis}.  In Section
\ref{sec:finalbeta} we identify the values for ambiguous graphs. This
yields a renormalizable theory with a finite $\beta$-function, which
is potential and free of a supercusp. The more systematic discussion
of these ambiguities is postponed to Section \ref{sec:ambiguities}.
We derive the $\beta$-function and in Section \ref{sec:fixedpoints}
present physical results, exponents and universal amplitudes to
$O(\epsilon^2)$. Some of these quantities are new, and have not yet
been tested numerically. In Section \ref{sec:ambiguities} we enumerate
all the methods which aim at lifting ambiguities and explain in
details several of them which gave consistent results. In Section
\ref{sec:correlations} we detail the proper definition and calculation
of correlation functions.  In Appendix \ref{sec:symm} and
\ref{sec:nona} we present two methods which seem promising but {\it do
not} work, in order to illustrate the difficulties of the problem.  In
Appendix \ref{sec:finiteT} we present a summary of all one and 2-loop
corrections including finite temperature.  In Appendix
\ref{sec:sloopBC} we give details of calculations for what we call the sloop
elimination method.

The reader interested in the results can skip Section \ref{sec:model}
and Section \ref{sec:renprog} and go directly to Section
\ref{sec:fixedpoints}.  The reader interested in the detailed
discussion of the problems arising in this field theory should read
Section \ref{sec:ambiguities}.


\section{Model and perturbation theory}
\label{sec:model}

\subsection{Replicated action and effective action}
\label{sec:effective} We study the static equilibrium problem using
replicas, i.e.\ consider the partition sum in presence of sources:
\begin{equation}\label{lf6}
{\cal Z}[j] = \int \prod_a {\cal D}[u_a]\, \exp\left(- {\cal S}[u] +
\int_x \sum_a j_x^a u_x^a\right) \ ,
\end{equation}
from which all static observables can be obtained. The
action ${\cal S}$ and
replicated Hamiltonian corresponding to (\ref{ham}) are
\begin{eqnarray}\label{lf7}
 {\cal S}[u] = \frac{{\cal H}[u]}{T} &=& \frac{1}{2T} \int_x \sum_a
[(\nabla u^a_x)^2
+ m^2 u^a_x] \nonumber \\
&& - \frac{1}{2 T^2} \int_x \sum_{a b} R(u^a_x - u^b_x)
\ . \label{action}
\end{eqnarray}
$a$ runs from 1 to $n$ and the limit of zero number of replicas $n=0$
is implicit everywhere. We have added a small mass which confines the
interface inside a quadratic well, and provides an infrared cutoff. We
are interested in the large scale limit $m \to 0$. We will denote
\begin{eqnarray}\label{intq}
\int_q &:=& \int \frac {\rmd ^d q}{(2 \pi)^d}
\label{intx}\\
\int_{x}&:=&\int \rmd^{d}x\ .\label{lf96}
\end{eqnarray}
For periodic systems the integration is over the first Brillouin
zone. A short-scale UV cutoff is implied at $q \sim \Lambda$, but for
actual calculations we find it more convenient to use dimensional
regularization.  We also consider the effective action functional
$\Gamma[u]$ associated to ${\cal S}$. It is, as we recall
\cite{Zinn,ItzyksonZuber}, the Legendre transform of the generating
function of connected correlations ${\cal W}[j] = \ln {\cal Z}[j]$,
thus defined by eliminating $j$ in $\Gamma[u] = j u - {\cal W}[j]$,
${\cal W}'[j] = u$.

If we had chosen non-Gaussian disorder additional terms with free
sums over $p$ replicas (called $p$-replica terms) corresponding to
higher cumulants of disorder would be present in (\ref{action}),
together with a factor of $1/T^p$. These terms are generated in
the perturbation expansion, i.e.\ they are present in $\Gamma[u]$.
We do not include them in (\ref{action}) because, as we will see
below, these higher disorder cumulants are not relevant within
(conventional) power counting, so for now we ignore them. The
temperature $T$ appears explicitly in the replicated action
(\ref{action}), although we will focus on the $T=0$ limit.

Because the disorder distribution is translation invariant, the
disorder term in the above action is invariant under the so called
statistical tilt symmetry
\cite{SchulzVillainBrezinOrland1988,HwaFisher1994b} (STS), i.e.\ the
shift $u^a_{x} \to u^a_{x} + g_x$. One implication of STS is that the
1-replica replica part of the action (i.e.\ the first line of
\ref{action}) is uncorrected by disorder, i.e.\ it is the same in
$\Gamma[u]$ and ${\cal S}[u]$ \cite{stsproof}.  Since the elastic
coefficient is not renormalized, we have set it to unity.

\subsection{Diagrammatics, definitions} 
\begin{figure}
\fig{7cm}{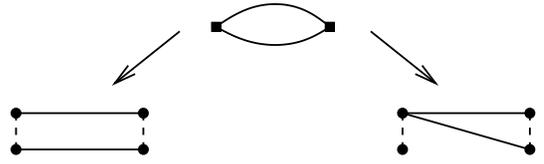} \caption{Each diagram with unsplitted vertices
contains several diagrams with splitted vertices: here the 1-loop
unsplitted diagram (top) generates three possible topologically
distinct splitted diagrams, two (shown here, bottom) are 2-replica
terms, the third one, i.e.\ (a) in Fig.\ (\protect\ref{f:sloops}) is a
three replica term} \label{f:1loopexpstat}
\end{figure}

We first study perturbation
theory, its graphical representation and power counting. Everywhere in
the paper we denote the exact 2-point correlation by $C_{ab}(x-y)$,
i.e.\ in Fourier:
\begin{equation} \label{exactcorr}
 \langle u^a_{q} u^b_{q'} \rangle = (2 \pi)^d \delta^d(q+q') C_{ab}(q)
\end{equation}
while the free correlation function (from the elastic term)
used for perturbation theory in the disorder is denoted by
$G_{ab}(x-y) = \delta_{ab} G(x-y)$ and reads in Fourier:
\begin{eqnarray}\label{lf8}
\langle u^a_{q} u^b_{q'} \rangle_0 &=& (2 \pi)^d \delta^d(q+q') G_{ab}(q) \\
 G_{ab}(q) &=& \frac{T}{q^2 + m^2} \delta_{ab}\ , \label{lf97}
\end{eqnarray}
which is represented graphically by a line:
\begin{equation}\label{lf98}
_{a} \diagram{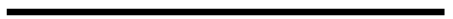}_{b} ~= \frac{T\delta_{ab}}{q^{2}+m^{2}}\ .
\end{equation}
Each propagator thus carries one factor of $G(q)=T/(q^2+m^2)$. Each
disorder interaction vertex comes with a factor of $1/T^2$ and gives
one momentum conservation rule. Since each disorder vertex is a
function, an arbitrary number of lines can come out of it.  $k$ lines
coming out of a vertex result in $k$ derivatives $R^{(k)}$ after Wick
contractions
\begin{equation}\label{lf99}
\diagram{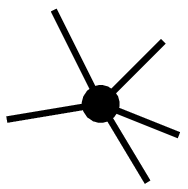}= R^{(k)}\ .
\end{equation}
Since each disorder vertex contains two replicas it is sometimes
convenient to use ``splitted vertices'' rather than ``unsplitted
ones''.  Thus we call ``vertex'' an unsplitted vertex and we call a
``point'' the half of a vertex.
\begin{equation}\label{lf100}
{a\atop b}\diagram{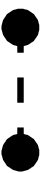}  = \sum_{ab} \frac{R (u_{a}-u_{b})}{2 T^{2}}\ .
\end{equation}
Each unsplitted diagram thus gives rise to several
splitted diagrams, as illustrated in Fig.\ \ref{f:1loopexpstat}

\begin{figure}
\fig{8cm}{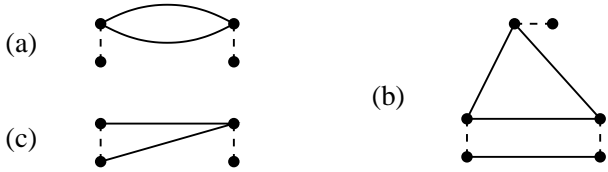} \caption{Graphs (a) (a 1-loop diagram) and (b) (a
2-loop diagram) each contains three connected components. Since each
contain one ``sloop'' they are both three replica terms proportional
to $T$.  The left vertex on diagram (c) is ``saturated'' : replica
indices are constrained to be equal and thus the diagram does not
depend on the left space point.}  \label{f:sloops}
\end{figure}

\begin{figure}
\Fig{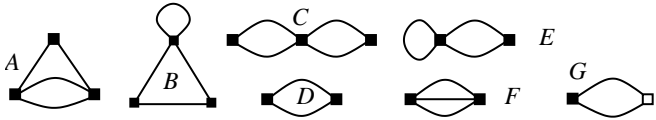}
\caption{unsplitted diagrams to one loop D, one loop with inserted 1-loop
counter-term G and 2-loop diagrams A, B, C, E and F.}
\label{alldiag2loop}
\end{figure}

One can define the number of connected components in a graph with
splitted vertices. Since each propagator identifies two replicas, a
$p$-replica term contains $p$ connected components.  When the 2-points
of a vertex are connected, this vertex is said to be ``saturated''. It
gives a derivative evaluated at zero $R^{(k)}(0)$. Standard momentum
loops are loops with respect to unsplitted vertices, while we call
``sloops'' the loops with respect to points (in splitted
diagrams). This is illustrated in Fig.\ (\ref{f:sloops})
The momentum 1-loop and 2-loop diagrams which correct the disorder at
$T=0$ are shown in Fig.\ \ref{alldiag2loop} (unsplitted
vertices). There are three types of 2-loop graphs A, B and C. Since
they have two vertices (a factor $R/T^2$ each) and three propagators
(a factor of $T$ each) the graphs E and F lead to corrections to
$R$ proportional to temperature and will not be studied here (see
however Appendix \ref{sec:finiteT}).

It is important to distinguish between {\it fully saturated diagrams}
and {\it functional diagrams}. The FS diagrams are those needed
for a full average, e.g.\ a correlation function.  There all
fields are contracted and one is only left  with the space dependence.
These are the standard diagrams in more conventional polynomial field
theories such as $\phi^4$. Then all vertices are evaluated at $u=0$,
yielding products of derivatives $R^{(k)}(0)$. These are also the
graphs which come in the standard expansion of $\Gamma[u]$ in powers
of $u$ which generate the ``proper'' or ``renormalized'' vertices,
i.e.\ the sum over all 1-particle irreducible graphs with some
external legs, from which all correlations can be obtained. Note that
in the fully saturated diagrams there can be no free point, all points
in a vertex have to be connected to some propagator (and to some
external replica) otherwise there is a free replica sum yielding a
factor of $n$ and a vanishing contribution in the limit of $n=0$.

However, since we have to deal with a function $R(u)$ we will more
often consider functional diagrams. A functional diagram still depends
on the field $u$. It can depend on $u$ at several points in space
(multi-local term), as for example:
\begin{equation}\label{lf101}
\rule[-1.5ex]{0mm}{2mm}_{x}\diagram{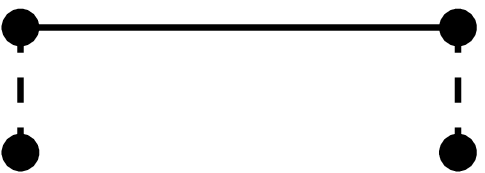}_{y} \sim \sum_{abc}
\frac{R'(u^a_x - u^b_x)}{2 T^2} \frac{ R'(u^a_y - u^c_y)}{2 T^2} T
G(x-y)
\end{equation}
Such a graph with $p$ connected components corresponds to a $p$
replica functional term. Or it can represent the projection of such a
term onto a local part, as arises in the standard operator product
expansion (OPE):
\begin{equation}\label{lf102}
\diagram{func1} \sim \sum_{abc}
\frac{R'(u^a_x - u^b_x)}{2 T^2} \frac{R'(u^a_x - u^c_x)}{2 T^2} T
\int_y G(x-y)\ .
\end{equation}
Typically using functional diagrams we  want to compute the
effective action functional $\Gamma[u]$, or its local part, i.e.\ its
value for a spatially uniform mode $u^a_x=u^a$, which includes the
corrections to disorder. Specifying the two replicas on each connected
component, one example of a 1-particle irreducible diagram producing
corrections to disorder is
\begin{equation}\label{lf103}
\diagram{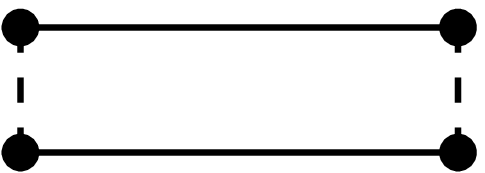} \sim \frac{T^2}{T^4}
R''(u^a - u^b) R''(u^a - u^b) \int_q G(q)^2\ .
\end{equation}
The complete analysis of these corrections will be made in Section
(\ref{sec:renprog}).  Finally, note that functional diagrams may
contain saturated vertices, whose space and field dependence
disappears (such as (c) in Fig.\ \ref{f:sloops}) and that the limit $n
\to 0$ does not produce constraints. An example is the calculation of
$\Gamma[u]$ since one can always attach additional external legs to
any point by taking a derivative with respect to the field $u$.

\subsection{Dimensional reduction}
\label{sec:dimred}
\begin{figure}
\Fig{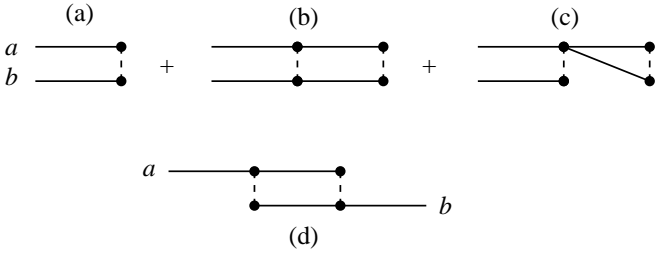}
\caption{Calculation of the 2-point
function for analytic $R(u)$. Due to DR only the first diagram (a)
survives. Diagrams (b) and (c) cancel because by shifting the line one
gets a minus sign. Diagram (c) is proportional to $R'''(0)^2$ and
vanishes in an analytic theory. Similar cancellations occur to all
orders.} \label{f:dimred}
\end{figure}

If we consider fully saturated diagrams and analytic $R(u)$ we find
trivial results. This is because at $T=0$ the model exhibits the
property of dimensional reduction
\cite{EfetovLarkin1977,NattermannBookYoung,AharonyImryShangkeng1976,Grinstein1976,ParisiSourlas1979,Cardy1983}
(DR) both in the statics and dynamics.  Its ``naive'' perturbation
theory, obtained by taking for the disorder correlator $R(u)$ an {\it
analytic function } of $u$ has a triviality property.  As is easy to
show using the above diagrammatic rules (see a typical cancellation
due to the ``mounting'' construction in Fig.\ \ref{f:dimred}, see also
Appendix D in Ref.~\cite{ChauveLeDoussal2001}) the perturbative
expansion of any correlation function $\langle \prod_i
u^{a_i}_{x_i}\rangle _S$ (of any {\it analytic} observable) in the
derivatives $R^{(k)}(0)$ yields to all orders the same result as that
obtained from the Gaussian theory setting $R(u) \equiv R''(0) u^2/2$
(the so called Larkin random force model). The 2-point function thus
reads to all orders:
\begin{equation}\label{lf9}
C(q)_{ab}^{\mathrm{DR}} = \frac{- R''(0)}{(q^2 + m^2)^2}
\ .
\end{equation}
(at $T=0$ correlations are independent of the replica indices
$a_i$). This dimensional reduction results in a roughness exponent
$\zeta=(4-d)/2$ which is well known to be incorrect.  One physical
reason is that this $T=0$ perturbation theory amounts to solving in
perturbation the zero force equation
\begin{equation} \label{lf10}
(- \nabla^2 + m^2) u + F(x,u) = 0\ .
\end{equation}
This, whenever more than one solution exists (which certainly happens
for small $m$) is clearly not identical to finding the lowest energy
configuration \footnote{One can easily see that the DR result
(\ref{lf9}) arises if one averages over multiple solutions $u_\alpha$
with some random weights $W_\alpha \sim |\det( \nabla^2 + m^2 +
F'_u(x,u_\alpha)|$ (then using the representation of the delta
function $\exp(i \int_x \hat u ((- \nabla^2 + m^2) u + F(x,u)))$ and
averaging over disorder using (\ref{corrstat}).  Summing over multiple
solutions $u_\alpha$ requires instead to include the crucial weight
$\exp(- \beta H(u_\alpha))$ in order to select the true ground state.}.
Curing this problem within the field theory, is highly non-trivial.
Coarse graining within the FRG up to a scale at which the renormalized
disorder correlator $R(u)$ becomes {\em non-analytic} (which includes
some of the physics of multiple extrema) is one possible route,
although understanding exactly how this cures the problem within the
field theory is a difficult open problem.

It is important to note that dimensional reduction is not the end of
perturbation theory, since saturated diagrams remain non-trivial at
finite temperature, so one way out is to study $T>0$. This is not the
route chosen here, instead we will attempt to work at $T=0$ with a
non-analytic action and focus on functional diagrams which remain
non-trivial.

\subsection{Power counting}
\label{sec:power}
Let us now consider power counting. Let us recall the conventional
analysis within e.g.\ the Wilson scheme
\cite{DSFisher1986,BalentsDSFisher1993}.  The elastic term is
invariant under $x \to b x$, $u \to b^\zeta u$ and $T \to b^\theta T$,
with $\theta = d - 2 + 2 \zeta$. $\zeta$ is for now undetermined.
Under this transformation the disorder function $R$ is multiplied by
$b^{d - 2 \theta} = b^{4 - d + 2 \zeta}$.  It becomes relevant for $d
< 4$, provided $\zeta < (4-d)/2$ which is physically expected (for
instance in the random periodic case, $\zeta=0$ is the only possible
choice, and for other cases $\zeta = O(\epsilon)$). The rescaled
dimensionless temperature term scales as $-m \partial_m \tilde T = -
\theta \tilde T$ (see below) and is  formally irrelevant near four
dimension. In the end $\zeta$ will be fixed by the disorder
distribution at the fixed point.

To be more precise, we want to determine in the field theoretic
framework the necessary counter-terms to render the theory UV finite
as $d \to 4$. The study of superficial divergences usually involves
examining the irreducible vertex functions (IVF):
\begin{equation} \label{gammaexp}
 \Gamma_{u \dots u}(q_i) = \prod_{i=1}^{E_{u}} \frac{\delta}{\delta u_{q_i} }
\Gamma[u] \ts _{u=0}
\end{equation}
with $E_{u}$ external fields $u$ (at momenta $q_i$, $i=1,..E_{u}$).
The perturbation expansion of a given IVF to any given order in the
disorder is represented by a set of 1-particle irreducible (1PI)
graphs (in unsplitted diagrammatics).  Being the derivative of the
effective action  they are the important physical objects
since all averages of products of fields  $u$ can be  expressed as tree
diagrams of the IVF. Finiteness of the IVF thus implies finiteness of
all such averages.

However since  $\Gamma[u]$ is non-analytic in some directions
(e.g.\ for a uniform mode $u_x^a=u^a$), derivatives such as
(\ref{gammaexp}) may not exist at $q=0$, and  we have  to be more general
and consider  {\em functional diagrams}. The (disorder part of the)
effective action is the sum of $k$-replica terms, noted $\Gamma_k[u]$
\begin{equation}\label{lf14}
 \Gamma[u] =  \sum_{k \geq 2} T^{-k} \Gamma_k[u]\ .
\end{equation}
Each $\Gamma_k[u]$ is the sum over 1PI graphs with $k$ connected
components (using splitted vertices), and itself depends on
$T$ as
\begin{equation}\label{lf15}
 \Gamma_k[u] = \sum_{l \geq 0} T^{l} \Gamma_{k,l}[u]\ ,
\end{equation}
where $l$ is the number of sloops. Thus at $T=0$ there are no sloops
and $\Gamma_k[u] = \Gamma_{k,l=0}[u]$ is the sum over 1PI {\em tree}
graphs with $k$ connected components (trees in replica-space, not
position-space).

Let us compute the superficial degree of UV divergence $\delta$ of a
functional graph entering the expansion of the local part of the
effective action.  We denote  $v$ the number of unsplitted disorder
vertices, $I$ the number of internal lines (propagators), $L$ the
number of loops and $l$ the number of sloops. One has the relations
\begin{eqnarray}
2 v  + l &=& k + I\label{lf12} \\
 v + L &=& 1 + I\label{lf13}\ .
\end{eqnarray}
The total factors of $T$ are $T^{ I - 2 v}= T^{l - k}$. At $T=0$
($l=0$) the superficial degree of UV divergence is thus
\begin{equation}\label{lf104}
\delta = d L - 2 I = d - k(d-2) + (d-4) v\ .
\end{equation}
Thus in $d=4$ the only graphs with positive superficial degree of
divergence are for $k=1$ (quadratic $\sim \Lambda^2$), and $k=2$ (log
divergence). $k=1$ corresponds  to a constant in the free energy.
Because of STS all single replica terms are uncorrected and  there
is no wave-function renormalization in this model.

Thus to renormalize the $T=0$ theory we need a priori to look only at
graphs with $p=2$ connected components, which by definition are those
correcting the second cumulant $R(u)$, compute their divergent parts,
and construct the proper counter-term to the function $R(u)$.  As
mentioned above, higher cumulants are irrelevant by power counting,
and are superficially UV-finite. The graphs which contribute to the
2-replica part $\Gamma_2[u]$ have $L$ loops with $L=1 + v + l$. At
zero temperature, $l=0$, thus $L=1+v$. The loop expansion thus
corresponds to the expansion in power of $R(u)$ and, as we will see
below, to an $\epsilon$-expansion.  More generally using the above
relation one has, schematically
\begin{equation}\label{lf16}
 \Gamma_{k,l}[u] = \sum_{L \geq \max(1,2+l-k)}  \partial^{(4 L - 4 + 2
k - 2 l)}_u
R^{(L-1+k-l)} \ ,
\end{equation}
where the number of internal lines gives the total number of
derivatives acting on an argument $u$ of the functions $R$.  For
instance, the 2-replica part at $T=0$ is a sum over  $L$-loop graphs of
the type
\begin{equation}\label{lf17}
 \Gamma_{k=2,l=0}[u] = \sum_{L \geq 1} \partial^{4 L}_u R^{L+1}\ .
\end{equation}
If one now considers $T>0$ one finds that $\delta = d - k(d-2) + (d-4)
v + (d-2) l$. Each additional power of $T$ yields an additional
quadratic divergence, more generally a factor of $T \Lambda^{d-2}$. Thus
to obtain a theory where observables are finite as $\Lambda \to
\infty$ one must start from a model where the initial temperature
scales with the UV cutoff as
\begin{eqnarray}
T = \hat{T} \Lambda^{2-d} \label{rescaled} \ .
\end{eqnarray}
This is similar to $\phi^4$-theory where it is known that a
$\phi^6$ term can be present and yields a finite UV limit (i.e.\
does not spoil renormalizability) only if it has the form $g_6
\phi^6/\Lambda^{d-2}$. Such a term, with precisely this cutoff
dependence, is in fact usually present in the starting bare model,
e.g.\ in lattice spin models.  It then produces only a finite
shift to $g_4$ without changing universal properties \footnote{We
thank E. Brezin for a discussion on this point.}. Here each factor
of $\hat{T}$ comes with a factor of $\Lambda^{2-d}$ which
compensates the UV divergence from the graph.  Thus the finite-$T$
theory may also be renormalizable. Computing the resulting shift
in $R(u)$ to order $R^2$ by resumming the diagrams $E$ and $F$ of
Fig.\ \ref{alldiag2loop} and all similar diagrams to any number of
loops has not been attempted here (see however Appendix
\ref{sec:finiteT}). The ``finite shift'' here is, however, much
less innocuous than in $\phi^4$-theory since it smoothes the cusp.
The effects of a non-zero temperature are explored in
\cite{BalentsLeDoussal2002a,BalentsLeDoussal2002b,LeDoussalWiese2001,LeDoussalWieseinprep}.

One can use the freedom to rescale $u$ by $m^{-\zeta}$. The
dimensionless temperature $\tilde{T} = T m^{\theta}$ is then
defined. The disorder term in $\Gamma[u]$ is then is as in
(\ref{action}) with $R(u)$ replaced by $m^{\epsilon - 4 \zeta}
\tilde{R}(u m^{\zeta})$ in terms of a dimensionless rescaled
function $\tilde{R}$ of a dimensionless rescaled argument. This
will be further discussed below.

\section{Renormalization program} \label{sec:renprog} In this section
we  compute the effective action to 2-loop order at $T=0$.  We are
only interested in the part which contains UV divergences as $d \to
4$.  We know from the  analysis of the last section that we only need to
consider the local $k=2$ 2-replica  part, i.e.\ the corrections to
$R(u)$. These $L=1$ and $L=2$ loop corrections contain $v=L+1$
vertices. Higher $v$ yields higher number of replicas.

\subsection{1-loop corrections to disorder} \label{sec:oneloop} 
\begin{figure}
\fig{6cm}{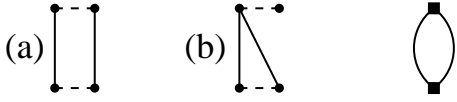} \caption{The two 1-loop diagrams with splitted vertices
and the corresponding diagram with standard (i.e.\ unsplitted)
vertices.}  \label{oneloop}
\end{figure}

To one
loop at $T=0$ there is only one unsplitted diagram $v=2$,
corresponding to two splitted diagrams (a) and (b) as indicated in figure
\ref{oneloop}.  Both come with a combinatorial factor of $1/2!$ from
Taylor-expanding the exponential function and $1/2$ from the action. (a)
has a combinatoric factor of $2$ and (b) of $4$. Together, they add up
to the 1-loop correction to disorder
\begin{eqnarray}
\delta^{1} R(u) &=&  \left[ \frac{1}{2} R''(u)^2 - R''(0) R''(u)
\right]  I_{1} \label{1loop}\\
 I_1^{m}=I_{1}& :=& \int_q \frac{1}{(q^2+m^2)^2} \nonumber\\
&=& \Gamma \left( 2 - \frac{d}{2}\right) m^{-\epsilon} \int_q \rme^{-q^2} \nonumber \\
&=& \frac{1}{(4 \pi)^{d/2}} \Gamma \left( 2 - \frac{d}{2}\right) m^{-\epsilon}=
O\left(\frac{1}\epsilon\right)\ .\label{I1}\qquad
\end{eqnarray}
Note that (b) has a saturated vertex, hence the
factor $R''(0)$. This does not lead to ambiguities in the  1-loop
$\beta$-function, since the FRG to one loop yields a discontinuity only in
the third derivative and $R''(u)$ remains continuous.

\subsection{2-loop corrections to disorder }
\label{sec:twoloop}
\begin{figure}[b]
\centerline{\fig{6cm}{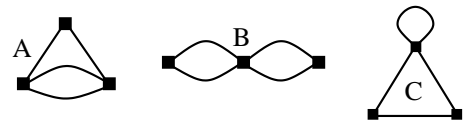}}
\caption{The 3 possible 2-loop unsplitted graphs correcting
disorder at $T=0$}
\label{f:twoloop}
\end{figure}
There are only three  graphs correcting disorder at
$T=0$ with $L=2$ loops and $v=3$ vertices. They are denoted
$A$, $B$ and $C$ and we will examine each of them.

We begin our analysis with class A.

\subsubsection{Class A} \label{static-2-loop-DO-classes}
\begin{figure}[h] \centerline{\fig{7cm}{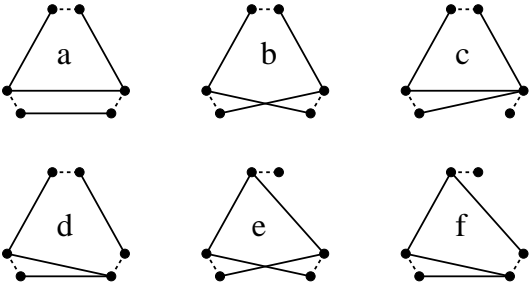}}
\caption{Graphs at 2-loop order in the form of a hat (class A
in figure \ref{static-2-loop-DO-classes}) contributing to 2-replica
terms.}
\label{static-2-loop-DO}
\end{figure}
The possible diagrams with splitted vertices of
type A are diagrams (a) to (f) given in Fig.\ \ref{static-2-loop-DO}.
The resulting correction to $R(u)$ is written as:
\begin{eqnarray}\label{n1}
\delta^{2} R(u) &=&  \frac{1}{3!} \frac{2}{2^3} 3 (2^3)
\sum (\mathrm{a} + \mathrm{b}  +
\mathrm{c}+\mathrm{d}+\mathrm{e}+\mathrm{f} )  \nonumber \label{corr}
\\
&=& \sum (\mathrm a + \mathrm b +
\mathrm{c}+\mathrm{d}+\mathrm{e}+\mathrm{f} )  \ ,
\end{eqnarray}
where the combinatorial factors are: $1/3!$ from the Taylor-expansion
of the exponential function, $2/2^3$ from the explicit factors of
$1/2$ in the interaction, a factor of 3 to chose the vertex at the top
of the hat, and a factor of 2 for the possible two choices in each of
the vertices. Furthermore below some additional combinatorial factors
are given: A factor of $2$ for generic graphs and $1$ if it has the
mirror symmetry with respect to the vertical axis. Each diagram symbol
denotes the diagram including the symmetry factor.  The first two
graphs are:
\begin{eqnarray}
 a &=& - R''(0) R'''(u)^2  I_A\label{lf19} \\
b &=& R''(u) R'''(u)^2 I_A\label{lf20}\ .
\end{eqnarray}
To obtain the sign one can choose an ``orientation'' in each vertex
($u_a - u_b$), the final result does not depend on the choice. The
minus sign in $a$ comes because the two legs enter on opposite points
in the top vertex. Define the 2-loop momentum integral (see
Appendix A in Ref.~\cite{LeDoussalWieseChauve2002})
\begin{eqnarray}
I_A&:=&\int_{q_{1}}\int_{q_{2}}
\frac{1}{q_1^2+m^2}\frac{1}{q_2^2+m^2} \frac{1}{(
(q_1{+}q_2)^2+m^2)^2}\nn\\
&=& \left(\frac1{2\E^2} + \frac1{4\E} +O(\E^2) \right) (\E I_1)^2
\ . \label{lf21}
\end{eqnarray}
Graphs $a$ and $b$ are non-ambiguous.  They are the only contributions
in an analytic theory. The other graphs are
\begin{eqnarray}
c &=& 2 \lambda_c R'''(0) R''(0) R'''(u) I_A \label{lf22}\\
 d &=& 2 \lambda_d R'''(0) R''(u) R'''(u) I_A \label{lf23}\\
 e &=& - \lambda_e (R'''(0+))^2 R''(u) I_A \label{lf24}\\
 f &=& 2 \lambda_f R'''(0)^2 R''(u) I_A\label{lf25}
\end{eqnarray}
and vanish if $R(u)$ is analytic (since then $R'''(0)=0$) but a priori
should be considered when $R(u)$ is non-analytic.  We have indicated
their ``natural'' sign and amplitude (e.g.\ symmetry factor setting
$\lambda_i=1$) but have introduced factors $\lambda_{i}$ to recall
that they are {\it ambiguous}: since $R'''(0^+) = - R'''(0^-)$ one is
confronted to a choice each time one saturates a vertex and there is
no obvious way to choose the sign at this stage.  We recall that we
have defined {\em saturated} vertices as vertices evaluated at $u=0$
while {\em unsaturated} vertices still contain $u$ and do not lead to
ambiguities.

At this stage we will not discuss in detail how to give a definite
values to these contributions to disorder.  This will be done in
Section \ref{sec:ambiguities}. We will just use the most reasonable
assumptions, which will be reevaluated, and justified later.
A natural step is  to set
\begin{equation}\label{lf1}
 c = d = 0\ ,
\end{equation}
since  these graphs cannot correct $R(u)$ as they are
{\it odd} functions of $u$, which yields no contribution when inserted
into the action $\sum_{ab} R(u_a - u_b)$.

\subsubsection{Class B}\label{s:ClassB} 
\begin{figure}[t]
\Fig{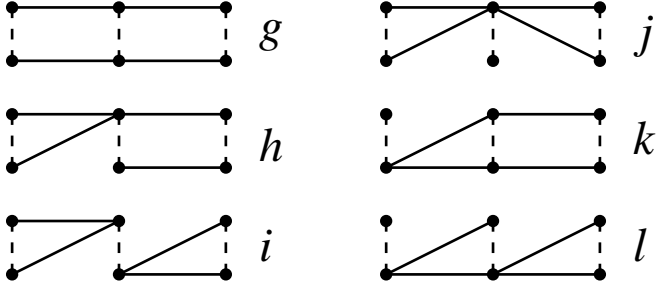}
\caption{2-loop diagrams of class B }\label{banana}
\end{figure}

We now turn to graphs of type B
(bubble-diagrams), $g$ to $l$ represented in Fig.\ \ref{banana}.
We use the same convention as in (\ref{corr}), and start
with the combinatorics.  There are 3 ways to choose the vertex in the
middle.  Upon splitting the vertices, for $i$ and $j$ there are only
two choices at the middle vertex whereas for $g$ there are four
choices. There are also four choices for $h$, $k$ and $l$.  There, one
must also choose the rightmost vertex, leading to an extra factor of
$2$.  The final result is
\begin{eqnarray}
 g &=& \frac{1}{2} R''(u)^2 R''''(u) I_1^2 \label{lf26}\\
 h &=& - R''(u) R''''(u) R''(0) I_1^2 \label{lf27}\\
 i &=& j = \frac{1}{4} R''''(u) R''(0)^2 I_1^2 \label{lf28}\\
 k &=&  - \lambda_k R''(u) R''(0) R''''(0) I_1^2 \label{lf29}\\
 l &=&   \lambda_l R''(u) R''(0) R''''(0) I_1^2 \label{lf30}\ .
\end{eqnarray}
Only $k$ and $l$ are ambiguous but it is also natural to set:
\begin{equation}
\label{lf2}
k+l=0\ ,
\end{equation}
which we do for now, and  discuss  later.

\subsubsection{Class C}
\label{s:ClassC}
\begin{figure}[t]
\centerline{\fig{6cm}{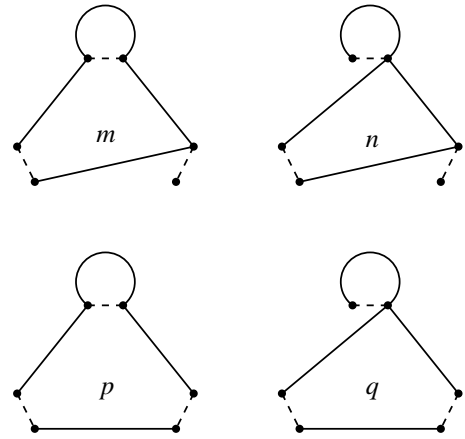}}
\caption{2-loop diagrams of class C}
\label{f:bonetdane}
\end{figure}
Diagrams   $m$, $n$, $p$, $q$ of class C are represented
in Fig.~\ref{f:bonetdane}
\begin{eqnarray}
m &=& c_1 \lambda_m R''(0) R''''(0) R''(u) I_t I_T\label{lf31} \\
n &=& - c_1 \lambda_n R''(0) R''''(0) R''(u) I_t I_T \label{lf32}\\
p &=& c_2 \lambda_p R''''(0) R''(u)^2 I_t I_T \label{lf33}\\
q &=&  - c_2 \lambda_q R''''(0) R''(u)^2 I_t I_T\label{lf34}
\end{eqnarray}
with
\begin{eqnarray}\label{It}
I_t &=& \int_q \frac{1}{q^2 + m^2}\\
\label{IT}
I_T &=& \int_q \frac{1}{( q^2 + m^2)^3} \ .
\end{eqnarray}
There it is natural to assume
\begin{eqnarray}
 m+n&=&0 \label{lf35}\\
p+q&=&0 \label{lf36} \ ,
\end{eqnarray}
which we do for now and discuss it later. This leaves no correction to
disorder from graphs C, as is the case for depinning
\cite{LeDoussalWieseChauve2002}. This is fortunate, since the integral
$I_t$ has a quadratic UV-divergence in $d=4$, while $I_T$ is
UV-finite.  Physically, it is unlikely that these could enter physical
observables as the tadpole divergence can usually be eliminated by
proper field reordering (normal-ordering) or vacuum subtraction.

To summarize, for the equilibrium statics at $T=0$ in perturbation of
$R \equiv R(u)$, the contributions to the disorder to one and two
loops, i.e.\ the corresponding terms in the effective action
$\Gamma[u,\hat u]$ are
\begin{eqnarray}
 \delta^{1} R (u) &=&
\left[ \frac{1}{2} R''(u)^2 - R''(0) R''(u) \right] I_{1} \qquad
\quad \label{d21} \\
\delta^{2} R(u) &=& \left[ R'''(u)^2 ( R''(u) - R''(0) )  \right] I_A
\nn  \\
&& + \frac{1}{2} \left[ (R''(u) - R''(0))^2 R''''(u) \right] I_1^2\nn
\\
&& - \lambda R'''(0^+)^2 R'' (u) I_A    \label{d22}
\ .
\end{eqnarray}
We have allowed for a yet undetermined constant $\lambda = \lambda_e -
2 \lambda_f$.  We now show that requiring renormalizability allows to
fix $\lambda$.

\subsection{Renormalization method to two loops and calculation of
counter-terms} \label{sec:rgdis} Let us now recall the method, also
used in our study of depinning \cite{LeDoussalWieseChauve2002}, to
renormalize a theory where the interaction is not a single
coupling-constant, but a whole function, the disorder-correlator
$R(u)$. We denote by $R_0$ the bare disorder --~this is the object in
which perturbation theory is carried out~-- i.e.\ one considers the
bare action (\ref{action}) with $R \to R_0$.  We denote here by $R$
the renormalized dimensionless disorder i.e.\ the corresponding term
in the effective action $\Gamma[u]$ is $m^{\epsilon} R$ (i.e.\ the
local 2-replica part of $\Gamma[u]$). Symbolically, we can write
\begin{eqnarray}\label{lf105}
 {\cal S}[u]  &\leftrightarrow& R_0 \\
 \Gamma[u] &\leftrightarrow& m^\epsilon R\ .\label{lf106}
\end{eqnarray}
We define the dimensionless symmetric bilinear 1-loop and trilinear
2-loop functions (see (\ref{d21}) and (\ref{d22})) such that
\begin{eqnarray}
 \delta ^{(1)}(R, R) &=& m^{\epsilon} \delta^1 R\label{lf37} \\
 \delta ^{(2)}(R, R, R) &=& m^{\epsilon} \delta^2 R \label{lf38}
\end{eqnarray}
They can be extended to non-equal argument using $f(x,y) :=\frac12
\left[ f(x+y,x+y)-f(x,x)-f(y,y)\right]$ and a similar expression for
the trilinear function. Whenever possible we will use the shorthand
notation $\delta ^{(1)}(R)=\delta ^{(1)}(R, R)$ and $\delta
^{(2)}(R)=\delta ^{(2)}(R, R,R)$.  The expression of $R$ obtained
perturbatively in powers of $R_0$ at 2-loop order reads:
\begin{equation} \label{lf39}
R = m^{-\epsilon} R_0 + \delta ^{(1)}(m^{-\epsilon} R_0)
+ \delta ^{(2)}(m^{-\epsilon} R_0) +
O(R_0^4)\ .
\end{equation}
It contains terms of order $1/\epsilon$ and $1/\epsilon^2$. This is
sufficient to calculate the RG-functions at this order. In principle,
one has to keep the finite part of the 1-loop terms, but we will work
in a scheme, where these terms are exactly 0, by normalizing all
diagrams by the 1-loop diagram.  Inverting (\ref{lf39}) yields
\begin{equation} \label{do}
R_0 = m^{\epsilon} \left[ R - \delta ^{(1)}(R) - \delta
^{(2)}(R) + \delta^{(1,1)}(R)+ \ldots \right]
\ ,
\end{equation}
where $\delta^{(1,1)} (R)$ is the 1-loop repeated counter-term:
\begin{equation}\label{lf41}
\delta^{(1,1)}(R) = 2 \delta ^{(1)}(R, \delta ^{(1)}(R, R))
\ .
\end{equation}
The $\beta$-function is by definition the derivative of $R$ at fixed
$R_0$. It reads
\begin{eqnarray}
-m \partial_{m}R\ts_{R_0} &=& \epsilon  \Big[ m^{-\epsilon}R_0
+ 2\delta ^{(1)} (m^{-\epsilon}R_0 )\nn
\\
&& \ \ \ \ + 3 \delta ^{(2)} (m^{-\epsilon}R_0) + \dots \Big]\ .\qquad
 \label{betar0}
\end{eqnarray}
Using the inversion formula (\ref{do}), the $\beta$-function can be
written in terms of the renormalized disorder $R$:
\begin{eqnarray}
-m\partial_{m}R\ts_{R_{0}} &=&
\epsilon \Big[ R+\delta ^{(1)} (R)
\nonumber \\
&&\quad+2\delta ^{(2)} (R)-\delta ^{(1,1)} (R) + \dots \Big]\ .
\qquad\qquad
\label{beta-renor-DO}
\end{eqnarray}
In order to proceed, let us calculate the repeated 1-loop counter-term
$ \delta^{1,1} (R)$. We start from the 1-loop counter-term
(\ref{d21}), which has the bilinear form
\begin{eqnarray}
\delta^{(1)} (f,g) = - \frac12 \Big[ f''(u) g''(u) &-& f''(0)
g''(u) \nn\\ &-& f''(u) g''(0) \Big] \tilde I_1  \qquad \qquad \label{lf42}
\end{eqnarray}
with the dimensionless integral $\tilde I_{1}:=I_{1}\ts _{m=1}$; we
will use the same convention for $\tilde I_{A}:=I_A\ts_{m=1}$.  Thus $
\delta^{1,1} (R)$ reads
\begin{eqnarray}
&&\hspace{-1cm} \delta^{(1,1)}( R(u) ) =  2
\delta^{(1)}\left(R,\delta^{(1)}(R)\right)\nn\\
\qquad  &=& \Big[ (R''(u) - R''(0)) R'''(u)^2 \nn\\
&&  + (R''(u) - R''(0))^2 R''''(u) - R'''(0^+)^2 R''(u) \Big]
\tilde I_1^2 \nn \ .\\
\label{1-loop-rep-CC}
\end{eqnarray}
In the course of the calculation the only possible ambiguity could
come from
\begin{eqnarray}
 g''(0) &=& \Big[ \frac{1}{2} R''(u)^2 - R''(0) R''(u) \Big]''\Big|_{u \to
0} \nonumber \\
&=& \Big[ R'''(u)^2 - R''''(u) ( R''(u) - R''(0)) \Big]\Big|_{u \to 0}
\nonumber \\ 
&=& 
R'''(0^+)^2 \label{lf43}
\end{eqnarray}
but there is {\it no ambiguity} since the function $R'''(u)^2$ is
continuous at $u=0$ with value $R'''(0^+)^2 = R'''(0^-)^2$. This is
exactly the same calculation as is done to one loop when computing the
non-trivial fixed point for the pinning force correlator
$\Delta(u)=-R''(u)$ yielding $0= (\epsilon - 2 \zeta)
\tilde{\Delta}(0) - \Delta'(0^+)^2 $.  Thus there is no doubt that the
graph $G$ with the 1-loop counter-term inserted in a 1-loop diagram is
{\it non-ambiguous}.

\subsection{Final $\beta$-function, renormalizability and
potentiality} \label{sec:finalbeta} The 2-loop $\beta$-function
(\ref{beta-renor-DO}) then becomes with the help of
(\ref{1-loop-rep-CC}) \begin{eqnarray}
-m \partial_m {{R}}(u) &=&  \epsilon R(u) \nn\\
&& + \left[ \frac{1}{2} {R}''(u)^2 -  {R}''(0) {R}''(u) \right] (\epsilon
\tilde I_{1})  \nonumber \\&&
+  \left[ ({R}''(u) - {R}''(0)) {R}'''(u)^2 \right] \,\epsilon\!
\left( 2\tilde I_{A}-\tilde I_{1}^{2} \right)\nn\\
&& - ({R}'''(0^+))^2 {R}''(u) \, \epsilon\!
\left( 2 \lambda \tilde I_{A}-\tilde I_{1}^{2} \right)
\label{rgdisorderunrescaled}\ .
\end{eqnarray}
The first result is that, apart from the last ``anomalous'' term, the
$1/\E^2$-terms cancel in the corrections to disorder.  In the terms
coming from graphs A this works because, as we recall, $\tilde I_{A} =
(\frac{1}{2 \epsilon^2} + \frac{1}{4 \epsilon^2} + O(\epsilon^2))
(\epsilon \tilde I_{1})^{2}$ so that the combination $\epsilon ( 2
\tilde I_{A}-\tilde I_{1}^{2} )$ is finite. Graphs B cancel completely
since we have chosen as counter-term the full 1-loop graph. So for an
analytic theory the above $\beta$-function would be finite. This
however is incomplete, since the flow of such a $\beta$-function leads
to a non-analytic ${R}(u)$ above the Larkin scale.

Thus we must consider the last, ``anomalous'' term in
(\ref{rgdisorderunrescaled}). It clearly appears that the only value
of $\lambda$ compatible with the cancellation of the $1/\E^2$ poles is
\begin{equation}\label{lf44}
 \lambda = 1\ ,
\end{equation}
leading to a finite $\beta$-function. Thus the requirement that the
theory  be renormalizable (i.e.\ yield universal large scale
results independent of the short-scale details) fixes the value
$\lambda=1$. Note  that the cancellation of the  graphs B also
works thanks to (\ref{lf2}).

It is interesting to compare with what happens at depinning.  There
the cancellation of the $1/\E^2$-terms in the anomalous part is more
complicated but automatic. It requires a consistent evaluation of all
anomalous non-analytic diagrams. In the depinning theory the
cancellation was unusual: a non-trivial bubble diagram (called $i_3$
in \cite{LeDoussalWieseChauve2002}) was crucial in achieving the
cancellation. In the statics the 2-loop bubble diagrams of type B
appear to be simply the square of the 1-loop ones which is the usual
situation. This however is clearly a consequence of (\ref{lf2}) so the
previous experience with depinning indicates that care is required and
we will discuss some justification for (\ref{lf2}) below.

In the search for a fixed point it is convenient to write the
$\beta$-function for the rescaled function $\tilde{R}(u)$ defined
through
\begin{eqnarray}
R(u) = \frac1{\E \tilde I_{1}} m^{- 4 \zeta} \tilde R(u m^{\zeta})
\label{resc} \ ,
\end{eqnarray}
which amounts to rescale the fields $u$ by $m^{\zeta}$. Note that this
is a simple field rescaling and different from standard wave-function
renormalization, since as mentioned above there is none in this theory
due to STS. We have also included the 1-loop integral factor to
simplify notations and further calculations (equivalently it can be
absorbed in the normalization of momentum or space integrals). With
this, the $\beta $-function takes the simple form:
\begin{eqnarray}
-m \partial_m \tilde{R}(u) &=&  (\epsilon - 4 \zeta) \tilde{R}(u)
+ \zeta u \tilde{R}'(u) \nn\\
&& + \left[ \frac{1}{2} \tilde{R}''(u)^2 -  \tilde{R}''(0)
\tilde{R}''(u) \right] \nonumber \\&&
+ \frac{1}{2} X \left[ (\tilde{R}''(u) - \tilde{R}''(0))
\tilde{R}'''(u)^2 \right]  \nn\\
&& - \frac{\lambda}{2} X (\tilde{R}'''(0^+))^2 \tilde{R}''(u)
\label{rgdisorder}\ . \\
 \lambda = 1 \ , &&\qquad X=1\ . \label{lf45}
\end{eqnarray}
We have left a $\lambda$ for future use, but its value in the theory
we study here is set to $1$. Also for convenience we have introduced
\begin{equation}\label{Xdefb}
X= \frac{2 \, \epsilon ( 2 I_A - I_1^2) }{ (\epsilon I_1)^2 }
\end{equation}
which is $X = 1 + O(\epsilon)$ in the $\epsilon$-expansion studied
here, but has a different value for LR elasticity, see below. In fact
it is shown in Appendix \ref{app:universal} that $\lim_{\epsilon \to
0} X$ is independent of the particular infrared cutoff procedure (here
a massive scheme). Although the global rescaling factor of
$\tilde{R}$, $\epsilon \tilde I_1$, has $O(\epsilon)$ corrections
which depend on the infrared cutoff chosen, the FRG equation above
does not depend on it.  Note that the above equation remains true in
fixed dimension, with the appropriate value for $X$, up to terms of
order $\tilde{R}^4$.

We will see that the value $\lambda = 1$ in (\ref{rgdisorder}) has
other highly desirable properties. First this value is {\it the only
one} which guarantees that the non-analyticity in $\tilde{R}(u)$ does
not become {\it more severe} at two loops than it is  at one loop. Let us
take one derivative of (\ref{rgdisorder}) and take $u \to 0^+$. One
finds:
\begin{equation}\label{lf3}
 -m \partial_m \tilde{R}'(0^+) = (\epsilon - 3 \zeta) \tilde{R}'(0^+)
+ \frac{1}{2} (1 - \lambda) \tilde{R}'''(0^+)^3 \ .
\end{equation}
Thus if $\lambda \neq 1$ the cusp in $\tilde{R}''$ and the resulting
finite value of $R''' (0^{+})$ immediately creates a cusp in
$\tilde{R}'$. The singularity has become worse! We call this a
supercusp. It must be avoided in the statics (see also discussion in
Section \ref{sec:ambiguities}).  Interestingly it {\it does occur in
the driven dynamics}, where it is a physical signature of
irreversibility.

Indeed this property is intimately related to another highly desirable
property of the statics: {\it potentiality}.  This property is more
conveniently described by considering the flow equation for the
(rescaled) {\it correlator of the pinning force} $\tilde \Delta(u) = -
\tilde{R}''(u)$, the second derivative of (\ref{rgdisorder}):
\begin{eqnarray}
-m \partial_m \tilde{\Delta}(u) &=& (\epsilon - 2 \zeta)
\tilde{\Delta}(u)
+ \zeta u \tilde{\Delta}'(u) \nn\\
&& - \frac{1}{2} \left[ (\tilde{\Delta}(u)^2 - \tilde{\Delta}(0))^2
\right]'' \nonumber \\&&
+ \frac{1}{2} \left[ (\tilde{\Delta}(u) - \tilde{\Delta}(0))
\tilde{\Delta}'(u)^2 \right]'' \nn\\ 
&& - \frac{\lambda}{2} (\tilde{\Delta}'(0^+))^2 \tilde{\Delta}''(u)
\label{rgdisorderdelta}\ .
\end{eqnarray}
Formally, this equation could have been obtained directly from a study
of the dynamical field theory. Such an equation was indeed obtained at
depinning but with a different value of $\lambda$:
\begin{equation}\label{lf46}
 \lambda_{\mathrm{dep}} = - 1\ ,
\end{equation}
which shows that statics and dynamics differ not at one, but at two
loops. Integrating the equation for $\Delta(u)$ once yields a non-zero
fixed point value for $\int \Delta(u)$ unless
$\lambda=1$. Potentiality on the other hand requires that the force
remains the derivative of a potential and that, for short-range
disorder (e.g.\ RB for interface) one must have $\int \Delta(u) =
0$. While violating potentiality is desirable at depinning where
irreversibility is expected, this would be physically incorrect in the
statics, and thus again points to the value $\lambda=1$ as the
physically correct one.

Thus we will for now assume that this is the correct theory of the
statics and explore its consequences in the next section. In section
\ref{sec:ambiguities} we will provide better justifications, and
explain our understanding of the tantalizing problem of ambiguous
diagrammatics in the non-analytic theory of pinned disordered
systems. Especially we will present methods, which satisfy all the
above constraints of renormalizability, absence of a supercusp and
potentiality up to 3-loop order \cite{LeDoussalWiesePREPb}.

\section{Analysis of fixed points and physical results}
\label{sec:fixedpoints} The FRG-equation derived above describes
several different physical situations, and admits a small number of
fixed-point functions $\tilde R^*(u)$ describing a few universality
classes. The fixed point associated to a periodic disorder correlator
describes single component periodic systems (such as charge density
waves).  The fixed point associated to a short-range (exponentially
decaying) correlator $\tilde R(u)$ describes a class of systems with
so called random bond disorder.  There is also a family of fixed
points associated to long range, i.e.\ algebraic, correlations. This
includes, as one particular example, the random field disorder, which
will be discussed separately.

We now give the results for these fixed points, first for short-range
elasticity, then for LR elasticity, and compare with available
numerical and exact results. The most important quantity to compute is
the roughness exponent $\zeta$. Since we have shown that $X$ in
(\ref{rgdisorder}) is universal to dominant order this proves
universality of $\zeta$ to the order in $\epsilon$ studied here (i.e.\
$O(\epsilon^2)$).  For LR disorder and for periodic fixed points we
can also compute the universal amplitudes for the correlation function
of displacements, and discuss their dependence on 
large scale boundary conditions. Anticipating a bit, let us summarize the general
result that we use in that case, which is derived in Section
\ref{sec:correlations}.  The $T=0$ disorder-averaged 2-point function
for $q \to 0$, $q/m$ fixed, reads for any dimension $d$, in Fourier
\begin{eqnarray} \label{lf107}
\overline{u_{q} u_{q'}} &=& (2 \pi)^d \delta^d(q+q') C(q) \\
 C(q) &=& C(q=0) F_d(q/m) \label{lf108} \\
 C(q=0) &=& \tilde{c}(d) m^{-d - 2 \zeta} \label{lf109} 
\end{eqnarray}
The amplitude $\tilde{c}(d)$ is given by the relation (exact to all
orders in the present scheme):
\begin{eqnarray}
 \tilde c(d) &=& - \frac{1}{(\epsilon \tilde{I}_1)} \tilde{R}^{*
\prime \prime}(0) \ , 
\label{ctd}
\end{eqnarray}
It is found to be universal only for long range and periodic
disorder. The scaling function, computed in Section \ref{sec:correlations}
for SR and LR elasticity, is always universal (independent of
short scale details) and satisfies $F_d(0) = 1$ and
\begin{eqnarray} \label{lf110}
F_d(z) &\sim& B z^{-(d + 2 \zeta)}  \qquad \mbox{for}~z\to\infty\\
 B &=& 1 + b \epsilon + O(\epsilon^2) \label{lf111}
\end{eqnarray}
where $b$ is computed in Section \ref{sec:correlations}. This gives us
all we need for a calculation to $O(\epsilon^2)$ of the universal
amplitude, e.g for the propagator in the massless limit $m \ll q$:
\begin{eqnarray}
 C(q) &=& c(d) q^{-(d+2 \zeta)} \label{fdmassless} \\
 c(d) &=& \tilde c(d) (1 + b \epsilon + O(\epsilon^2)) \label{cd}
\end{eqnarray}
The result for $C(q=0)$ in presence of a mass is also interesting
since it gives the fluctuations of the center of mass coordinate for
an interface physically confined in a quadratic well. Although that
situation would be interesting to study numerically, most numerical
results are for finite size systems of volume $L^d$ (and $m \to
0$). We thus also define in that case:
\begin{equation}\label{finitesizeg}
  C_L(q) = c'(d) q^{-(d+2 \zeta)} g_d(q L)
\end{equation}
with $\lim_{z \to \infty} g_d(z) = 1$. For {\it periodic} boundary
conditions $q=2 \pi n/L$, $n \in \mathbb{Z}^d$ and $n \neq 0$. The
prime indicates that the value of this amplitude depends on the large
scale boundary conditions, i.e.\ it depends on whether e.g.\ a mass is
used or periodic boundary conditions as an infrared cutoff. The ratio,
computed in Section \ref{sec:correlations} for short range elasticity,
\begin{equation}\label{ratfin}
\frac{c'(d)}{c(d)} = 1 - 1.46935 \zeta  + O(\epsilon^2)\ ,  
\end{equation}
is unity only for periodic disorder, in which case the amplitude is
independent of both large and small scale details.

Before studying the different fixed points, let us mention an
important property, valid under all conditions: If $\tilde{R}(u)$ is
solution of (\ref{rgdisorder}), then
\begin{equation}\label{DeltaRescale}
\hat{R} (u):= \kappa ^{4} \tilde R (u/\kappa )
\end{equation}
is also a solution (for $\kappa$ a constant independent of $m$).  We
can use this property to fix $\tilde{R}(0)$ or $\tilde R''(0)$ in the
case of non-periodic disorder.  (For periodic disorder the solution is
unique, since the period is fixed.)

\subsection{Non-periodic systems: Random bond disorder}
\label{sec:rbsystems} Let us now look for a solution of our 2-loop FRG
equation which decays exponentially fast at infinity as expected for
SR random-bond disorder.  To this aim, we have to solve order by order
in $\epsilon$ the fixed-point equation (\ref{rgdisorder}) numerically.
Making the ansatz
\begin{eqnarray}\label{ansatzRB}
\tilde{R} (u) &=& \epsilon r_{1} (u) + \epsilon^{2} r_{2} (u) + \dots \\
\zeta &=& \epsilon \zeta_{1} + \epsilon^{2}\zeta_{2} + \dots\ ,
\label{lf112}
\end{eqnarray}
the partial differential equation to be solved at leading order is
\begin{figure}
\Fig{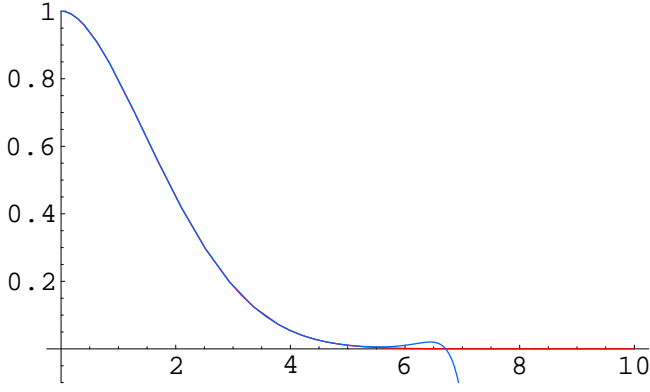} \caption{The fixed-point function $r_{1} (u)$ at
1-loop order. We have plotted a numerical solution (red) as well as
the Taylor-expansion (\ref{taylor1}) about 0 up to order 25 (blue).}
\label{f:Rranbond1loop}
\end{figure}%
\begin{eqnarray}
0&=& (1-4 \zeta_{1})r_{1} (u) + \zeta_{1} u r'_{1} (u) +\frac{1}{2}r''_{1}
(u)^{2}-r''_{1} (u)r''_{1} (0)\nn  \\
1&=& r_{1} (0) \ ,
\label{RRB1loop}
\end{eqnarray}
where we have used our freedom to normalize $\tilde{R} (0):=\epsilon$.
(\ref{RRB1loop}) has a solution for any $\zeta_{1}$, but only for one
specific value of $\zeta_{1}$ does this solution decay exponentially
fast to 0, without crossing the axis, see figure
\ref{f:Rranbond1loop}.  The strategy is thus the following: One guesses
$\zeta_{1}$, and then integrates (\ref{RRB1loop}) from 0 to
infinity. In practice, however, there are numerical problems for small
$u$. One strategy, which we have adopted here, and which works very
well, is to use the value of $\zeta_{1}$, to generate a
Taylor-expansion about 0. This Taylor-expansion is then evaluated at
0.5, where the numerical integration of (\ref{RRB1loop}) is started,
both forwards to infinity (which in practice is chosen to be 25) and
backwards to 0. This enables to control the accuracy of both the
Taylor-expansion and the numerical integration. The result for the
best value
\begin{equation}\label{lf48}
\zeta_{1} = 0.20829806 (3)
\end{equation}
is given on figure \ref{f:Rranbond1loop}. (Note that in
\cite{DSFisher1986} only the first four digits were given.)  On this
scale, Taylor-expansion and numerical integration are
indistinguishable. The error-estimate on the last digit comes from
moving the starting-point of the numerical integration (which was 0.5
above) up to 1, which allows for a crude estimate of the error.  We
also reproduce the Taylor-expansion up to order 25 below:
\begin{eqnarray}\label{taylor1}
r_{1} (u) &=& 1{-}0.288797\,u^2{+}0.0967487\,u^3{-}0.0109959\,u^4\nn \\
&&{+}0.000197282\,u^5{+}0.0000162077\,u^6\nn \\
&&{+}1.37054\,{10}^{{-}6}\,u^7{+}1.06127\,{10}^{{-}7}\,u^8\nn \\
&&{+}5.84538\,{10}^{{-}9}\,u^9{-}1.50021\,{10}^{{-}10}\,u^{10}\nn \\
&&{-}1.19821\,{10}^{{-}10}\,u^{11}{-}2.52931\,{10}^{{-}11}\,u^{12}\nn \\
&&{-}3.93584\,{10}^{{-}12}\,u^{13}{-}4.90717\,{10}^{{-}13}\,u^{14}\nn \\
&&{-}4.49154\,{10}^{{-}14}\,u^{15}{-}1.21758\,{10}^{{-}15}\,u^{16}\nn \\
&&{+}6.77579\,{10}^{{-}16}\,u^{17}{+}2.11465\,{10}^{{-}16}\,u^{18}\nn \\
&&{+}4.19348\,{10}^{{-}17}\,u^{19}{+}6.49482\,{10}^{{-}18}\,u^{20}\nn \\
&&{+}7.78044\,{10}^{{-}19}\,u^{21}{+}5.52691\,{10}^{{-}20}\,u^{22}\nn \\
&&{-}4.37557\,{10}^{{-}21}\,u^{23}{-}2.72231\,{10}^{{-}21}\,u^{24}\nn \\
&&{-}6.74331\,{10}^{{-}22}\,u^{25} + O (u^{26})\ .
\end{eqnarray}
At second order in $\epsilon $, we have to solve
\begin{eqnarray}\label{lf49}
0&=&  r_{2} (u) - 4 \zeta_{2}r_{1} (u)  -4 \zeta_{1} r_{2} (u) + u
\zeta_{2}r_{1}' (u) + u \zeta_{1} r_{2}' (u) \nn \\
&& + r_{1}'' (u)r_{2}'' (u)
-  r_{1}'' (0)r_{2}'' (u)- r_{1}'' (u)r_{2}'' (0) \nn\\
&& + \frac{1}{2} \left(r_{1}'' (u)-r_{1}'' (0) \right) r_{1}'''
(u)^{2} -\frac{1}{2} r_{1}'' (u)r_{1}''' (0^{+})^{2}\\
0&=& r_{2} (0)\ , \label{lf113}
\end{eqnarray}
where the last equation reflects our choice of $\tilde{R} (0) =
\epsilon$.  Note that to solve the 2-loop order equation, one has to
feed in the solution at 1-loop order, both the Taylor-expansion about
0 and the numerically obtained solution for larger $u$.
\begin{figure}
\Fig{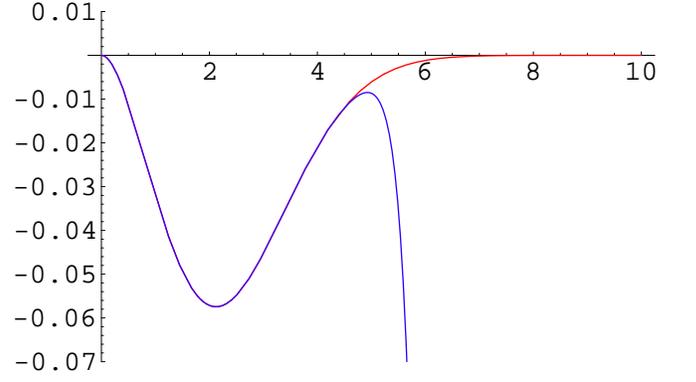} \caption{The fixed-point function $r_{2} (u)$ at
2-loop order. We have plotted a numerical solution (red) as well as
the Taylor-expansion (\ref{taylor2}) about 0 up to order 25 (blue).}
\label{f:Rranbond2loop}
\end{figure}%
Again $\zeta_{2}$ is determined from the condition that
the solution decays at infinity. Following the same procedure as at
1-loop order, we find
\begin{equation}\label{zeta2}
\zeta_{2} = 0.006858 (1)\ .
\end{equation}
The function $r_{2}$ is plotted on figure~\ref{f:Rranbond2loop}. The
Taylor-expansion up to order 25 about 0 reads
\begin{eqnarray}\label{taylor2}
r_{2} (u)&=&
{-}0.0604942\,u^2{+}0.0345276\,u^3{-}0.00628098\,u^4\nn \\
&& {+}0.000239628\,u^5{+}0.000019823\,u^6\nn \\
&&{+}1.42202\,{10}^{{-}6}\,u^7{+}5.17941\,{10}^{{-}8}\,u^8\nn \\
&&{-}8.64456\,{10}^{{-}9}\,u^9{-}2.72755\,{10}^{{-}9}\,u^{10}\nn \\
&&{-}4.78607\,{10}^{{-}10}\,u^{11}{-}6.23531\,{10}^{{-}11}\,u^{12}\nn \\
&&{-}5.49541\,{10}^{{-}12}\,u^{13}{-}8.78473\,{10}^{{-}15}\,u^{14}\nn \\
&&{+}1.30232\,{10}^{{-}13}\,u^{15}{+}3.60568\,{10}^{{-}14}\,u^{16}\nn \\
&&{+}6.7239\,{10}^{{-}15}\,u^{17}{+}9.51299\,{10}^{{-}16}\,u^{18}\nn \\
&&{+}9.06111\,{10}^{{-}17}\,u^{19}{-}9.06201\,{10}^{{-}20}\,u^{20}\nn \\
&&{-}2.59561\,{10}^{{-}18}\,u^{21}{-}7.67911\,{10}^{{-}19}\,u^{22}\nn \\
&&{-}1.53922\,{10}^{{-}19}\,u^{23}{-}2.36569\,{10}^{{-}20}\,u^{24}\nn \\
&&{-}4.42973\,{10}^{{-}21}\,u^{25}+O (u^{26})\ .
\end{eqnarray}
One observes that $\zeta_{\mathrm{SR}}$ is necessarily bounded
from above by $\epsilon/4$ as no SR solution can cross this value
(to any order) without exploding. This reflect the exact bound for
SR disorder $\theta < d/2$, which simply means that optimization
of energy must lower energy fluctuations compared to a simple sum
of random numbers. Equality is obtained for the trivial constant
eigenmode $\tilde{R}(u)=\tilde{R}(0)$ corresponding to
$\zeta=\epsilon/4$, associated to the fluctuation of the zero mode
of the random potential.

We can now discuss our results for the roughness exponent. These are
summarized in Table \ref{table} and compared to numerical simulations
in $d=3,2$ and the exact result for the directed polymer in $d=1$. A
first observation is that the corrections compared to the 1-loop
result have the correct sign and, further, that they improve the
precision of the 1-loop result. Given the difficulties associated with
this theory, this is a significant achievement. Second, the error bars
given in Table \ref{table} are estimated as half the 2-loop
contribution, which should not be taken too literally, as it is
difficult to obtain a good precision from only two terms of the series
and no currently available information about the large order behavior
of this novel $\epsilon$-expansion. Third, one may try to improve the
precision using the exact result $\zeta =2/3$ in $d=1$. Estimating the
third order correction in the three possible Pade's in order to match
$\zeta=2/3$ for $\epsilon=3$, we obtain consistently the values quoted
in the fourth column of Table \ref{table}. We hope that these
predictions can be tested in higher precision numerics soon.

\begin{figure*}[bt]
\begin{tabular}{||c|c|c|c|c|c||}
\hline
$\zeta _{\rm eq}$ & one loop & two loop & estimate & improved estimate &
simulation and exact\\
\hline \hline $d=3$  & 0.208 &  0.215  & $0.215 \pm 0.004$  &
$0.214$ &
$0.22\pm 0.01$ \cite{Middleton1995}  \\
\hline
$d=2$ &0.417 &0.444 &$0.444 \pm 0.015$ & $0.438$ & $0.41\pm 0.01$ \cite{Middleton1995} \\ 
\hline
$d=1$ & 0.625 & 0.687 & $0.687 \pm 0.03$ & 2/3 & 2/3 \cite{KardarHuseHenleyFisher1985} \\
\hline
\end{tabular}
\vspace{1mm} \caption{First column: Exponents obtained by setting
$\epsilon =4-d$ in the 1-loop result. Second column: Exponents
obtained by setting $\epsilon=4-d$ in the 2-loop result. Third column:
errors bars are estimated as half the 2-loop contribution. Fourth
column: Improved estimates using the exact result $\zeta _{\rm
eq}=2/3$ in $d=1$ (see text).} \label{table}
\end{figure*}

\subsection{Non-periodic systems: random field disorder}
\label{sec:rfsystems}

Let us first recall that at the level of the {\em bare} model the
static random field disorder correlator obeys $\tilde R(u) \sim -
\tilde{\sigma} |u|$ at large $|u|$
\cite{DSFisher1986,ChauveLeDoussal2001}, where $\tilde{\sigma}=
(\epsilon \tilde I_1) \sigma$ is proportional to the amplitude of the
random field. 

If one studies the large $u$ behavior in the FRG equation
(\ref{rgdisorder}) one clearly sees that the non-linear terms do not
contribute, thus one has:
\begin{equation}\label{lf50}
- m \partial_m \tilde \sigma = (\epsilon - 3 \zeta) \tilde \sigma\ .
\end{equation}
Thus for a RF fixed point to exist,  the $O(\epsilon^2)$
correction to $\zeta$ has to vanish.
\begin{equation}\label{lf51}
\zeta_{\mathrm{RF}} = \epsilon/3\ .
\end{equation}
This will presumably hold to all orders. Indeed it is clear that if
there is a similar $\beta$-function to any order, since each $R$
carries at least two derivatives and at least one must be evaluated at
$u \neq 0$, the sum of all non-linear terms to a given {\it finite}
order decreases at least as $R''(u) \sim 1/u$.  (This does not
strictly excludes that summing up all orders may yield a slower decay,
although it appears far fetched and does not occur in the
non-perturbative large-$N$ limit.) The above value of $\zeta$ ensures
that $m^\epsilon R(u) \sim - \sigma |u|$ in the effective action,
i.e.\ non-renormalization of $\sigma$.

Note that this argument based on long-range large $u$ behavior is a
priori valid for any $\lambda$. Since it is made on the $R$ equation
(no such argument can be made on the equation for $\Delta$) it uses
the property of potentiality. However, from (\ref{lf3}) with
$\zeta=\epsilon/3$ one sees that $\lambda \neq 1$ is incompatible with
the existence of a fixed point, even of a fixed point with a
supercusp. Thus, the only way to satisfy potentiality for the static
random field problem seems to have $\sigma$ unrenormalized,
$\zeta=\epsilon/3$ and $\lambda=1$ (the previous discussion of
potentiality in Section III.D assumed short-range disorder).

This must be contrasted with the theory of depinning, where we found
that:
\begin{equation}\label{lf69}
 \zeta_{\mathrm{dep}} = \frac{\epsilon}{3} (1 + 0.14331 \epsilon)
\end{equation}
following from $\lambda_{\mathrm{dep}}=-1$ in
(\ref{rgdisorderdelta}). Since in that case the RG-flow is
non-potential, it is clear that no similar argument as above exists to
protect the value $\zeta = \epsilon/3$. (The force correlator is short
range).  The conjecture of \cite{NarayanDSFisher1993a} thus appears
rather unphysical in that respect.

\subsubsection{Fixed-point function} We first study the fixed-point
equation for
\begin{eqnarray}\label{lf114}
 \tilde \Delta(u) &=& - \tilde R''(u) = \frac{\E}{3} y(u)  \\
 y(0)&=& 1\label{lf115}
\end{eqnarray}
and later use the rescaling freedom to tune the solution to the
correct value of $\sigma$ at large scale $u$.

The 2-loop FRG equation (\ref{rgdisorderdelta}) becomes ($\lambda=1$):
\begin{eqnarray} \label{lf116}
0 &=& (u y)' - \frac{1}{2} ((y-1)^2)'' \\
&& + \frac{\epsilon}{3} \left[
\frac{1}{2} ({y'}^2 (y-1))'' - \frac{1}{2} y'(0^+)^2 y''\right] \nonumber \ .
\end{eqnarray}
One can then integrate once with respect to $u$:
\begin{eqnarray}\label{lf117}
0 &=& u y - y' (y-1)  \\
&& + \frac{\epsilon}{3} \left[
\frac{1}{2} ({y'}^2 (y-1))' - \frac{1}{2} y'(0^+)^2 y' \right] \nonumber\ .
\end{eqnarray}
There is no integration constant here because the second line
precisely vanishes at $u=0^+$ (absence of supercusp).

The 1-loop solution involves the first line only. Dividing by $y$
and integrating over $u$ yields:
\begin{equation} \label{IV.12}
\frac{u^2}{2} = y_1 - 1 - \ln y_1\ ,
\end{equation}
i.e.\ an implicit equation for $y$, which defines $y=y_1(u)$. It satisfies
\begin{eqnarray}
y_{1} (0)&=&1\ , \qquad y_{1}' (0^{+})= -1\nonumber \\
 y_{1}'' (0^{+})&=&
\frac{2}{3}\ , \qquad  y_{1}''' (0^{+})= -\frac{1}{6}\ .
\label{yDerivatives}\label{IV.13}
\end{eqnarray}
We can put the 2-loop solution under a similar form. Making the ansatz
\begin{equation}\label{k1}
\frac{u^2}{2} = y - 1 - \ln y - \frac{\epsilon}{3} F(y)\ ,
\end{equation}
one obtains
\begin{equation}\label{lf55}
F (y (\bar u)) = \frac{1}{2}\int_{0}^{\bar u} \frac{\rmd  u}{y }
\left(y'^{2} (y-1)-y \right)'  \ .
\end{equation}
At this order, one can replace $y$ by $y_1$, i.e.\ use $ u y =
y'(y-1)$ to eliminate $y'$. This gives,
changing variables from $u$ to $y$:
\begin{equation}\label{lf57}
F(\bar y) = \half \int_1^{\bar y} \rmd  y\, \frac{1}{ y}
\frac{\rmd}{\rmd  y} \left(\frac{ y^2 [u( y)]^2}{ y-1} -y \right)\ .
\end{equation}
The last term in the brackets is easily integrated. For the remaining
terms, we integrate by part and use (\ref{k1}) to replace $u^{2}/2$ by
$y-1-\ln y$:
\begin{equation}\label{lf58}
F(\bar y) =  \int_1^{\bar y} \rmd  y\, \frac{y-1-\ln y}{y-1} +
\frac{\bar y \left(\bar y -1 -\ln \bar y \right)}{\bar y-1} -\half \ln
\bar y\ .
\end{equation}
This yields the final result
\begin{eqnarray} \label{fy}
F (y) &=& 2 y -1 +\frac{y\ln y}{1-y} -\frac{1}{2} \ln y +
\mbox{Li}_{2} (1-y)\qquad \label{lf118} \\
\mbox{Li}_{2} (z)&:=& \int_{z}^{0} \rmd t\, \frac{\ln (1-t)}{t} =
\sum_{k=1}^{\infty} \frac{z^{k}}{k^{2}} \ . \label{lf119}
\end{eqnarray}
We find:
\begin{equation}\label{lf59}
 F(y) = \frac{2}{3} (y-1)^2 - \frac{13}{36} (y-1)^3 + O((y-1)^4)
\end{equation}
has a quadratic behavior around $y=1$, similar to the 1-loop result,
and corrects the value of the cusp.

\subsubsection{Universal amplitude} \label{ss:Universal amplitude}
Since we know the exact fixed point function up to a scale factor, we
can now fix the scale by fitting the exact large $|u|$ behavior to
$R(u) \sim - \sigma |u|$ where $\sigma$ is the amplitude of the random
field. The general fixed point solution reads:
\begin{equation}\label{lf120}
\tilde \Delta(u) = \frac{\epsilon}{3} \xi^2 y(u/\xi)\ ,
\end{equation}
where $\xi$ can be related to $\sigma$ as:
\begin{equation}\label{lf121}
\tilde \sigma = \int_0^\infty \rmd u\, \tilde \Delta(u) =
\frac{\epsilon}{3} \xi^3 I_y\ .
\end{equation}
We need
\begin{eqnarray}\label{lf122}
 I_y &=& \int_0^\infty \rmd u\, y(u) = \int_0^1 \rmd y\, u(y)\nonumber  \\
&& = \gamma_1 + \epsilon \gamma_2 \label{lf123}\\
\gamma_1 &=& \int_0^1 \rmd y\, \sqrt{2 ( y - 1 - \ln y) }\nonumber \\
& =& 0.775304245188  \label{lf124}\\
 \gamma_2 &=& - \int_0^1 \rmd y\, \frac{F(y)}{3 \sqrt{2 (y - 1 - \ln
y)}} \nonumber \\
&=& - 0.13945524\ .\label{lf125}
\end{eqnarray}
One can now express
\begin{equation}\label{lf126}
\tilde \Delta^*(0) = \frac{\epsilon}{3} \xi^2 = \frac{\epsilon}{3}
\left(\frac{3 \tilde \sigma}{\epsilon}\right)^{2/3} I_y^{- 2/3}
\end{equation}
and thus compute, using (\ref{ctd}) the universal amplitude
(\ref{lf109}) associated to the mode $q=0$ in presence of a confining
mass:
\begin{eqnarray}\label{lf127}
\tilde c(d) &=& \sigma^{\frac{2}{3}}
\left(\frac{\epsilon}{3}\right)^{\frac{1}{3}} (\gamma_1 + \epsilon
\gamma_2)^{-\frac{2}{3}}
(\epsilon \tilde{I}_1)^{- \frac{1}{3}} \\
& =& \left(\frac{\epsilon}{3}\right)^{\frac{1}{3}} \left(\gamma_1 +
\epsilon \gamma_2\right)^{-\frac{2}{3}} \left[\frac{\epsilon
\Gamma(\frac{\epsilon}{2})}{ (4 \pi)^{d/2}} \right]^{- \frac{1}{3}}
\sigma^{\frac{2}{3}} \nonumber \ ,\label{lf128}
\end{eqnarray}
where one has restored the factors $\epsilon \tilde{I}_1$ absorbed in
$\tilde \Delta$ and $\tilde \sigma$. Expanding all factors in a series
of $\epsilon$ one finds:
\begin{equation}\label{lf70}
\tilde c(d) = \epsilon^{1/3} ( 3.52459 - 0.725079 \epsilon +
O(\epsilon^2)) \sigma^{2/3} \ ,
\end{equation}
The lowest order was obtained in Ref.~\cite{ChauveLeDoussal2001} and
we have obtained here the next order corrections. It is interesting to
compare our result with the exact result in $d=0$, which is
\cite{LeDoussalMonthus2003}:
\begin{equation}\label{lf129}
\tilde c(d=0) = 1.05423856519\dots  \sigma^{2/3}\ .
\end{equation}
While the simple extrapolation setting $\epsilon=4$ of (\ref{lf70}) to
one loop $\tilde c(d=0) = 5.59 \sigma^{2/3}$ is very far off, to two
loop it gives $\tilde c(d=0) = 0.99 \sigma^{2/3}$, surprisingly close
to the exact result. It was noted in Ref.~\cite{ChauveLeDoussal2001}
that extrapolation of the 1-loop result could be considerably improved
by not expanding (\ref{lf127}) in $\epsilon$ but instead directly
setting $\epsilon=4$ (with $\gamma_2 = 0$) in (\ref{lf127}). That
gives $\tilde c_1(d=0) = 0.821 \sigma^{2/3}$, an underestimate already
reasonably close from the exact result. We extend this procedure to
two loop by truncating the $\epsilon$ expansion of $I_y^{-2/3}$ to
second order in (\ref{lf127}), and then set $\epsilon=4$. This yields
$\tilde c_2(d=0) = 1.22 \sigma^{2/3}$, and the exact result is then
halfway between $\tilde c_1(d=0)$ and $\tilde c_2(d=0)$.  To
summarize, our 2-loop corrections (\ref{lf128}) have the correct sign
and order of magnitude to improve the agreement with the exact result
in $d=0$.

The universal amplitude for the massless case (\ref{fdmassless}) (or
$q \gg m$) is obtained from (\ref{cd}) with $b=-1/3$ from Section
\ref{sec:correlations} as:
\begin{eqnarray}\label{lf131a}
c(d) &=& \tilde{c}(d) \left[1 - \frac{1}{3} \epsilon +
O(\epsilon^2)\right] \nonumber   \\
&=& \epsilon^{1/3} \left[  3.52459 - 1.89994 \epsilon + O(\epsilon^2) \right]
\sigma^{2/3}\ ,\qquad 
\end{eqnarray}
and writing $c(d)=\tilde{c}(d)/(1 + \frac{1}{3} \epsilon)$ should
provide a reasonable extrapolation to low dimensions. Finally,
we recall that for random field disorder, this coefficient is
different for different large scale boundary conditions. The result
for periodic boundary conditions can be obtained from formula (\ref{ratfin}). 

In Ref.~\cite{ChauveLeDoussal2001}, the 1-loop result was compared to
the result of the Gaussian Variational Method (GVM). It is instructive
to pursue this comparison to two loops. We get from
\cite{ChauveLeDoussal2001}:
\begin{eqnarray}\label{rfvar}
 \tilde c_{\mathrm{GVM}}(d) & = & \left(2
\frac{\epsilon}{\pi}\right)^{\frac{1}{3}} \left[\frac{\epsilon
\Gamma(\frac{\epsilon}{2})}{
(4 \pi)^{d/2}} \right]^{- \frac{1}{3}} \frac{1}{1 - \frac{\epsilon}{12}} 
\sigma^{\frac{2}{3}}  \\
& = &\epsilon^{1/3} ( 3.69054 - 0.894223 \epsilon + O(\epsilon^2) )
\sigma^{2/3} \nonumber \\ 
\frac{c_{\mathrm{GVM} }(d)}{\tilde c_{\mathrm{GVM}}(d)} &=& \left( 1 -
\frac{\epsilon - 2 \zeta}{2} \right)\!\! \left( 1 - \frac{\epsilon - 2
\zeta}{4} \right)
\frac{\pi \left(\epsilon - 2 \zeta\right)/2}
{\sin(\pi (\epsilon - 2 \zeta)/2)}  \nonumber \\ 
&=& 1 - \frac{\epsilon}{4} + O(\epsilon^2)
\end{eqnarray}
where in the last line we have inserted $\zeta=\epsilon/3$ and
performed the $\epsilon$ expansion. Thus one finds, quite generally
that $b_{var} = 3 b/4$. As noted in \cite{ChauveLeDoussal2001} to one
loop the FRG and the GVM give rather close amplitudes (differing by
about 5 per cent). We see here that to two loop, i.e.\ next order in
$\epsilon$, the difference increases. Finally,
\begin{equation}\label{lf131b}
 c_{\mathrm{GVM}}(d) = \epsilon^{1/3} ( 3.69054 - 1.81686 \epsilon +
O(\epsilon^2) ) \sigma^{2/3} \nonumber
\end{equation}
and the coefficient remains rather close to the one in (\ref{lf131a}). 

\begin{figure}[t]
\centerline{\Fig{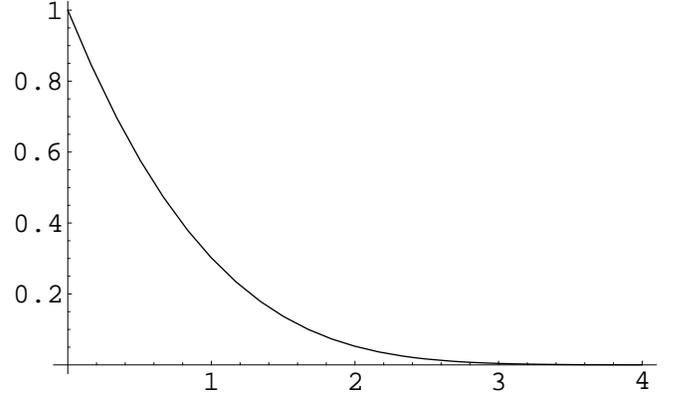}}
\caption{The fixed-point function $y_{1} (u)$ at 1-loop order for non-periodic
disorder.}
\label{fig:RFFP1loop}
\end{figure}%

\subsection{Generic long range fixed points} \label{sec:genlongrange}
There is a family of fixed points such that
\begin{equation}\label{lf132}
\tilde R(u) \sim  |u|^{2 (1-\gamma)}
\end{equation}
associated with
\begin{equation}\label{lf133}
\zeta = \frac{\epsilon}{2 (1 + \gamma)} \ .
\end{equation}
These fixed points where found for infinite $N$ in any $d$ in
Ref.~\cite{MezardParisi1991,LeDoussalWiese2001} (we use the same
notations). They were studied to first order in $\epsilon$ for any $N$
in \cite{BalentsDSFisher1993}, and argued to be stable only for
$\gamma < \gamma^*(d)$ the value of the crossover to short range
identified in \cite{BalentsDSFisher1993} as
$\zeta_{\mathrm{SR}}=\zeta_{\mathrm{LR}}(\gamma^*(d))$.

Here, we have not studied these fixed points in detail but we note
that the 2-loop corrections do not change $\zeta$, by the same
discussion as for the random field case $\gamma=1/2$. They will
however affect the amplitudes.

\subsection{Periodic systems}
\label{sec:periodic}

\subsubsection{Fixed point function} For periodic $R(u)$ as e.g.\ CDW
there is another fixed point of (\ref{rgdisorder}). It is sufficient
to study the case where the period is set to unity, all other cases
are easily obtained using the reparametrization invariance of equation
(\ref{DeltaRescale}).  No rescaling is possible in that direction, and
thus the roughness exponent is
\begin{equation} \label{lf52}
\zeta =0\ .
\end{equation}
The fixed-point function is then periodic, and can in the interval
$\left[0,1 \right]$ be expanded in a Taylor-series in $u (1-u)$. Even
more, the ansatz
\begin{equation} \label{lf53}
 \tilde R (u) = (a_{1}\epsilon +a_{2}\epsilon ^{2}+ \dots ) +
\left(b_{1}\epsilon +b_{2}\epsilon ^{2}+\dots \right) u^2 (1-u)^2
\end{equation}
allows to satisfy the fixed-point equation (\ref{rgdisorder}) to order
$\E^2$ and will presumably work to all orders. For a more general case
of this see Ref.~\cite{LeDoussalWiese2002a}.

To gain insight into the more general case, let us write the fixed
point for (\ref{rgdisorder}) with arbitrary $\lambda$:
\begin{eqnarray} \label{DeltaPeriodicXalpha}
\tilde R^*(u) &=&
 \frac{\epsilon}{2592} + (3 - 2 \lambda) \frac{\epsilon^2}{7776} \nonumber \\
&& + (\lambda - 1) \frac{\epsilon^2}{432} u (1-u)  \nonumber \\
&& - \left( \frac{\epsilon}{72} + \frac{\epsilon^2}{108} \right) u^2 (1-u)^2\ .
\end{eqnarray}
One can see on this solution that $\lambda=1$ is the only value which
avoids the appearance at two loops of the supercusp, i.e.\ a cusp in
the potential correlator $\tilde R(u)$ rather than in the force
correlator $\tilde \Delta(u)$.

The same discussion can be made on the the flow equation of $\tilde
\Delta(u)$ by taking two derivatives of (\ref{rgdisorder}). One finds
that there is a priori an unstable direction corresponding to a
uniform shift in $\tilde \Delta(u) \to \tilde \Delta(u) + cst$.  While
this is natural in e.g.\ depinning, it is here forbidden by the
potential nature of the problem which requires
\begin{equation}\label{IV.34}
\int_{0}^{1} \rmd u\, \tilde \Delta (u) = 0\ ,
\end{equation}
since in a potential environment, the integral of the force over one
period must vanish. This is indeed satisfied  for the fixed point
for $\tilde \Delta (u)$
\begin{eqnarray}\label{lf134}
 \tilde \Delta^*(u)&=& - \tilde R^{* \prime \prime}(u) \nonumber \\
&=& 
 \frac{\epsilon}{36} + \frac{\epsilon^2}{54} \left(1 + \frac{\lambda-1}{4}
\right)  - \left(\frac{\epsilon}{6} + \frac{\epsilon^2}{9}\right) u (1 -
u)\nonumber \\
&& \label{lf135}
\end{eqnarray}
only if $\lambda=1$: 
\begin{equation}\label{IntDelta}
\int_{0}^{1}\rmd u\, \tilde \Delta^{*} (u) = \frac{\epsilon^{2}}{216}
(\lambda -1)
\end{equation}
The values for depinning are obtained by setting
$\lambda=-1$: in that case, the problem becomes non-potential
at large scales.

\subsubsection{Universal amplitude} This fixed point implies for the
amplitude of the zero mode in presence of an harmonic well,
defined in (\ref{lf109}), using (\ref{ctd}):
\begin{eqnarray}\label{ua1}
\tilde c(d) &=& \frac{(4 \pi)^{d/2}}{\epsilon
\Gamma(\frac{\epsilon}{2})}
\left(  \frac{\epsilon}{36} + \frac{\epsilon^2}{54} + O(\epsilon^3)
\right)\nonumber   \\
&= & 2.19325 \epsilon - 0.680427 \epsilon^2 + O(\epsilon^3)
\end{eqnarray}
In the other limit $m \ll q$ one obtains the amplitude (using $b=-1$
from Section \ref{sec:correlations}):
\begin{eqnarray}
c(d) &=& \frac{(4 \pi)^{d/2}}{\epsilon \Gamma(\frac{\epsilon}{2})}
\left( \frac{\epsilon}{36} - \frac{\epsilon^2}{108} + O(\epsilon^3)
\right) \label{quasi} \\ 
&= & 2.19325 \epsilon - 2.87367 \epsilon^2 + O(\epsilon^3) \label{s1}
\end{eqnarray}
Note that we prove in Section (\ref{sec:correlations}) that
this amplitude is {\it independent} of large scale boundary conditions,
and is thus identical for e.g. periodic boundary conditions and
in presence of a mass. As can be seen from (\ref{ratfin})
this is a consequence of $\zeta$ being zero.

This can be compared to the GVM method
\cite{GiamarchiLeDoussal1995,GiamarchiLeDoussal1994}:
\begin{eqnarray}\label{lf136}
c_{\mathrm{GVM}}(d) & = & (4-d) 2^{d-3} \pi^{\frac{d}{2}-2}
\Gamma\left(\frac{d}{2}\right) \\
& = & 2 \epsilon - 2.9538 \epsilon^2 + O(\epsilon^3)\label{lf138}
\end{eqnarray}
with coefficients surprisingly close to the $\epsilon$-expansion.

It is interesting to compare predictions in $d=3$. We recall that we
are studying a problem where the period is unity, the general case
being obtained by a trivial rescaling in $u$.  Since (\ref{s1}) has a
poor behavior (and so does (\ref{lf138}) which resums into
(\ref{lf136})), it is better to use instead (\ref{quasi}).  It was
indeed noted in \cite{GiamarchiLeDoussal1995,GiamarchiLeDoussal1994}
that the improved 1-loop prediction $c_1(d=3)$ obtained by setting
$\epsilon=1$ and ignoring the $\epsilon^2/108$ term in (\ref{quasi})
yields a value rather close to the prediction of the GVM:
\begin{eqnarray}\label{lf139}
c_1(d=3) &=& 2 \pi/9 = 0.6981 \\
c_{\mathrm{GVM}}(d=3) &=& 1/2\ .
\end{eqnarray}
Including the 2-loop $\epsilon^2/108$ term now gives $c_2(d=3) =
0.4654$ and $c_2(d=3) = 0.5235$ for the two Pades respectively. This
type of extrapolation makes the GVM and FRG predictions get closer
when including the 2-loop corrections. On the other hand, comparison
of (\ref{s1}) and (\ref{lf138}) suggests that $c(d) >
c_{\mathrm{GVM}}(d)$.

This is in reasonable agreement with the numerical results of
Middleton et al. \cite{McNamaraMiddletonChen1999}. They obtained good
evidence for the existence of the Bragg glass (i.e.\ its stability
with respect to topological defects predicted in
\cite{GiamarchiLeDoussal1995,GiamarchiLeDoussal1994}).  They measure
directly the correlation (\ref{fdmassless}) and obtain strong evidence
for the behavior (\ref{quasi}) (as well as the correct correction to
scaling behavior) with
\begin{equation}\label{lf60}
 2 c(d=3) \approx 1.04\ ,
\end{equation}
(their amplitude $A$ is twice our $c(d)$) which lies in between the
GVM and the 1-loop FRG. (More precisely two different discretizations
gave $2 c(d=3) = 1.01 \pm 0.04$ and $2 c(d=3) = 1.08 \pm 0.05$).

Another interesting observable is the slow growth of displacements
characteristic of the Bragg glass:
\begin{eqnarray}\label{lf142}
\overline{(u_x-u_0)^2} &=& \tilde A_d \ln|x| 
\end{eqnarray}
at large $x$. Performing the momentum integral from
(\ref{fdmassless}), one obtains:
\begin{eqnarray}
\tilde A_d &=& \frac{4}{(4 \pi)^{d/2} \Gamma (d/2)} c(d) \nonumber \\
& = & \frac{4 \sin(\pi \epsilon/2)}{\pi \epsilon (1 -
\frac{\epsilon}{2}) } \left[ \frac{\epsilon}{36} -
\frac{\epsilon^2}{108} + O(\epsilon^3)\right] \label{amp}
\end{eqnarray}
If one expands each factor in $\epsilon$ it yields:
\begin{eqnarray}
\tilde A_d &=& \frac{\epsilon}{18} + \frac{\epsilon^2}{108} + O(\epsilon^3) 
\label{adeps} 
\end{eqnarray}
For comparison, the GVM gives
\begin{equation}\label{lf61}
\tilde A_{d,{\mathrm{GVM}}} = \frac{\epsilon}{2 \pi^2} \ .
\end{equation}
Here extrapolation directly setting $\epsilon=1$ in (\ref{adeps})
looks possible, and yields $\tilde A_3=0.0556$ to one loop increasing
to $\tilde A_3=0.0648$ to two loop.  On the other hand, setting
$\epsilon=1$ in (\ref{amp}) yields instead $\tilde A_3=0.0707$ to one
loop decreasing to $\tilde A_3=0.047$ at two loops.  The GVM gives the
result $A_{3,{\mathrm{GVM}}}=0.0507$.

Another interesting observable is:
\begin{eqnarray}
 \overline{w^2} &=& B_d \ln L \\
 w^2 &=& \frac{1}{L^d} \int_x u_x^2 - \left(\frac{1}{L^d} \int_x
u_x\right)^2\ ,
\end{eqnarray}
where $L$ is the linear system size.  In
Ref.~\cite{McNamaraMiddletonChen1999} it was assumed that $B_d =
\tilde A_d/2$ thus in $d=3$, $B_3 = c(3)/(2 \pi^2)$ yielding a value
of $c(d)$ consistent with the direct measurement of this quantity
\footnote{One can also comment on their result for the extremal
excursions $\Delta H = u_{\mathrm{max}} - u_{\mathrm{min}}$. If $u$
were a {\it Gaussian} variable with the same 2-point correlator, the
exact result for extrema of logarithmically correlated Gaussian
variables predicted in \cite{CarpentierLeDoussal2001} yields: $\Delta
H = \tilde{b} \ln L - \tilde{c} \ln (\ln L) + \tilde{a}$ where
$\tilde{b} = 4 \sqrt{\frac{3 b}{2}}$, $c=2 \gamma \sqrt{\frac{b}{6}}$,
$\gamma=3/2$ and $\tilde{a}$ a fluctuating constant of order $O(1)$
(in their notations $b$ is our $B_d$). Inserting the value obtained
numerically in \cite{McNamaraMiddletonChen1999} for $b$ yields
$\tilde{b} =0.795$ and $\tilde{c} = 0.20$. This is in reasonable
agreement with the measured values $\tilde{b} \approx .73$ quoted in
\cite{McNamaraMiddletonChen1999}. Since deviations from Gaussian are
not expected to be large, this agreement could probably be improved by
using the above form of finite size corrections (as was done in
\cite{CarpentierLeDoussal2001} for a one dimensional version where
much larger sizes had to be considered) rather than the simpler form
used in \cite{McNamaraMiddletonChen1999}.}.  This was also done in
\cite{NohRieger2001} where it was deduced from a measurement of $B_d$
that $0.98 < 2 c(d=3) < 1.11$ \footnote{They also measure the decay of
the correlation of $\exp(2 i \pi u)$, which, within a Gaussian
approximation for the distribution of $u$, yields the decay $L^{-
A_d}$, with the Bragg glass exponent $A_d= 2\pi^{2} \tilde A_{d}$.}.
Although this is a reasonable approximation, it is not exact. Indeed
the quantity $B_d$, contrary to $c(d)$, depends on the (large scale)
boundary conditions. It is of course universal, since it does not
depend on small scale details.  Its value can be computed e.g.\ for
periodic boundary conditions and pinned zero mode, and depends on the
whole finite size scaling function (\ref{finitesizeg}) computed in
Section \ref{sec:correlations}:
\begin{equation}
 \overline{w^2} = c(d) \sum_{q \neq 0} q^{-(d+2 \zeta} g_d(q L)
\end{equation}
As shown
recently, $w^2$ fluctuates from sample to sample and the full
distribution $P(w^2)$ averaged over disorder realizations was computed
for the depinning problem
\cite{RossoKrauthLeDoussalVannimenusWiese2003,LeDoussalWiese2003a}.

\subsection{Long range elasticity} \label{sec:lr} Let us now consider
the case of long range elasticity. There are physical systems where
the elastic energy does not scale with the square of the wave-vector
$q$ as $E_{\mathrm{elastic}}\sim q^{2}$ but as
$E_{\mathrm{elastic}}\sim |q|^{\alpha }$.  In this situation, the upper
critical dimension is $d_{c}=2 \alpha$ and we define:
\begin{equation}\label{lf144}
\epsilon:= 2 \alpha - d \ .
\end{equation}
The most interesting case, a priori relevant to model a contact line
is $\alpha =1$, thus $d_{c}=2$. For
calculational convenience, we choose the elastic energy to be
\begin{equation}\label{lf145}
E_{\mathrm{elastic}}\sim ( q^{2}+m^{2})^{\frac \alpha 2 }\ .
\end{equation}
This changes the free correlation to:
\begin{equation}\label{lf146}
G_{ab}(q) = \delta_{ab} \frac{T}{(q^2 + m^2)^{\frac{\alpha}{2}}}  \ .
\end{equation}
The energy exponent in that case is:
\begin{equation}\label{lf147}
\theta = \alpha - d + 2 \zeta\ .
\end{equation}
The changes are very similar to the case of
Ref.~\cite{LeDoussalWieseChauve2002} so we summarize them here only
briefly.  The $\beta$-function is still given by
(\ref{rgdisorderunrescaled}) but with the integrals replaced by:
\begin{eqnarray}\label{lf148}
 I_{1}^{(\alpha)} &=&
\int_{q} \frac{1}{(q^2 + m^2)^{\alpha}} = m^{-\epsilon}
\frac{\Gamma(\epsilon/2)}{\Gamma(\alpha)}  \int_q \rme^{- q^2}\\
I_A^{(\alpha) } &=&
\int_{q_1,q_2}
\frac{1}{(q_1^2 + m^2)^{\frac\alpha2} (q_2^2 + m^2)^{\alpha} ((q_1 +
q_2)^2 + m^2)^{\frac\alpha2}} \nonumber\\\label{lf149}
\end{eqnarray}
and thus the $\beta$-function is given by (\ref{rgdisorder}) with:
\begin{eqnarray}
 X \to {X^{(\alpha) }} &:=& \frac{2\, \epsilon ( 2 I_A^{( \alpha) } - (
I_1^{( \alpha)})^2) }{ (\epsilon I_1^{(\alpha )})^2 }\nn \\
& =&
\int_{0}^{1}\frac{\rmd t}t\, \frac{1+t^{\frac{\alpha }{2}} -
(1+t)^{\frac{\alpha }{2}}}{(1+t)^{\frac{\alpha }{2}}}+
\psi(\alpha )  - \psi(\frac {\alpha}{2}) \nn \\
&& +\, O(\epsilon)\ . \label{lf150}
\end{eqnarray}
(See appendix F of Ref.~\cite{LeDoussalWieseChauve2002}).  And of
course the relation (\ref{resc}) between $R$ and $\tilde R$ is
identical except that $\epsilon \tilde I_1$ must be replaced by
$\epsilon \tilde I_1^{( \alpha)}$.  Since $X^{(\alpha) }$ is finite,
the $\beta$-function is finite; this is of course necessary for the
theory to be renormalizable. For the cases of interest $\alpha =1$ and
$\alpha =2$, we find
\begin{eqnarray}\label{lf151}
 X^{(2)} &=& 1  \\
 X^{(1)} &=& 4 \ln 2
\ .\label{lf152}
\end{eqnarray}
The exponent $\zeta$ (as a function of $\epsilon$) and the fixed point
function is thus changed only at two loops.

Let us now give the results in the cases of interest:

\subsubsection{Random bond disorder} The solution of
(\ref{rgdisorder}) with $X \to X^{(\alpha) }$ can be written, to
second order in $\epsilon$ as:
\begin{eqnarray}\label{ansatzRB2}
\tilde{R} (u) &=& \epsilon\, r_{1} (u) + \epsilon^{2} X^{(\alpha) }
r_{2} (u) + \dots \\
\zeta &=& \epsilon\, \zeta_{1} + \epsilon^{2} X^{(\alpha) } \zeta_{2} +
\dots \label{lf47} \ ,
\end{eqnarray}
since the equation (\ref{lf49}) for $r_2(u)$ is linear. Thus one has
for any $\alpha$:
\begin{equation}\label{zetaalpha}
\zeta = 0.20829806 (3) \epsilon +  0.006858 (1) X^{(\alpha)}
\epsilon^{2} + O(\epsilon^3) \ .
\end{equation}
For the case of most interest $\alpha=1$, $X^{(1)}=4 \ln 2$ one finds:
\begin{eqnarray} \label{lf153}
 \zeta&=& 0.20829806 (3) \epsilon +  0.0190114 (3) \epsilon^2 \nonumber  \\
 \epsilon&=& 2 - d
\end{eqnarray}
and $\theta=2 \zeta$.

It would thus be interesting to perform numerical simulations in $d=1$
for the directed polymer with LR elasticity. This would be another non
trivial test of the 2-loop corrections. The 1-loop prediction is
$\zeta = 0.208$, significantly smaller than the roughness for SR
elasticity $\zeta=2/3$. The naive 2-loop result is (setting
$\epsilon=1$), $\zeta \approx 0.227 \pm 0.01$. Error bars are
estimated by half the difference between the 1-loop and 2-loop
results.  Note that the bound $\theta < d/2$ implies $\zeta < 1/4$ in
$d=1$, already rather close to the 2-loop result.

\subsubsection{Random field disorder}
The exponent is still
\begin{equation}\label{lf71}
\zeta= \frac{\epsilon}{3}
\end{equation}
and was indeed measured in experiments on an equilibrium contact line
\cite{PrevostThese}. It would be  of interest to measure the
universal distributions there, such as the one defined in
\cite{RossoKrauthLeDoussalVannimenusWiese2003,LeDoussalWiese2003a}.

The fixed point function is given by (\ref{k1}) and (\ref{fy}) upon
replacing $F(y) \to X^{(\alpha )} F(y)$.  The amplitude of the zero
mode in a well $c(d)$ is now given by:
\begin{equation}\label{lf72}
 \tilde c(d) = \sigma^{2/3} \left(\frac{\epsilon}{3}\right)^{1/3}
\left(\gamma_1 + \epsilon X^{(\alpha)} \gamma_2\right)^{-2/3}
(\epsilon \tilde{I}_1^{(\alpha)})^{- 1/3}
\end{equation}
and the amplitude of the massless propagator
\begin{equation}\label{lf73}
 c(d) = \tilde c(d) (1 + b_\alpha \epsilon)\ .
\end{equation}
where $ b_\alpha$ is given in (\ref{lf193}) setting $\zeta_1=1/3$. 

\subsubsection{Periodic disorder}
The fixed point becomes:
\begin{eqnarray} \label{lf154}
 \tilde \Delta^*(u) &=& - \tilde R^{* \prime \prime}(u) \nonumber \\
&=& \frac{\epsilon}{36} + \frac{\epsilon^2}{54} X^{(\alpha )} -
\left(\frac{\epsilon}{6} + \frac{\epsilon^2}{9} X^{(\alpha )}\right) u (1 - u)\
.\qquad \quad
\end{eqnarray}
For the periodic case, the universal amplitude reads:
\begin{equation}\label{lf62}
\tilde c(d) = \Gamma (\alpha )\frac{(4 \pi)^{d/2}}{\epsilon
\Gamma(\frac{\epsilon}{2})} \left( \frac{\epsilon}{36} +
\frac{\epsilon^2}{54} X^{(\alpha )} + O(\epsilon^3) \right)\
\end{equation}
and
\begin{equation}\label{lf63}
 c(d) =  \tilde c(d) (1 + b_\alpha \epsilon) \ .
\end{equation}
Setting $\zeta_1=0$ in (\ref{lf193}) yields
\begin{equation}\label{lf157}
 \tilde A_d = \frac{4}{(4 \pi)^{d/2} \Gamma ( d/2)} c(d) \ .
\end{equation}
Using $\epsilon =2\alpha -d$, this gives
\begin{equation}
\tilde A_{d}=\frac1 {18} \epsilon + \frac{4 X^{(\alpha)} +3 (\gamma
+\psi (\alpha )) + 6 b_\alpha}{108} \epsilon^2\ ,
\end{equation}
which in the case of $\alpha =1$ takes the simple form
\begin{equation}
\tilde A_{1}=\frac1 {18} \epsilon + \frac{4 X^{(\alpha)} + 6 b_\alpha}{108}
\epsilon^2\ .
\end{equation}

\section{Lifting ambiguities in non-analytic theory}
\label{sec:ambiguities}

\subsection{Summary of possible methods} \label{sec:list} As we have
seen above ambiguities arise in computing the effective action at the
level of 2-loop diagrams if one uses a non-analytic action. One can
see that these arise even at the 1-loop level for correlations (see
below Section \ref{sec:correlations}).  To resolve this issue, our
strategy has been to use physics as a guide and require the theory to
be renormalizable, potential and without supercusp.  This pointed to a
specific assignment of values to the ``anomalous'' graphs. The
physical properties of the ensuing theory, studied in the previous
section, were found to be quite reasonable. Of course, one would like
to have a better, more detailed justification of the used
``prescription''.  Although we do not know at present of a derivation
of this theory from first principles, we have developed a set of
observations and a number of rather natural and compelling ``rules''
which all lead to the same theory. We  describe below our successful
efforts in that direction as well as some unsuccessful ones, which
illustrate the difficulty of the problem.

A number of approaches can be explored to lift the ambiguities in
the non-analytic theory.  We here give  a list; some of the
methods will be detailed in the forthcoming sections.

\smallskip
{\bf 1) Non-zero temperature}: At $T>0$ previous Wilson 1-loop FRG
analysis\nopagebreak
\cite{Balents1993,ChauveGiamarchiLeDoussal1998,ChauveGiamarchiLeDoussal2000,ChauveLeDoussal2001}
found that the effective action {\it remains analytic} in a boundary
layer $u \sim \tilde{T}$.  However, since the rescaled temperature
(\ref{rescaled}) flows to zero as $\tilde{T} \sim m^\theta $ as $m \to
0$ (temperature being formally irrelevant) all (even) derivatives of
$R(u)$ higher than second grow unboundedly as $m \to 0$, for instance
$R''''(0) \sim R^{* \prime \prime \prime}(0^+)^2/\tilde{T}$ (in terms
of the zero temperature fixed point function).  On a {\it qualitative}
level one can thus see how finite $T$ diagrams such as $E$ in
Fig.\ \ref{alldiag2loop} yielding
\begin{equation}\label{lf158}
\sim T R''''(0) R''(u) \to R^{* \prime \prime \prime}(0^+)^2 R''(u)
\end{equation}
can build up ``anomalous'' terms in the $\beta$-function, hence
confirming what is found here\cite{ChauveLeDoussal2001}. However,
correctly and quantitatively accounting for higher loops is a
non-trivial problem as stronger blow-up in $1/\tilde{T}^k$ seem to
arise. In fact each new loop brings two derivatives and a
propagator, hence an additional factor $1/\tilde{T}$. Despite some
recent progress, a quantitative finite-temperature approach which
would reproduce and justify the present $\epsilon$ expansion has
proved
difficult\cite{BalentsLeDoussal2002a,BalentsLeDoussal2002b}. Not
only for technical reasons, as methods using exact RG where found
to be appropriate, but also for physical reasons, as an extension
to non-zero $T$ must also handle low-lying thermal excitations in
the system (e.g.\ droplets). A theory from first principles at
$T>0$ is thus presently not available and will not be further
addressed here. All other methods use a {\it non-analytic action}.

\smallskip
{\bf 2) Exact RG}: Exact RG methods directly at $T=0$ have been
studied to one loop \cite{ChauveLeDoussal2001,scheidl} and two
loops\cite{erg2b,scheidl2}. Although it does yield interesting
insights into the way to handle ambiguities (see below), and confirm
the present results, it suffers from basically the same problems as
described here.

\smallskip
{\bf 3) Direct evaluation of non-analytic averages}: In this approach
one attempts a direct evaluation of non-analytic averages (e.g.\ in
fully saturated diagrams).  For instance, expanding at each vertex the
disorder $R(u^x_a-u^x_b)$ in powers of $|u^x_a-u^x_b|$ using the
proper non-analytic Taylor expansion:
\begin{equation}\label{exp}
R'' (u)=R ''(0) + R'''(0^+) |u| + R''''(0^+) u^2 + \dots
\end{equation}
one can try to compute directly all averages in vertex functions and
correlations. After performing a few Wick contractions one typically
ends up with averages involving sign functions or delta
functions. These can be computed {\it in principle} using the free
Gaussian measure. For instance, using formulae such as:
\begin{equation}\label{lf74}
 \left< \sgn(u) \sgn(v) \right>_0 = \frac{2}{\pi}
\text{asin}\left(\frac{\left<uv\right>_0}{\sqrt{\left<uu\right>_0
\left<vv\right>_0 }}\right)\ .
\end{equation}
Although promising at first sight, the results are
disappointing. Averages over the thermal measure involve many
changes of signs which kill all interesting divergences indicating
that some physics is missing.  The method, briefly described in
Appendix \ref{sec:nona} is thus not developed further.  A dynamical
version of this method which is  similar in spirit
\cite{ChauveLeDoussalWiese2000a,LeDoussalWieseChauve2002}, did work for
depinning, although there it simply identified with another method
used below, the background field (which, for depinning is $u_{xt} \to
v t + u_{xt}$ see below).
\smallskip

{\bf 4) Calculation of $\Gamma(u)$ with excluded vertices and
symmetrization}: A valid, general and useful observation (not limited
to this method) is that if one uses the {\it excluded vertex}
\begin{equation}\label{lf75}
 \frac{1}{2 T^2} \sum_{a \neq b} R(u_a - u_b)\ ,
\end{equation}
then all Wick contractions can be performed {\it without
ambiguities}. The excluded vertex is as good as the non-excluded one
since one can always add a constant $- n R(0)$ to the action of the
model (\ref{action}). Thus one can compute {\it without any ambiguity} the
effective action $\Gamma(u)$ for an ``off-diagonal'' field
configuration
\begin{equation}\label{diag}
 u^a_x \quad \text{such that} \quad u^a_x \neq u^b_x \quad \text{for
all} \quad a \neq b\ ,
\end{equation}
since then no vertex is ever evaluated at $u=0$.  The drawback is that
one ends up with expressions containing terms such as
\begin{equation}\label{3rep}
 \sum_{a \neq b , a \neq c } R''(u_a - u_b) R'''(u_a - u_b) R'''(u_b -
u_c)\ ,
\end{equation}
which superficially looks like a three replica term, but due to the
exclusions, may in fact contain a 2-replica part which can in
principle be recovered from the above by adding appropriate diagonal terms,
using that  $p$-replica parts are properly defined as {\it free} replica
sums e.g.\ from a cumulant expansion. The 2-replica part of
(\ref{3rep})  thus naively is
\begin{eqnarray} \label{lf159}
&& - \sum_{a c }  R''(0) R'''(0) R'''(u_a - u_c) \nonumber \\
&& +
\sum_{a b }  R''(u_a - u_b) R'''(u_a - u_b) R'''(0) \label{lf160}
\end{eqnarray}
and one is again faced with the problem of assigning a value to
$R'''(0)$. The calculation with excluded vertices thus yields a sum of
$p$-replica terms with $p \geq 2$ and to project them onto the needed
2-replica part, one may need to continue these expressions to
coinciding arguments $u^a=u^b$.

The symmetrization method attempts to do that in the most ``natural''
way. Using the permutation symmetry over replicas and the hypothesis
of no supercusp yields a rather systematic method of
continuation. Surprisingly, it fails to yield a renormalizable theory
at two loops. We identified some difference with methods which do
work, but the precise reason for the failure in terms of continuity
properties remains unclear. It may thus be that there is a way to make
this method work but we have not found it. Being interesting in spirit
this method is reported in some details in Appendix \ref{sec:symm}.

If one renounces to the projection onto 2-replica terms one can, in
a certain sense, obtain renormalizability properties. This
generates an infinite number of different replica sums and seems to be
not promising, too. It is described in Appendix \ref{sec:finiteT}.

We now come to methods which were found to work, and which will be
described in detail in the next section. In all of them one performs
the Wick contractions in some given order (the order hopefully does not
matter) and uses at each stage some properties.  The fact that one can
order the Wick contractions stems from the identity, which we recall,
for any set of mutually correlated Gaussian variables $u_i$:
\begin{equation}\label{lf77}
\left< u_i W(u) \right> = \sum_j \left< u_i u_j \right> \left<
\partial_{u_j} W(u) \right>
\end{equation}
under very little analyticity assumption for $W(u)$, which can even be
a distribution. At each stage one can either use excluded or
non-excluded vertices as is found more convenient.\smallskip

{\bf 5) Elimination of sloops:}
We found another method, which seems rather compelling, to determine
the 2-replica part of terms such as (\ref{3rep}). It starts, as the
previous one, by computing (unambiguously) diagrams with the
excluded vertices. Then instead of symmetrization, one uses
identities derived from the fact that diagrams with free replica sums
and which contain sloops cannot appear in a $T=0$ theory and can thus
be set to zero. Further contracting such diagrams generates a set of
identities which, remarkably, is sufficient to obtain unambiguously the
2-replica projection without any further assumption. It works very nicely
and produces a renormalizable theory, as we have checked up to three
loops.  In some sense, it uses in a non-trivial way the constraint that we
are working with a true $T=0$ theory. This method is detailed below.
\smallskip

{\bf 6) Background field method:} This method is similar to method
number 3 except that the vertex $R(u)$ at point $x$ is evaluated
at the field $u^a_x = u^a + \delta u^a_x$, then expanded in
$\delta u^a_x$, which then are contracted in some order.  This
amounts to compute the effective action in presence of a uniform
background field which satisfies (\ref{diag}). Thanks to this
uniform background and upon some rather weak assumptions, the
ambiguities seem to disappear. The method is explained below.
\smallskip

{\bf 7) Recursive construction:} An efficient method is to
construct diagrams recursively. The idea is to identify in a first
step parts of the diagram, which can be computed without
ambiguity. This is in general the 1-loop chain-diagram
(\ref{1loop}). In a second step, one treats the already calculated
sub-diagrams as effective vertices. In general, these vertices
have the same analyticity properties, namely are derivable twice,
and then have a cusp. (Compare\ $R (u)$ with $( R'' (u)-R''
(0))R''' (u)^{2}-R'' (u)R''' (0^{+})^{2}$). By construction, this
method ensures renormalizability, at least as long as there is
only one possible path. However it is not more general than the
demand of renormalizability diagram by diagram, discussed below.
\smallskip

{\bf 8) Renormalizability diagram by diagram:} In Section III we
have used a {\it global} renormalizability requirement: The 1-loop
repeated counter-term being non-ambiguous one could fix all
ambiguities of the divergent 2-loop corrections. However, as will
be discussed in \cite{LeDoussalWiesePREPb}, this global constraint
appears insufficient at three loops to fix all ambiguities.
Fortunately, one notes that renormalizability even gives a
stronger constraint, namely renormalizability {\em diagram by
diagram}. The idea goes back to formal proofs of perturbative
renormalizability in field-theory, see e.g.\
\cite{BogoliubovParasiuk1957,Hepp1966,Zimmermann1969,BergereLam1975,DDG2,DDG4,WieseHabil,Collins}.
These methods define a  subtraction operator $\mathrm{\bf R}$.
Graphically it can be constructed by drawing a box around each
sub-divergence, which leads to a ``forest'' or ``nest'' of
sub-diagrams (the counter-terms in the usual language), which have
to be subtracted, rendering the diagram ``finite''. The advantage
of this procedure is that it explicitly assigns all counter-terms
to a given diagram, which finally yields a proof of perturbative
renormalizability. If we demand that this proof goes through for
the {\em functional} renormalization group, the counter-terms must
necessarily have the same functional dependence on $R(u)$ as the
diagram itself. In general, the counter-terms are less ambiguous,
and this procedure can thus be used to lift ambiguities in the
calculation of the diagram itself. By construction this procedure
is very similar to the recursive construction discussed under
point 7.

It has some limitations though. Indeed, if one applies this
procedure to the 3-loop calculation, one finds that it renders
unique all but one ambiguous diagram, namely
\begin{equation}\label{lf161}
\diagram{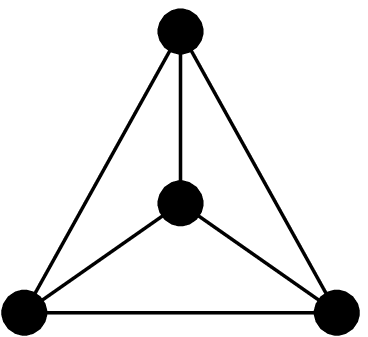}\ ,
\end{equation}
which has no subdivergence, thus there are no counter-terms, which
could lift the ambiguities. Thus this diagram must be computed
directly and we found that it can be obtained unambiguously by the
sloop elimination method\cite{LeDoussalWiesePREPb}.
\smallskip

{\bf 9) Reparametrization invariance:} From standard field theory,
one knows that renormalization group functions are not unique, but
depend on the renormalization scheme. Only critical exponents are
unique. This is reflected in the freedom to reparametrize the
coupling constant $g$ according $g  \longrightarrow  \tilde g
(g) $ where $\tilde g (g)$ is a smooth function, which has to be
invertible in the domain of validity of the RG $\beta$-function.

Here we have chosen a scheme, namely defining $R(u)$ from the
exact zero momentum effective action, using dimensional
regularization, and a mass. One could explore the freedom in
performing reparametrization. In the functional RG framework,
reparametrizations are also functional, of the form
\begin{equation}\label{lf163}
R (u) \ \longrightarrow \ \hat R (u) = \hat R[R] (u)\ .
\end{equation}
Of course the new function $\hat R (u)$ does not have the same
meaning as $R(u)$. Perturbatively this reads
\begin{equation}\label{lf164}
R (u) \ \longrightarrow \ \hat R (u) = R (u) + B (R,R) (u) +O
(R^{3}) \ ,
\end{equation}
where $B (R, R)$ is a functional of $R$. For consistency, one has
to demand that $B (R,R)$ has the same analyticity properties as
$R$, at least at the fixed point $\tilde R=\tilde R^{*}$, i.e.\ $B
( R,R)$ should as $R$ be twice differentiable and then have a
cusp. A specifically useful candidate is the 1-loop counter-term
$B(R,R) =\delta^{(1,1)}R$. One can convince oneself, that by
choosing the correct amplitude, one can eliminate all
contributions of class A, in favor of contributions of class B.
Details can be found in \cite{LeDoussalWiesePREPb}.\smallskip

Apart from methods 3 and 4 which did not work for reasons which remain
to be better understood, methods 2,5,6,7,8,9 were
all found to give consistent result, making us confident that the
resulting theory is sufficiently constrained by general arguments
(such as renormalizability) to be uniquely identified.
Let us now turn to actual calculations using these methods.

\subsection{Calculation using the sloop elimination method}
\label{sec:sloops}

\subsubsection{Unambiguous diagrammatics}

\label{sec:unambdiag}

Let us redo the calculation of Section \ref{sec:twoloop} using {\it
excluded vertices}. From now on we use sometimes the short-hand
notations
\begin{equation}\renewcommand{\arraystretch}{1.5}
\begin{array}{rclcrcl}
 u^{ab} &=& u^a - u^b \quad &,& \quad u^{ab}_x &=& u^a_x - u^b_x \\
 R_{ab} &=& R(u^a - u^b) \quad &,& \quad R^{(p)}_{ab} &=& R^{(p)}(u^a - u^b)
\end{array}
\label{lf165}
\end{equation}
whenever confusion is not possible.

\begin{figure}
\centerline{\Fig{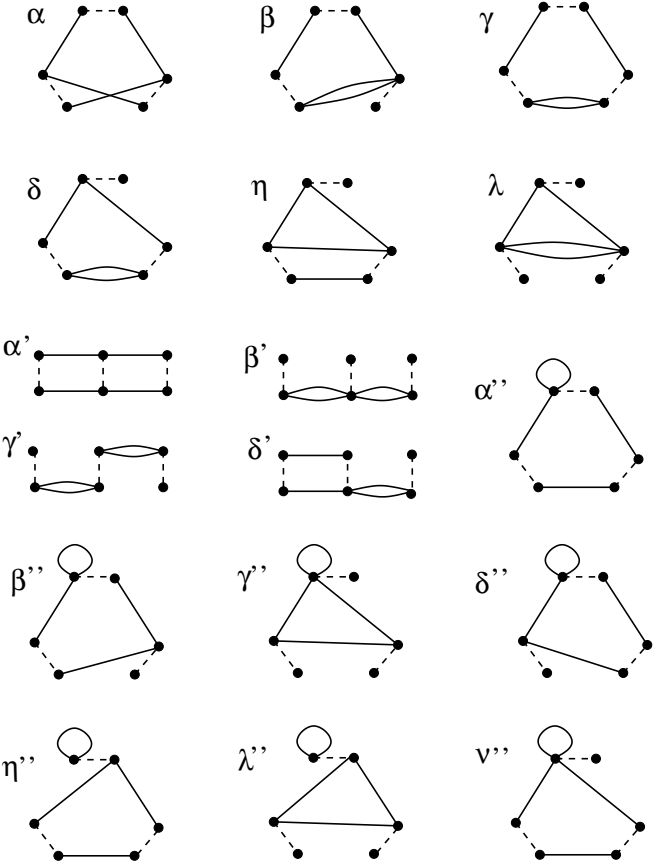}}
\caption{2-loop diagrams corresponding to Larkin's hat (top)
and banana (bottom). The graph
$\alpha$ is a 1-replica term, $\beta$, $\gamma$, $\delta$ and $\eta$ are
improper three replica terms and $\lambda$ is an improper four replica term.}
\label{excl}
\end{figure}

The resulting diagrammatics looks very different from the usual
unexcluded one. When making all four Wick contractions of the 2-loop
diagrams A, B and C in Fig.~\ref{f:twoloop} between three unsplitted
vertices one now excludes all diagrams with saturated vertices, but
instead has to allow for more than two connected components and for
sloops.  The splitted excluded diagrams corresponding to classes A, B
and C are given in Fig.\ \ref{excl}. There is an additional
multiplicative coefficient $1/(m_1 ! m_2 ! m_3 ! m_4 !)$ in the
combinatorics for each pair of unsplitted vertices (say $ab$ and $cd$)
linked by an internal line where $m_1$ propagators link $ac$, $m_2$
link $ad$, $m_3$ link $bc$, $m_2$ link $bd$.  (This is equivalent to
assigning a color to each propagator).

Let us denote by $\delta \Gamma = - \frac{1}{2 T^2} \delta^{(2)}_{A}
R$ the 2-loop contribution of  all diagrams of class A to
the effective action. One finds:
\begin{eqnarray}
 \delta^{(2)}_{A} R &=& \Big[ \sum_{a \neq b} R''_{ab} (R'''_{ab})^2
+ \sum_{a \neq b, a \neq c} R''_{ab} R'''_{ab} R'''_{ac} \nonumber
\\
&& - \frac{1}{2} \sum_{a \neq b, a \neq c , b \neq c} R''_{ab}
R'''_{ac} R'''_{bc}
+ \frac{3}{2} \sum_{a \neq b, a \neq c} R''_{ab} (R'''_{ac})^2  \nonumber
\\
&& + \frac{1}{2} \sum_{a \neq b, a \neq c , a \neq d} R''_{ab}
R'''_{ac} R'''_{ad} \Big] I_A
\label{deltaA}
\end{eqnarray}
coming respectively and in the same order from graphs $\alpha$,
$\beta$, $\gamma$, $\delta$+$\eta$ (they are equal) and $\lambda$ in
Fig.\ \ref{excl}. The only graph common to excluded and free-sum
diagrammatics is $\alpha$ which is graph $b$ of Fig.\
\ref{static-2-loop-DO}, since all the other graphs in Fig.\
\ref{static-2-loop-DO} have saturated vertices.

Similarly, the graphs of class B give a total contribution:
\begin{eqnarray}
 \delta^{(2)}_{B} R &=& \Bigg [ \frac{1}{2} \sum_{a \neq b} R''''_{ab}
R''_{ab} R''_{ab} +
\frac{1}{4} \sum_{a \neq b, a \neq c , a \neq d} R''''_{ab} R''_{ac}
R''_{ad} \nonumber \\
&& \hspace{-.5 cm}+ \frac{1}{4} \sum_{a \neq b, a \neq c , b \neq d} R''''_{ab}
R''_{ac} R''_{bd}
+ \sum_{a \neq b, a \neq c} R''''_{ab} R''_{ab} R''_{ac} \Bigg ] I_1^2 \nonumber \\
&& \label{deltaB}
\end{eqnarray}
coming respectively and in the same order from graphs $\alpha'$,
$\beta'$, $\gamma'$, $\delta'$ in Fig.\ \ref{excl}. Again, the only
graph common to excluded and free-sum diagrammatics is $\alpha'$ which
is graph $g$ of Fig.\ \ref{banana}, since all the other graphs in
Fig.\ \ref{banana} have saturated vertices.

The contribution $\delta^{(2)}_{C} R$ of the diagrams of class C 
is given in Appendix \ref{sec:classC}. Note that adding a
tadpole does not alter the structure of the summations in the
excluded-replica formalism, since a tadpole can never identify
indices on different vertices. This indicates that
class C does not contain a 2-replica contribution,
but starts with a 3-replica contribution
(times $T$). This is explained in more details in Appendix 
\ref{sec:sloopBC}

One can first check that when $R(u)$ is {\it analytic} one recovers
correctly the same result as (\ref{d22}) setting the last (anomalous)
term to zero. Adding and subtracting the excluded terms in
(\ref{deltaA}) to build free replica sums (using $R'''(0)=0$ in that
case), or equivalently lifting all exclusions but replacing
everywhere
\begin{equation}\label{lf78}
 R^{(p)}_{ab} \to R^{(p)}_{ab} (1 - \delta_{ab})
\end{equation}
and then expanding and selecting the 2-replica part, one finds the
 contributions
\begin{eqnarray}
\alpha &\to& R''(u) R'''(u)^2 \nonumber \\
 \gamma &\to& \frac{1}{2}
R''(0) R'''(u)^2 \nonumber \\
 \delta + \eta &\to& - \frac{3}{2} R''(0) R'''(u)^2\nonumber  \\
\beta &\to& 0 \nonumber \\
\lambda &\to& 0\ . \label{lf166}
\end{eqnarray}
Similarly in (\ref{deltaB}) one obtains
\begin{eqnarray}
 \alpha' &\to& \frac{1}{2} R''''(u) R''(u)^2 \nonumber \\
\beta' &\to& \gamma' \to
\frac{1}{2} R''''(0) R''(0) R''(u) +
\frac{1}{4} R''(0)^2 R''''(u) \nonumber \\
 \delta' &\to& - R''''(0) R''(0) R''(u)
- R''(0) R''''(u) R''(u) \nonumber\ .\\
&&\label{lf167}
\end{eqnarray}
We now want to perform the same projection for a non-analytic $R(u)$.

\subsubsection{The sloop elimination method}
The idea of the method is very simple. Let us consider the 1-loop
functional diagram (a) in Fig.\ \ref{f:sloops} which contains a
sloop. It is a three replica term proportional to the temperature. In a
$T=0$ theory such a diagram should not appear, so it can be
identically set to zero:
\begin{equation}\label{id1}
W:=\frac{1}{T^2} \sum_{abc} R''(u^{ab}_x) R''(u^{ac}_y)  \equiv 0\ .
\end{equation}
It is multiplied by $G(x-y)^2$, which we have not written. We will
also omit global multiplicative numerical factors. Projecting such
terms to zero at any stage of further contractions is very natural
in our present calculation (and also e.g.\ in the exact RG
approach, where terms are constructed recursively and such
forbidden terms must be projected out). It is valid only when (i)
the summations over replicas are free (ii) the term inside the sum
is non-ambiguous. These conditions are met for any diagram with
sloops, provided the vertices have at most two derivatives. (One
can in fact start from vertices which either have no derivative
or exactly two.)

Let us illustrate the procedure on an example. We want to contract $W$
with a third vertex $R$ at point $z$, i.e.\ we first write the
product:
\begin{equation}\label{id2}
W \frac{1}{T^2} \sum_{de} R_{de} = \frac{1}{T^4} \sum_{a \neq b ,
a \neq c ,de} R''_{ab} R''_{ac} R_{de} \equiv 0  \ ,
\end{equation}
where implicitly here and in the following the vertices are at
points $x,y,z$ in that order. We will contract the third vertex
twice, once with the first and once with the second , i.e.\ look
at the term proportional to $G(x-y)^2 G(x-z) G(y-z)$. Note that
since we will contract each vertex, we are always allowed to
introduce excluded sums (clearly the diagonal terms $a=b$, $a=c$
or $d=e$ give zero, since $R_{ab}$ and its two lowest derivatives
at $a=b$ are field independent constants). Performing the first
contraction (i.e.\ inserting $\delta_{ad} - \delta_{ae} -
\delta_{bd} + \delta_{be}$ multiplied by the exclusion factors
$(1-\delta_{ab})(1-\delta_{ac})(1-\delta_{de})$ yields (up to a
global factor of $2$):
\begin{equation}\label{id3}
\frac{1}{T^3} \Big[ \sum_{a \neq b , a \neq c , a \neq e} R'''_{ab}
R''_{ac}  R'_{ae}
- \sum_{a \neq b , a \neq c , b \neq e} R'''_{ab}  R''_{ac}  R'_{be} \Big]
\equiv 0  \ .
\end{equation}
Similarly, the second contraction then yields (up to a global factor of $4$):
\begin{eqnarray}
&& \!\!\!\frac{1}{T^2} \Big[ \frac{1}{2} \sum_{a \neq b , a \neq c , a \neq e}
R'''_{ab}  R'''_{ac}  R''_{ae}
+  \sum_{a \neq b , a \neq c } R'''_{ab}  R'''_{ac}  R''_{ac}  \nonumber
\\
&& +  \frac{1}{2} \sum_{a \neq b ,  b \neq e} R'''_{ab}  R'''_{ab}  R''_{ae}
- \frac{1}{2} \sum_{a \neq b ,  a \neq c , b \neq c} R'''_{ab}
R'''_{ac}  R''_{bc} \Big]  \equiv 0  \nonumber \ .
\\
&& \label{id4}
\end{eqnarray}
This non-trivial identity tells us that the sum of all terms (or
diagrams) generated upon contractions of diagram (a) of
Fig.~\ref{f:sloops} (i.e.\ the 1-loop sloop-diagram equivalent to
term $W$ in (\ref{id1})) with other vertices, must vanish. Stated
differently: A sloop, as well as the sum of all its descendents
vanishes. Note that this is {\em not} true for each single term,
but only for the sum.

A property that we request from a proper $p$-replica term is that upon
one self contraction it gives a $( p-1)$-replica term. It may also
give $T$ times a $p$-replica term (a sloop) but this is zero at $T=0$,
so we can continue to contract. Thus we have generated several
non-trivial projection identities. The starting one is that the
2-replica part of (\ref{id1}) is zero, since (\ref{id1}) is a
proper 3-replica term. Thus, (\ref{id2}) prior to the
exclusions, is a legitimate 5-replica term, and its 4-replica part is
zero. Upon contracting once we obtain that the 3-replica part of
(\ref{id3}) is zero.  The final contraction tells us that the
2-replica part of (\ref{id4}) is zero. This is what is meant by the
 symbol ``$\equiv$'' above and the last identity is the one we now use.

Indeed  compare (\ref{id4}) with (\ref{deltaA}). One
notices that all terms apart from the first in (\ref{deltaA}) appear
in (\ref{id4}), and with the same relative coefficients, apart from
the third one of (\ref{deltaA}). Thus one can use (\ref{id4}) to
simplify (\ref{deltaA}):
\begin{equation}\label{lf79}
\delta_{A}^{(2)} R = \Big[ \sum_{a \neq b} R''_{ab} (R'''_{ab})^2
+  \sum_{a \neq b, a \neq c} R''_{ab} (R'''_{ac})^2 \Big]
I_A  \ .
\end{equation}
The function $R'''(u)^2$, which appears in the last term, is
continuous at $u=0$. It is thus obvious how to rewrite this
expression using free summations and extract the 2-replica part
\begin{eqnarray}\label{delta2AR}
 \delta^{(2)}_{A} R(u) &=&  \Big [ (R''(u) - R''(0)) R'''(u)^2 \nonumber \\
&& - R'''(0^+)^2 R''(u) \Big] I_A\ ,
\end{eqnarray}
which coincides with the contribution of diagrams A in (\ref{d22})
with $\lambda=1$.

We can write diagrammatically the subtraction that has been performed
\begin{equation}\label{lf168}
\delta_{A}^{(2)} R = \diagram{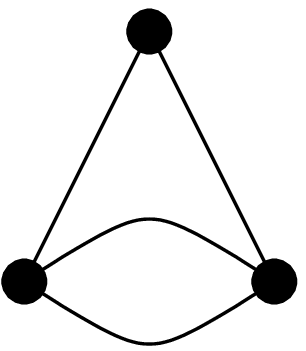}-\diagram{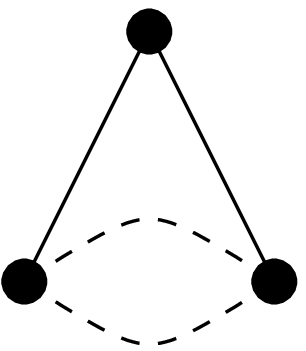} \ ,
\end{equation}
where the loop with the dashed line represents the sub-diagram with the
sloop, i.e.\ the term (\ref{id1}) (with in fact the same global
coefficient). The idea is of course
that subtracting sloops is allowed since they formally vanish.

There are other possible identities, which are descendants of
other sloops. For instance a triangular sloop gives, by a similar
calculation:
\begin{eqnarray}\label{lf169}
\diagram{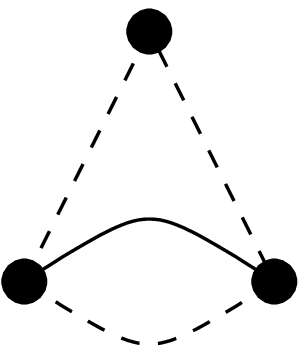} &=& R''(0) \sum_{a \neq b} (R'''_{ab})^2 + \sum_{a \neq b, a \neq c}
R''(0) R'''_{ab} R'''_{ac} \nonumber \\
&& + \sum_{a \neq b, b \neq c} R''_{bc} (R'''_{ab})^2
+ \sum_{a \neq c, b \neq c, c \neq d } R'''_{ac} R'''_{bc} R''_{cd}
\nonumber  \ . \\
&&
\end{eqnarray}
This however does not prove useful to simplify $\delta^{(2)}_{A} R$.

Since the above method generates a large number of identities, one
can wonder whether they are all compatible. We have checked a
large number of examples (see the 3-loop calculations in
\cite{LeDoussalWiesePREPb}) and found no contradictions, although
we have not attempted a general proof.

The diagrams B and C are computed in Appendix \ref{sec:sloopBC}. One finds by the same
procedure 
\begin{eqnarray}\label{lf170}
 \delta_{B} R &=& \frac{1}{2} R''''(u) (R''(u) - R''(0))^2 \\
 \delta_{C} R &=& 0\ ,\label{lf171}
\end{eqnarray}
confirming our earlier results in section \ref{s:ClassB} and
\ref{s:ClassC}.

\subsection{Background method}
\label{sec:background}
In the background method, one computes $\Gamma[u]$ to two loops
for a uniform background $u$ such that $u_{ab} \neq 0$ for
any $a \neq b$. We start from
\begin{equation}\label{lf80}
 \left< {\cal S}[u + v_x]^3 \right>_{\mathrm{1PI}}
\end{equation}
Taylor expand in $v_x$, and contract all the $v$ fields keeping
only 1PI-diagrams. This is certainly a correct formula for the
uniform (i.e.\ zero momentum) effective action.

Then one needs the small $|u|$-expansion of derivatives of $R$, i.e.\
(\ref{exp}) as well as
\begin{eqnarray}\label{R'''}
 R'''(u) &=& R'''(0^+) \text{sign}(u) + R''''(0^+) u + \dots \qquad  \\
 R''''(u) &=& 2 R'''(0^+) \delta(u) + R''''(0^+) + \dots \ . \label{lf172}
\end{eqnarray}
Let us start from:
\begin{equation}\label{lf81}
 \sum_{a b c d e f} R(u_{ab} + v^{ab}_x) R(u_{cd} + v^{cd}_y)
R(u_{ef} + v^{ef}_z) \nonumber \ .
\end{equation}
We expand in $v$ and of course in diagrams A one must handle terms
involving $R'''(0)$ and in diagrams B terms proportional to
$R''''(0)$. Let us start with diagrams A, which come from the
following term in the Taylor expansion:
\begin{equation}\label{lf82}
 \sum_{a b c d e f} R'''(u_{ab}) R'''(u_{cd})
R''(u_{ef}) \left<(v^{ab}_x)^3 (v^{cd}_y)^3 (v^{ef}_z)^2\right>
\end{equation}
Here and in the following, we will drop all combinatorial factors.
Note that the expectation-values  vanish at coinciding replicas
so there is no need to specify the values of $R'''(u_{ab})$
at $a=b$. Let us perform the first $xy$ contraction
\begin{equation}\label{lf83}
\sum_{a b c e f} R'''(u_{ab}) R'''(u_{ad}) R''(u_{ef})
\left<(v^{ab}_x)^2 (v^{ad}_y)^2 (v^{ef}_z)^2\right> \nonumber\ .
\end{equation}
If we now perform a second $xy$ contraction there is
a $\delta_{aa}$ term which is a sloop and thus should be discarded.
The $\delta_{ad} + \delta_{ba}$ terms build saturated vertices.
However the corresponding expectation values contain
\begin{equation}\label{lf84}
 R'''(u_{ad}) \left<(v^{ad}_y) \dots \right> |_{d \to a} = 0\ ,
\end{equation}
which is reasonably set to zero. Thus the first two contractions
have been performed with no ambiguity leading to
\begin{equation}\label{lf85}
 \sum_{a b e f} R'''(u_{ab}) R'''(u_{ab}) R''(u_{ef})
\left<(v^{ab}_x) (v^{ab}_y) (v^{ef}_z)^2\right>
\end{equation}
This term  is no more ambiguous. Expanding as in (\ref{R'''})
the potentially ambiguous part is
\begin{equation}\label{yuyu}
 R'''(0+)^2 \sum_{a b e f} R''(u_{ef}) \left<(v^{ab}_x)
(v^{ab}_y) (v^{ef}_z)^2\right>\ .
\end{equation}
clearly free of any ambiguity. It yields the result
(\ref{delta2AR}). The question arises, whether the result may
depend on the order. We found that when first contracting $xy$ and
$xz$, one reproduces the result  (\ref{delta2AR}). However when
one first contracts $xy$ and $yz$ (in any order) one encounters a
problem, if one wants to contract $yz$ again. The intermediate
result after the first two contractions is
\begin{equation}\label{lf173}
 \sum_{a b e f} R'''(u_{ab}) R'''(u_{ad}) R''(u_{af})
\left<(v^{ab}_x) (v^{ad}_y)^{2} (v^{af}_z)\right>
\end{equation}
The next contraction between $xy$ contains one term with a single
$R'''_{aa}$. One would like to argue that this term can be set to
0. Following this procedure however leads to problems. We
therefore adopt the rule, that whenever one arrives at a single
$R'''_{aa}$, one has to stop, and search for a different path.
Note that this equivalently applies to the recursive constructions
method.  In 2-loop order, one can always find a path, which is
unambiguous. It seems to fail at 3-loop order; at least we
have not yet been able to calculate
\begin{equation}\label{lf174}
\diagram{3loopi}
\end{equation} using any other than the sloop elimination method.
Whether some refinement of the background method can be constructed
there is an open question.

For diagrams of class B one expands as
\begin{equation}\label{lf175}
 \sum_{a b c d e f} R''(u_{ab}) R''''(u_{cd}) R''(u_{ef})
\left<(v^{ab}_x)^2 (v^{cd}_y)^4 (v^{ef}_z)^2\right> \nonumber\ .
\end{equation}
Again no need to attribute a value to $R''''(u_{cd})$
for $c=d$ since the summand vanishes there. Contract $xy$:
\begin{equation}\label{lf176}
 \sum_{a b d e f} R''(u_{ab}) R''''(u_{ad}) R''(u_{ef})
\left<(v^{ab}_x) (v^{ad}_y)^3 (v^{ef}_z)^2\right>\ .
\end{equation}
Contracting  $yz$ one gets
\begin{eqnarray}\label{lf177}
&& \sum_{a b f} R''(u_{ab}) R''''(u_{ad}) \nonumber \\
&&\qquad
\langle(v^{ab}_x) (v^{ad}_y)^2 (R''(u_{af}) v^{af}_z  - R''(u_{df}) v^{df}_z)\rangle  \nonumber\ .
\end{eqnarray}
Contracting next $xy$ the danger is the term  $\delta_{ad}$
yielding a saturated vertex in the middle. But,
again if one takes
\begin{equation}\label{lf86}
 R''''(u_{ad}) \left<(v^{ad}_y) ..\right>|_{d \to a} = 0  \ ,
\end{equation}
then one gets  unambiguously
\begin{equation}\label{lf87}
 \sum_{a b } R''(u_{ab}) R''''(u_{ab})
\left<(v^{ab}_y) (R''(u_{af}) v^{af}_z
 {-} R''(u_{bf}) v^{bf}_z)\right> \nonumber\ .
\end{equation}
The rest is straightforward. The backgound method thus seems to work
properly at two loop order.

\subsection{Renormalizability, diagram by diagram}\label{rd7}In section
\ref{sec:list} we have stated that renormalization diagram by diagram
gives a method to lift the ambiguity of a given diagram, as long as it
has sufficient sub-divergences. This method is inspired by formal
proofs of perturbative renormalizability; the reader may consult
\cite{BogoliubovParasiuk1957,Hepp1966,Zimmermann1969,BergereLam1975,DDG2,DDG4,WieseHabil,Collins}
for more details.  The key-ingredient is the subtraction operator
$\mathrm{\bf S}$, which acts on the effective action, i.e.\ all
terms generated in perturbation-theory, which contribute to the
renormalized $R$, and which subtracts the divergences at a scale $\mu
$. At 1-loop order, the renormalized disorder $R_{m}$ at scale $m$ is
symbolically (with $R_{0}$ the bare disorder)
\begin{equation}\label{lf93}
R_{m} = \left[ R_{0} + ( R''_{0})^{2} \diagram{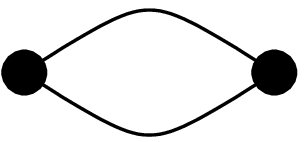} +\ldots
\right]\nonumber\ ,
\end{equation}
where of course the integral $\textdiagram{1loop}$ depends on $m$.
The operator ${\mathrm{\bf S}}$ rewrites this as a function of the
{\em renormalized} disorder $R_{\mu}$ at scale $\mu$
\begin{eqnarray}\label{lf94}
\!R_{m} &=& \mathrm{\bf S} \left[ R_{0 } + ( R''_{0})^{2}
\diagram{1loop} + \dotsb
\right]\nonumber \\
&:=& R_{\mu } + ( R''_{\mu })^{2}\left (\diagram{1loop} -
\fbox{\diagram{1loop}} + \dotsb \right)\qquad~
\end{eqnarray}
Here, the boxed diagram is defined as
\begin{equation}\label{lf178}
\fbox{\diagram{1loop}}  = \left.\diagram{1loop} \right|_\mu \ .
\end{equation}
The idea behind this construction is that at any order in perturbation
theory, any observable in the renormalized theory can be written as
perturbative expansion in the {\em bare} diagrams, to which one
applies $\mathrm{\bf S}$.  $\mathrm{\bf S}$ reorganizes the
perturbative expansion in terms of the {\em renormalized}
diagrams. The action of $\mathrm{\bf S}$ is to subtract divergencies,
which graphically is denoted by drawing a box around each divergent
diagram or sub-diagram, and to repeat this procedure recursively
inside each box.  The second line of (\ref{lf94}) is manifestly
finite, since it contains the diagram at scale $m$ minus the diagram
at scale $\mu$. This is easily interpreted as the 1-loop contribution
to the $\beta$-function.

The power of this method is not revealed before 2-loop order. Let us
give the contribution from the hat-diagram (class A):
\begin{equation}\label{lf255}
\delta R^{(2)}_{A} = \diagram{LH} R''_{0} (R'''_{0})^{2}
\end{equation}
Using $\mathrm{\bf S}$, this is rewritten as \begin{widetext}
\begin{eqnarray}\label{lf256}
\delta R^{(2)}_{A} &=& \mathrm{\bf S}\left[ R''_{0}
(R'''_{0})^{2}  \diagram{LH} \right] = R''
(R''')^{2} \left[\diagram{LH}-\,\fbox{\diagram{LH}}\,-
\diagram{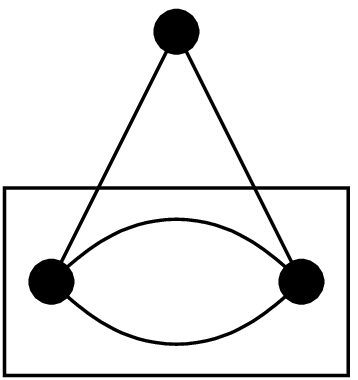}+\,\fbox{\diagram{LHsub}}\;\right]
\end{eqnarray}
\end{widetext}
Note that not only the global divergence is subtracted, but also the
sub-divergence in the bottom loop; and finally the divergence which remains,
after having subtracted the latter (last
term). Note the factor of $1=(-1)^{2}$ in front of the last diagram,
which comes with the two (nested) boxes.

Let us halt the discussion of the formal subtraction-operator at this
point, and not prove that the procedure renders all expectation
values finite; this task is beyond the scope of this article,
event hough it is not difficult to prove e.g.\ along the lines of
\cite{WieseHabil}, once the question of the ambiguities of a diagram
are settled. However let us discuss, what the subtraction procedure
can contribute to the clarification of the amgibuities.

In standard field-theory, the main problem to handle is the
cancellation of divergences, whereas the combinatorics of the vertices
is usually straightforward. This means that the sum of the
integrals, represented by the diagrams  in the brackets on the
r.h.s.\ of (\ref{lf256}) are finite. This ensures of course
renormalizability, subject to the condition that all diagrams have the
same functional dependence on $R$. Here, the factor $R'' (
R''')^{2}$ should more completely read
\begin{equation}\label{rd1}
( R'' (u)-R' (0)) R'''(u)^{2} -R'' (u)R''' (0^{+})^{2}\ .
\end{equation}
For the first term, there was no problem. However, we have seen that
the last term was more difficult to obtain.
If we  {\em demand renormalizability diagram by diagram}, all
diagrams have to give the {\em same} factor (\ref{rd1}). Thus, if {\em at
least one of them} can be calculated without ambiguity, we have an
unambiguous procedure to calculate {\em all of them}. We now
demonstrate that $\textdiagram{LHsub}$ is unambiguous. To this aim, we
detail on the subtraction operator $\mathrm{\bf
S}$, whose action is represented by the box. This box tells us to
calculate the divergent part of the sub-diagram in the box, and to
replace everything in the box by the counter-term, which here is
\begin{equation}\label{rd2}
\fbox{\diagram{1loop}}= I_{1}^{\mu} \left( R'' (u)^{2}-2 R''
(u)R'' (0) \right)\ .
\end{equation}
In a second step, one has to calculate the remaining
diagram, which is obtained by treating the box as a point, i.e.\ as a
local vertex. The idea is of course, that the sub-divergence comes from
parts of the integral, where the distances in the box are {\em much}
smaller than all remaining distances, such that this replacement is
justified. Graphically this can be written as
\begin{equation}\label{rd3}
\diagram{LHsub}= I_{1}^{\mu } ~ \times  \diagram{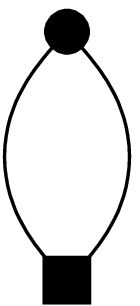}
\end{equation}
Remains to calculate the rightmost term, i.e.\ to calculate the 1-loop
diagram, from one vertex $R (u)$ and a second vertex $V
(u):= R'' (u)^{2}-2 R'' (u)R'' (0) $.
The result is in straightforward generalization of  (\ref{1loop})
\begin{equation}\label{rd4}
\diagram{1loopwithCTins}= {I_{1}} \Big[  R'' (u)
V'' (u)- R'' (0)V'' (u)- R'' (0)V'' (0) \Big]
\end{equation}
We need
\begin{equation}\label{rd5}
V'' (u) = R''' (u)^{2} + \dots
\end{equation}
The omitted terms are proportional to $R''''R''$, and contribute to
class B. We could have  avoided their appearance altogether, but
this would have rendered the notation unnecessarily heavy. The term
which contributes to (\ref{rd4}) is $V'' (u)=R''' (u)^{2}$. It has the
same analyticity properties as $R ''(u)$, especially can unambiguously
be continued to $u=0$, i.e.\ $V'' (0) = R''' (0^{+})^{2}$.
Expression (\ref{rd4}) becomes
\begin{equation}\label{rd6}
\!\!\diagram{1loopwithCTins}\!\! = {I_{1}} \Big[  R'' (u)
R''' (u)^{2}- R'' (0)R''' (u)^{2}- R'' (u)R''' (0^{+})^{2} \Big]
\end{equation}
without any ambiguity. \footnote{The same procedure can be applied to
the dynamics at the depinning transition. Care has to be taken there,
since it exists an additional 1-loop counter-term, which is an
asymmetric function with a vanishing integral. The repeated
counter-term at 2-loop order (integrated over all positions) therefore
also vanishes; however it gives a non-zero contribution both to
classes A and B (chains and hat-diagrams), of which the sum
vanishes. In order to ensure finiteness diagram by diagram, these
contributions may not be neglected. This is discussed in
\cite{LeDoussalWieseChauve2002}.}

To summarize: Using ideas of perturbative renormalizability diagram by
diagram, we have been able to compute unambiguously one of the terms in
(\ref{lf256}) and can use this information to make the functional
dependence of the whole expression unambiguous. If we were to chose any other
prescription, a proof of perturbative renormalizability is doomed to
fail, a scenario which we vehemently reject.

\subsection{Recursive construction}
\label{sec:recursive}
This method is very similar in spirit to the one of section
\ref{rd7}. There we had first calculated a subdiagram, and then
treated the result as a new effective vertex. This procedure can be
made a prescription, which insures renormalizability and potentiality,
since the 1-loop diagram insures the latter. Only at 3-loop order
appears a new diagram, (\ref{lf174}), which can not be handled that
way, but the procedure, which is otherwise very economic, can handle
again most diagrams at 4-loop order, using the new 3-loop diagram
(\ref{lf174}).

\section{Correlation functions} \label{sec:correlations} Here we
address the issue of the calculation of correlation functions.  We
note that it has not been examined in detail in previous works on the
$T=0$ FRG.  Usually correlations are obtained from tree diagrams using
the proper or renormalized vertices from the polynomial expansion of
the effective action. Thus in a standard theory one could check at
this stage that correlation functions are rendered finite by the above
counter-terms, compute them and obtain a universal answer.  In a more
conventional theory that would be more or less automatic

Here, as we point out, it is not so easy. Indeed, as we show below if
one tries to compute even the simplest 2-point correlation at non-zero
momentum one finds ambiguities already at one loop. This is because
the effective action (the counter-term) is non-analytic.

Again the requirement of renormalizability and independence of
short-scale details guide us toward a proper definition of the
correlation functions that we can compute. Interestingly, this
definition is very similar as the one obtained from an exact solution
in the large $N$ limit in \cite{LeDoussalWiese2001}. Let us illustrate
this on the 2-point function, and at the same time, derive the (finite
size) scaling function for any elasticity (not done in
\cite{LeDoussalWieseChauve2002}) for massive and finite size scheme.

\subsection{2-point function}
\label{s:2pointfunction}
We want to compute the 2-point correlation function at
$T=0$. In Fourier-representation it is given by (\ref{lf107}) with:
\begin{eqnarray}\label{lf179}
C(q) &=& (\Gamma^{(2)}(q))^{-1}_{a b}\label{lf180}
\end{eqnarray}
in terms of the quadratic part of the effective action,
which reads at any $T$
\begin{eqnarray}\label{lf181}
 \Gamma^{(2)}(q)_{a b} &=& \frac{q^2 + m^2}{T} \delta_{ab} +
\Gamma^{(2)}_{\mathrm{OD}}(q) \\
  \Gamma^{(2)}_{\mathrm{OD}}(q =0) &: =& m^\epsilon \frac{R''(0)}{T^{2}}\ .
\label{lf182}
\end{eqnarray}
i.e.\ by construction here $R''(0)$ gives the exact off-diagonal
element of quadratic part of the effective action.  Inverting the
replica matrix gives the relation, exact to all orders:
\begin{eqnarray}\label{lf183}
 C(q=0) &=& - m^\epsilon \frac{R''(0)}{m^4}\nonumber \\
&=& - \frac{1}{\epsilon \tilde{I}_1}
(\tilde R'')^*(0) m^{-d + 2 \zeta}\ .\label{lf184}
\end{eqnarray}
$R(u)$ is exactly the function entering the $\beta $-function (in the
rescaled form $\tilde{R} (u)$). In the second line we have inserted
the fixed point form, which thus gives exactly the $q=0$ correlations
in the small $m$ limit (i.e.\ up to subdominant terms in $1/m$) which
are bounded because of the small confining mass.

\subsubsection{Calculation of scaling function} We now compute $C(q)$
for arbitrary but small wave vector $q$, and to one loop, i.e.\ to next
order in $\epsilon$. One expects the scaling form (\ref{lf108}) and
that the scaling function is independent of the short-scale UV details
(i.e.\ universal), if the theory is renormalizable. It satisfies
$F(0)=1$ and, from scaling should satisfy $F(z) \sim B/z^{d + 2
\zeta}$ at large $z$.  In $d=4$ one has $F_4(z) = 1/(1 + z^2)^2$ and
we want to obtain the scaling function to the next order,
i.e.\ identify $b$ in $B=1 + b \epsilon + O(\epsilon^2)$.

Let us use straight perturbation theory with $R_0$, defined as in
Section (\ref{sec:rgdis}), including the 1-loop diagrams. This amounts
to attach two external legs to the 1-loop diagrams in
Fig.~\ref{oneloop}, and use a non-analytic\footnote{A subtle point in
that construction is that if one defines $R_0$ perturbatively from $R$
to a given order, then $R_0$ is not the original bare action (which is
analytic); thus there is no contradiction in $R_0$ being non
analytic. In a sense, introducing $R_0$ is just a trick commonly used
in field theory to express a closed equation for the flow of $R$ to
the same order. The (perturbative) exact RG method introduced in
\cite{SchehrLeDoussal2003} does that automatically without the need to
introduce $R_0$.} $R_0$. Our result is:
\begin{eqnarray}\label{lf185}
 (q^2 + m^2)^2 C(q)  &=& - T^2 \Gamma^{(2)}(q =0)\nonumber  \\
& =&  - R_0''(0) - R_0'''(0^+)^2 I(q) \label{cr0} \qquad \\
I(q)&=&\int_p \frac1{(p^2 + m^2)((p+q)^2 + m^2)}\ .\qquad \nonumber
\end{eqnarray}
There is however an ambiguity in this calculation, i.e.\ again it is
not obvious, a priori, how to interpret the $R_0'''(0)^2$ which
appear.  If one computes the one loop correction using (\ref{exp}),
one must evaluate:
\begin{equation}\label{lf186}
 R'''(0^+)^2 \langle  u^a_x u^b_y \int_{z t} \sum_{cd} |u^c_z - u^d_z|
|u^c_t - u^d_t| \rangle  G(z-t)^2  \ . 
\end{equation}
One notes that at the very special point $z=t$ there is no ambiguity,
as the interaction term is analytic to this order. Then performing the
average amounts to take two derivatives $\partial_{u_a}
\partial_{u_b}$ of
\begin{equation}\label{lf187}
 \sum_{cd} R_0''(u_c - u_d)^2  = R_0'''(0^+)^2 |u_c - u_d|^2 + O(u^3) \nn\ ,
\end{equation}
which, to this order, is analytic. In this case this is exactly the
same calculation as for the repeated 1-loop counter-term. However, the
full expression (\ref{lf186}) integrated over $z,t$ is, itself,
ambiguous.  Interestingly, this simple ambiguity already to one loop
has never been discussed previously.

Let us first show that {\it renormalizability} fixes the form to be
the one written in (\ref{cr0}). Indeed, let us re-express (\ref{lf185})
in terms of the renormalized dimensionless disorder in (\ref{do}) and
(\ref{d21})
\begin{equation}\label{subst}
R''_{0} (0)=m^\epsilon R''(0) - R'''(0^+)^2 m^{2 \epsilon}
I(0)
\ .
\end{equation}
As discussed in Section (\ref{sec:rgdis}), no ambiguity arises when
taking two derivatives of (\ref{do}) at $u=0^+$, i.e.\ the 1-loop
counter-term is unambiguous. This gives
\begin{eqnarray}\label{lf188}
&&\!\!\! (q^2 + m^2)^2 C(q)  =  \nn \\
&&\qquad = - m^\epsilon \Big[ R''(0) + R'''(0^+)^2 m^\epsilon (I(q) -
I(0)) \Big] \nn
\end{eqnarray}
Thus the substitution (\ref{subst}) acts as a counter-term which
exactly subtracts the divergence as it should. The result is finite,
as required by renormalizability, only with the above choice
(\ref{cr0}). Stated differently, the $q=0$ calculation of (\ref{cr0})
fixes the ambiguity. We show below that the methods described in the
previous Section also allow to obtain this result
unambiguously. Before that, let us pursue the calculation of the
scaling function.

Upon using (\ref{resc}) and the fixed point equation, we obtain:
\begin{equation}
 F_d\!\left(\frac{q}{m}\right) = \frac{m^4}{(q^2 + m^2)^2}\left[ 1 -
(\epsilon - 2 \zeta) \frac{1}{\epsilon \tilde{I}_1} m^\epsilon (I(q) -
I(0))  \right]\ . 
\end{equation}
Apart from the dependence on $\zeta$, the calculation of the scaling
function is very similar to the one given in
\cite{LeDoussalWieseChauve2002}. We perform here a more general
calculation which also contains the case of elasticity of arbitrary
range
\begin{equation}\label{lf189}
 q^2 + m^2 \to (q^2 + m^2)^{\frac{\alpha}{2}}\ ,
\end{equation}
and expand in $\epsilon=2\alpha -d$. Using that, in that case
\begin{widetext}
\begin{eqnarray}\label{lf190}
 I(q)
&=& \frac{1}{\Gamma (\frac{\alpha}{2})^2 }
\int_p \int_{s,t>0}  (s t)^{\frac{\alpha}{2}-1} \rme^{ - s (p+q/2)^2 - t
(p-q/2)^2 - (s+t) m^2} \nonumber \\
&=& \frac{1}{\Gamma (\frac{\alpha}{2})^2 } \int_p \rme^{-p^2}
\int_{s,t>0} (s t)^{\frac{\alpha}{2}-1} (s+t)^{-d/2} \rme^{  - q^2
\frac{s t}{s+t}
- (s+t) m^2}\nonumber  \\
&=& \frac{1}{\Gamma (\frac{\alpha}{2})^2 }\int_p \rme^{-p^2}
m^{-\epsilon} \Gamma \left(\frac{\epsilon}{2}\right) \int_{t>0}
t^{\frac{\alpha}{2}-1} (1+t)^{-d/2} \left[(1+t) + \frac{t}{1+t}
\frac{q^2}{m^2}\right]^{-\epsilon/2} \ .\qquad
\end{eqnarray}
Defining the 1-loop value of $\frac{\zeta}{\epsilon} = \zeta_1 +
O(\epsilon)$, we obtain, to the same accuracy, the scaling function in
the form ($z=|q|/m$):
\begin{eqnarray}\label{lf191}
F_d(z) &=& \frac1{(1 + z^2)^{\alpha}} \left\{ 1 - (1-2 \zeta_1)
\frac{\Gamma (\alpha)}{\Gamma (\frac{\alpha}{2})^{2}} \int_0^{\infty}
\rmd t\, t^{\frac{\alpha}{2}-1} (1+t)^{- {\alpha}} \left[ \left(1+
\frac{t
z^2}{(1+t)^2}\right)^{-\epsilon/2} - 1\right]  \right\}  \nonumber \\
 &=& \frac1{(1 + z^2)^{\alpha}} \left\{ 1 + \frac{\epsilon}{2} (1-2
\zeta_1) \frac{\Gamma (\alpha)}{\Gamma (\frac{\alpha}{2})^{2}}
\int_0^{1} \rmd s\,
[s (1-s) ]^{\frac{\alpha}{2}-1}  \ln (1+ s (1-s) z^2)  \right\}  \nonumber \\
\end{eqnarray}
(We have used the variable transformation $s=1/(1+t)$). To obtain $b$,
we need the large $z$ behavior of the scaling function:
\begin{equation}
F_d(z) \stackrel{z \to \infty }{-\!\!\!\longrightarrow} \frac{1}{z^{2
\alpha}}\Bigg\{1+ (\epsilon - 2 \zeta) \left[ \ln z + \psi \left(
\frac{\alpha}{2} \right)-\psi(\alpha) \right] \Bigg\}
\end{equation}
\end{widetext}
We want to match at large $z$:
\begin{eqnarray}\label{lf192}
 F_d(z) &=& \frac1{z^{2\alpha }} (1 + b \epsilon + O(\epsilon^2))
z^{\epsilon - 2 \zeta} \nonumber  \\
 &=& \frac1{z^{2\alpha }} \left[ 1 + (\epsilon - 2 \zeta) \ln z + b
\epsilon + O(\epsilon^2) \right]\ .\qquad \label{equiv}
\end{eqnarray}
The above result yields
\begin{eqnarray}\label{lf193}
 b = b_\alpha &=& (1 - 2 \zeta_1) \left[ \psi\!\left(\frac{\alpha}{2}
\right)-\psi \left( \alpha \right)\right]\nonumber \\
&=& \left\{
\begin{array}{lcc}
 - 2 (1 - 2 \zeta_1) \ln 2 &~\mbox{for}~&\alpha =1\\
 -  (1 - 2 \zeta_1)  &~\mbox{for}~&\alpha =2
\end{array} \right.
\end{eqnarray}

\subsubsection{Lifting the ambiguity} Let us now present two
additional methods to lift the ambiguity in the 1-loop correction to
the 2-point function and recover (\ref{lf185}).

In the background method of Section \ref{sec:background} one performs
this calculation in presence of a background field, i.e.\ considering
that the field $u^a_x$ has a uniform background expectation value:
\begin{equation}\label{lf89}
 u^a_x = u^a + v^a_x
\end{equation}
with $u_a \neq u_b$ for all $a \neq b$, and contracting the
$v^a_x$. Then at $T=0$ the sign of any $u_a - u_b$ is determined, and
the above ambiguity in (\ref{lf186}) is lifted (contracting further
the $v^a_x$ yields extra factors of $T$ and thus is not needed
here). Using the background method is physically natural as it amounts
to compute correlations by adding a small external field which splits
the degeneracies between ground states whenever they occur, as was
also found in \cite{LeDoussalWiese2001}. On the other hand, performing
the calculation in the absence of a background field, in perturbation
theory directly of the non-analytic action yields a different result,
detailed in Appendix \ref{sec:nona}, which appears to be
inconsistent. It presumably only captures correlations within a single
well.

The second method is  sloop elimination. We want to
compute contractions of:
\begin{equation}\label{ooo}
\frac{1}{8 T^4} u^a_x\, u^a_y \sum_{cdef} R (u^{c}_{z}-u^{d}_{z}) R
(u^{e}_{w}-u^{f}_{w}) 
\end{equation}
where the two disorder are at points $z$ and $w$ respectively.  Let us
restrict to the part proportional to $G_{xz} G_{zw}^2 G_{wy}$ which
gives the $q$ dependent part of the two point function.  Since $a$ is
fixed, we need to extract the ``0-replica part'' of the expression
after contractions (which will necessarily involve excluded
vertices). Starting by contracting twice the two $R$'s we get
\begin{eqnarray}\label{u1}
&&\frac{1}{4 T^2} \Big[ \sum_{c\neq d} R'' (u_{z}^{c}-u_{z}^{d}) R''
(u_{w}^{c}-u_{w}^{d}) \nonumber  \\ 
&&\qquad +\!\!\! \sum_{c\neq d,c\neq e} R'' (u_{z}^{c}-u_{z}^{d}) R''
(u_{w}^{c}-u_{w}^{e}) \Big]
\end{eqnarray}
Subtracting the sloop $W$ from (\ref{id1}) gives (up to terms which do
not depend on both $w$ and $z$, and which thus disappear after the two
remaining contractions)
\begin{equation}\label{u2}
\frac{1}{4 T^2} \sum_{c\neq d} R'' (u_{z}^{c}-u_{z}^{d}) R''
(u_{w}^{c}-u_{w}^{d})
\end{equation}
Contracting the external $u$'s with (\ref{u2}) we obtain
(restoring the correlation-functions) 
\begin{equation}\label{u3}
\int_{z,w} G_{wx} G_{wy} G_{xy}^{2}
\sum_{a\neq b} (R'''_{ab})^2 
\end{equation}
The excluded sum can be rewritten as the sum minus the term with
coinciding indices. Only the latter is a 0-replica term, which gives
the desired result:
\begin{equation}\label{u4}
- R''' (0^+)^{2} \int_{z,w} G_{xz} G_{zw}^{2} G_{wy} 
\end{equation}
This result can also be obtained writing directly the
graphs with excluded vertices and eliminating the descendants of
the sloop. 

\subsubsection{Massless finite size system with periodic boundary
conditions} The FRG method described here can also be applied to a
system of finite size, with e.g.\ periodic boundary conditions
$u(0)=u(L)$, and zero mass, which are of interest for numerical
simulations. The momentum integrals in all diagrams are then replaced
by discrete sums with $q=2 \pi n/L$,  $n \in \mathbb{Z}^d$.  One must
however be careful in specifying the mode $q=0$, i.e.\
$\left<u\right>=\frac{1}{L^d} \int_x u_x$. The simplest choice is to
constrain $\left<u\right>=0$ in each disorder configuration, which we
do for now. Since the zero mode is forbidden to fluctuate sums over
momentum in each internal line exclude $q=0$.

One then finds that the 2-loop FRG equation remains identical to
(\ref{rgdisorder}), the only changes being that
\begin{enumerate}
\item $-m \partial_m \tilde{R}$ has to be replaced by $L
\partial_L \tilde{R}$.
\item  $m \to
1/L$ in the definition of the rescaled disorder.
\item The 1-loop integral $I_{1}=\int_{k}\frac{1}{(k^{2}+m^{2})^{2}}$
entering into the definition of the rescaled disorder has to be
replaced by its homologue for periodic boundary conditions:
\begin{equation}
I_1 \to I_1' = L^{-d} \sum_{n \in \mathbb{Z}^d, n \neq 0} \frac{1}{(2
\pi/L)^4 (n^2)^2}
\end{equation}
used below.
\end{enumerate}
Here and below we use a prime to distinguish the different IR schemes.
As we have seen $X$ is, to dominant order, independent of the IR
cutoff procedure.

Thus we can now compute the 2-point function. Following the same
procedure as above, we find:
\begin{eqnarray}\label{lf188b}
C(q)  &=&  \frac{1}{q^4} L^{2 \zeta-\epsilon} 
\frac{- (\tilde R'')^*(0)}{\epsilon \tilde{I}'_1}\times \nonumber \\
&&\quad 
\left[ 1 -
(\epsilon - 2 \zeta) \frac{1}{\epsilon \tilde{I}'_1} (\tilde I'(q) -
\tilde I'(0))  \right]\ 
\end{eqnarray}
with $\tilde I'(0)=\tilde I_1'$, and, for $q=2 \pi n/L$:
\begin{equation}
\tilde I'(q)= \sum_{m \in \mathbb{Z}^d, 
m \neq 0, n+m \neq 0} \frac{1}{(2 \pi)^4 m^2 (n+m)^2 }
\end{equation}
Thus one finds the finite size scaling function (defined in
(\ref{finitesizeg}))
\begin{align}
&c'(d) g_d(q L) = q^{d+2\zeta}C (q) = \nonumber \\
&\qquad  (q L)^{ 2 \zeta -\epsilon} 
\frac{- (\tilde R'')^*(0)}{\epsilon \tilde{I}'_1}
\left[ 1 -
(\epsilon - 2 \zeta) \frac{1}{\epsilon \tilde{I}'_1} (\tilde I'(q) -
\tilde I'(0))  \right]\ 
\end{align}
as function of $q L=2 \pi n$. The asymptotic behaviour is
\begin{align}
& \left[ 1 - (\epsilon - 2 \zeta) \frac{1}{\epsilon \tilde{I}'_1} 
(\tilde I'(q) - \tilde I'(0))  \right] \nonumber \\
&\qquad  \stackrel{q \to \infty }{-\!\!\!\longrightarrow}
\left[ 1 + b' \epsilon + (\epsilon - 2 \zeta) \ln(q L) \right] \ .
\end{align}
which defines $b'$. The corresponding equation 
(\ref{equiv}), when regularizing with a mass, holds.
Taking the difference between the two
equations yields
\begin{equation}\label{bpmp}
(b-b')\epsilon =\lim_{q\to\infty}\frac{\epsilon
-2\zeta}{\epsilon}\left[\frac{\tilde I' (q)}{\tilde I'
(0)}-\frac{\tilde I (q)}{\tilde I (0)} \right]
\end{equation}
where $\tilde I(q) := I(q)|_{m=1}$. 
To leading order in $1/\epsilon $, $\tilde I' (q)=\tilde I' (0)=\tilde
I (q)=\tilde I (0)$, such that this difference takes the simpler form
\begin{eqnarray}
(b-b')\epsilon =\frac{\epsilon -2\zeta}{\epsilon\tilde I
(0)}&\!\bigg\{\!&\!\lim_{q\to\infty}\left[ \tilde I' (q) {-}\tilde I
(q) \right]
+ \left[\tilde I (0){-}\tilde I' (0) \right]\!\bigg\}\nonumber \\
&& + O
(\epsilon^{2})\label{bmbpsimp}
\end{eqnarray}
Now observe that for large $q$ the first integral can be bounded by 
\begin{equation}\label{Null!}
\left| \tilde I' (q) - \tilde I (q) \right| <\frac{\mbox{const}}{Lq} \
,
\end{equation}
which is obtained by estimating the maximal difference of integral and
discrete sum in each cell 
(defined by the discreteness of the sum), and then integrating. The difference
$\tilde I (0)-\tilde I' (0)$ is finite and can be evaluated in $d=4$
dimensions.
We need the following formulae 
\begin{eqnarray}\label{form1}
\int_{0}^{\infty} \rmd s\,s\, \rme^{-s ( n^{2}+\tau )} &=& 
\frac{1}{(n^{2}+\tau )^{2}}\\ 
\label{form3}
\int_{-\infty }^{{\infty }} \rmd n\, \rme^{-s n^{2}} &=& \sqrt{\frac{\pi}{s}}
\\ 
\sum_{n\in\mathbb{Z}} \rme^{-s n^{2} } &=& \Theta_{3,0}
(\rme^{-s} ) \label{form2} \ ,
\end{eqnarray}
where $\Theta_{3,0} (t)$ is the elliptic $\Theta$-function. 
This allows to write sum and integral as
\begin{eqnarray}\label{sumint1}
\sum_{n\in \mathbb{Z}^{4}, n\neq 0} \frac{1}{(n^{2})^{2}}
&=&\int_{0}^{\infty}\rmd s\, s \left[\left(\Theta_{3,0}  (\rme^{-s}
) \right)^{4}-1\right] \qquad \\
\label{sumint2} \int \rmd^{4} n \frac{1}{(n^{2}+1)^{2}} &=&
\int_{0}^{\infty}\rmd s\, s\,\frac{\pi^{2}}{s^{2}} \,\rme^{-s}
\end{eqnarray}
The difference in question is integrated numerically:
\begin{eqnarray}\label{asfkl}
\tilde I (0)-\tilde I' (0) &=& \frac{1}{(2\pi)^{4}}
\int\limits_{0}^{\infty}\rmd s\,
s\left[\frac{\pi^{2}}{s^{2}}\rme^{-s}-\left(\Theta_{3,0}
(\rme^{-s}) \right)^{4}+1\right]\nonumber \\
&=& -\frac{14.5019}{(2\pi)^{4}} =-0.00930479\ .
\end{eqnarray}
We thus arrive at ($\epsilon \tilde I (0) = 1/ (8 \pi ^{2}) +O
(\epsilon )$)
\begin{equation}\label{bdiffinal}
b'-b = \frac{14.5019}{(2\pi)^{4}} \frac{1-2\zeta_{1}}{\epsilon \tilde
I (0)} =  0.734676 (1-2\zeta_{1})
\end{equation}
Since the FRG equation and fixed point value $(\tilde{R}'')^*(0)$ is
universal to two loops, the final result for the amplitude ratio
between periodic and massive boundary conditions is 
\begin{eqnarray}
 \frac{c'(d)}{c(d)} &=& 1 - 2 \zeta \frac{\tilde I' (0)-\tilde I
(0)}{\epsilon \tilde I (0)} + O(\epsilon^2) \nonumber \\
& =& 1 - 1.46935 \zeta  + O(\epsilon^2)  \label{ratfinal}
\end{eqnarray}

\subsection{4-point functions and higher} 

Let us now show how to compute higher
correlation-functions with no ambiguity using the sloop-method.
Let us illustrate the method on e.g.\ the 4-point function:
\begin{equation}\label{fp1}
\left< u^{a} (w) u^{a} (x) u^{a} (y) u^{a} (z)
\right>^{c}\ .
\end{equation}
The following class of diagrams contributes:
\begin{equation}\label{fp2}
\rule[-7.2mm]{0mm}{16mm}^{x,a}_{w,a}\diagram{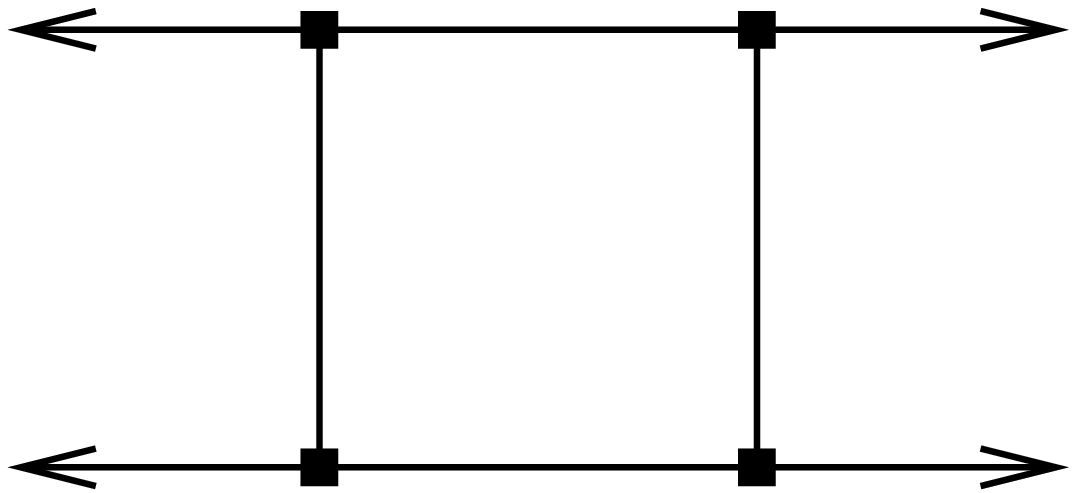}
\rule[-7.2mm]{0mm}{16mm}^{y,a}_{z,a}
\end{equation}
An arrow indicates contraction towards an external field, with
position and replica-index as indicated.  The combinatorial factors
are: $\frac{1}{4!}$ from the 4 $R$'s.  $\frac{1}{2^{4}}$, the
prefactor of the $R$'s.  $4!$ the possibilities to connect the $u$'s
to the $R$'s.  3 for the ways to make the loop of $R$'s. When
contracting first the $u$'s, there is another $2^{4}$ for the
possibilities, to attach the $u$'s to the two replicas of
$R$. Therefore only the factor of 3 remains, which is the
combinatorial factor for ordering 4 points on an unoriented ring.

We start by contracting the four $u$'s with four $R$'s, schematically:
\begin{equation}\label{fp3}
\rule[-11.9mm]{0mm}{25.5mm}^{x,a}_{w,a}\diagram{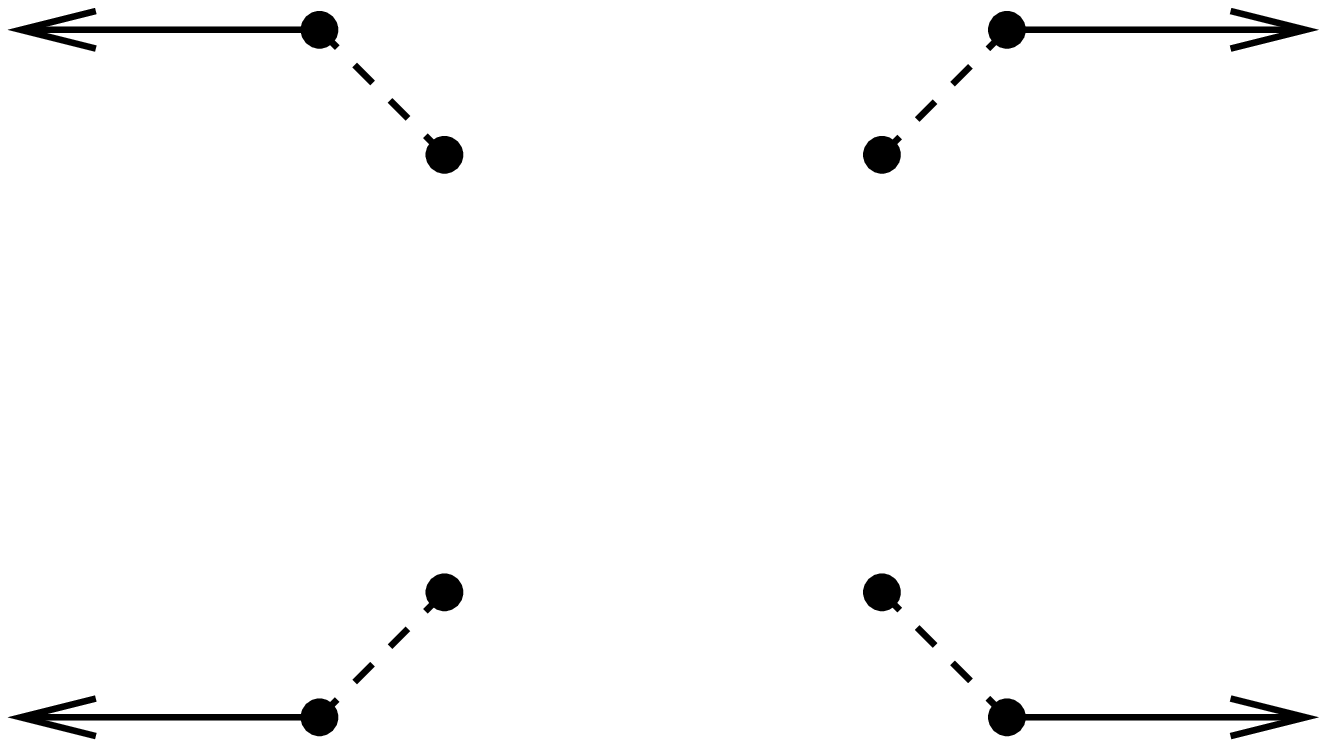}
\rule[-11.9mm]{0mm}{25.5mm}^{y,a}_{z,a}
\end{equation}
and then we perform the four other contractions. Since 
exclusions at each vertex can be introduced early on, the
number of possibilities is not too high and one easily obtains:
\begin{eqnarray}\label{fp4}
F&:=&5{R'''_{ab}}^4+4{R'''_{ab}}^3R'''_{ac}+2{R'''_{ab}}^2{R'''_{ac}}^2
+4R'''_{ab}R'''_{ac}{R'''_{ad}}^2\nonumber \\
&& +R'''_{ab}R'''_{ac}R'''_{ad}R'''_{ae}
\end{eqnarray}
where all terms have to be summed over with excluded replicas at each vertices. Due to
the factors of $R'''_{ab}$ with an odd power, it is not trivial how to
project this expression onto the space of 0-replica terms to yield the
desired expectation value (as in the previous Section
$a$ is fixed and thus no free replica sum should remain in the final result). 

To perform this projection we will first simplify the above expression using
sloops. There are a number of possible sloops which can be subtracted.
The first one is obtained by starting from
\begin{equation}\label{fp5}
\diagram{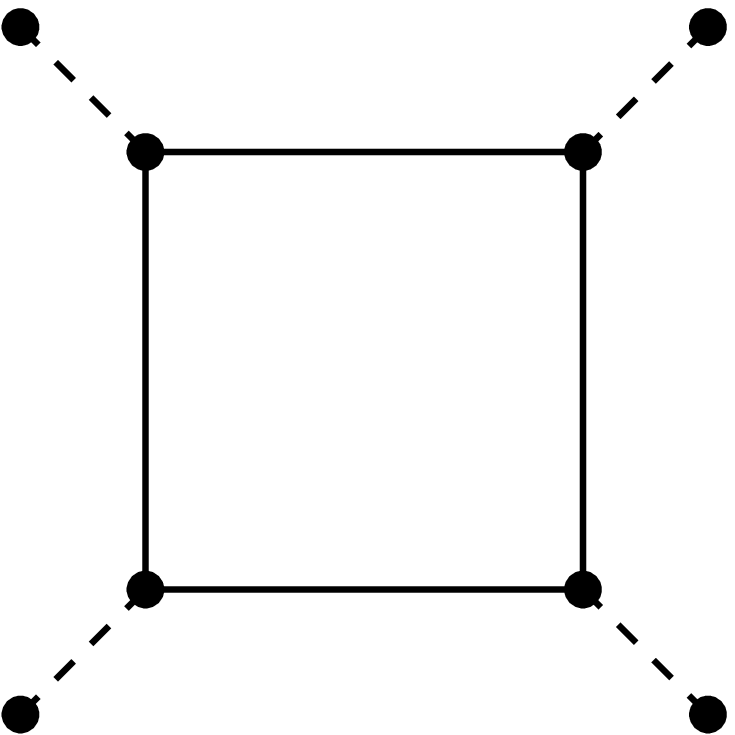}\ .
\end{equation}
It reads
\begin{equation}\label{fp6}
{S}_{1}:={R'''_{ab}}^4+R'''_{ab}R'''_{ac}R'''_{ad}R'''_{ae} \equiv 0
\end{equation}
The next sloop is 
\begin{equation}\label{fp7}
\diagram{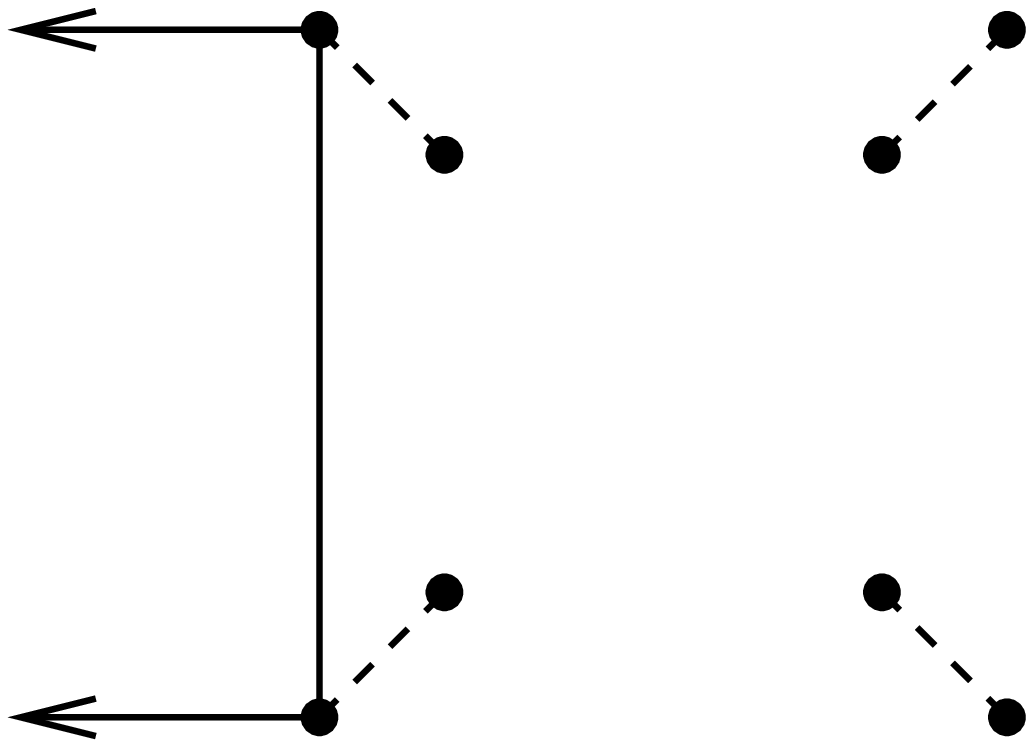}\ ,
\end{equation}
generating
\begin{eqnarray}\label{fp8}
S_{2}&:=&{R'''_{ab}}^4+2{R'''_{ab}}^3R'''_{ac}+{R'''_{ab}}^2{R'''_{ac}}^2
\nonumber \\
&&+3R'''_{ab}R'''_{ac}{R'''_{ad}}^2
+R'''_{ab}R'''_{ac}R'''_{ad}R'''_{ae} \equiv 0 \ .\qquad
\end{eqnarray}
The last needed sloop is  
\begin{equation}\label{fp9}
\diagram{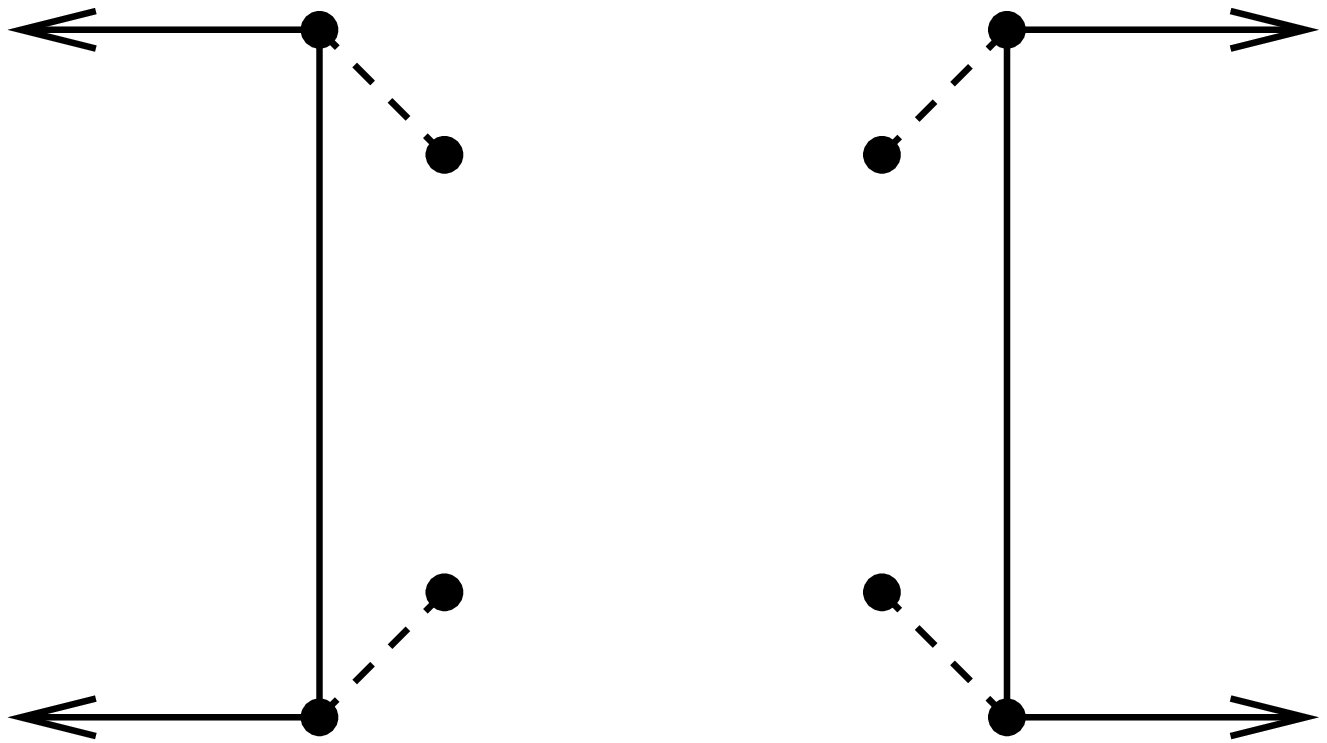}\ ,
\end{equation}
leading to
\begin{equation}\label{fp10}
S_{3}:={R'''_{ab}}^2{R'''_{ac}}^2+2R'''_{ab}{R'''_{ac}}^2R'''_{ad}
+R'''_{ab}R'''_{ac}R'''_{ad}R'''_{ae} \equiv 0
\end{equation}\begin{figure}
\Fig{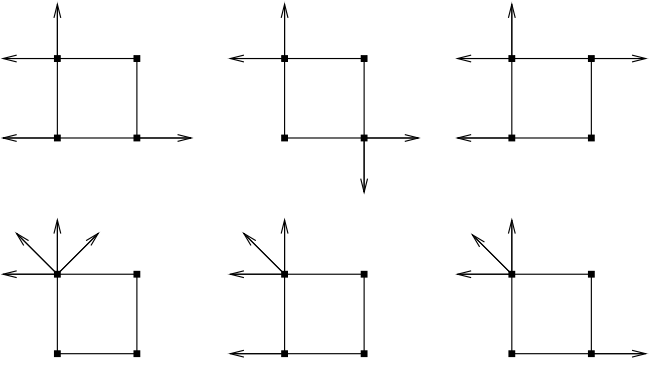}
\caption{Other possible contributions to the connected 4-point
function. They all are higher replica-terms, and thus do not
contribute. Arrows indicate contraction towards external points.}
\label{other4point}
\end{figure}
The simplest combination out of $F$, $S_{1}$, $S_{2}$ and $S_{3}$ is
\begin{equation}\label{fp11}
F- 2 S_{2} + S_{3} =
3{R'''_{ab}}^4+{R'''_{ab}}^2{R'''_{ac}}^2
\end{equation}
This expression is unambiguous because only squares of
$R'''(u)$ appear and it is easily projected onto the 0-replica part
\begin{equation}\label{fp12a}
- 2 R''' (0^{+})^{4} \ .
\end{equation}
e.g.\ one can replace $R'''_{ab} \to (1-\delta_{ab}) R'''_{ab}$
in (\ref{fp11}) and use free summations. 

Other possible contributions are given on figure \ref{other4point}.
However none of these diagrams contributes. The reason is that they are all
descendents of a sloop. We start by noting that
\begin{equation}\label{fp12}
\rule[-3.7mm]{0mm}{9.1mm}^{x,a}_{w,a}\diagram{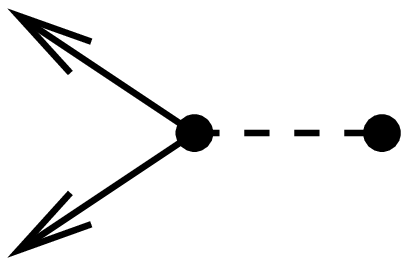}\sim \sum_{b}R''_{ab}
\end{equation}
is a true 1-replica term, i.e.\ a sloop. When constructing a diagram
on figure \ref{other4point}, {\em each} of the terms in the excluded
replica formalism is proportional to (\ref{fp12}), thus descendant of
a sloop. This means, that to any order in perturbation theory, at
$T=0$, no diagram contributes to a connected expectation value (of a
single replica), which has two lines parting from one $R$ towards
external points.

Thus the leading contribution in $R$ to the connected 4-point
function, as determined by the sloop method is the 1-loop diagram
\begin{eqnarray}\label{fp13}
&&\!\!\!\!\!\left< u^{a} (w) u^{a} (x) u^{a} (y) u^{a} (z)
\right>^{c} \nonumber \\ 
&& \!\!= -2 R'''(0^{+})^{4} \, \int_{rstu} G_{rs}G_{st}G_{tu}G_{ur} 
\nonumber \\
&& \!\!\times (G_{wr}G_{xs} G_{yt}G_{zu} +G_{ws}G_{xr}
G_{yt}G_{zu} + G_{wr}G_{xt} G_{ys}G_{zu} )\nonumber \\
&&
\end{eqnarray}
If one expressed this result in terms of the force correlator
$R'''(0^{+})^{4} = \Delta''(0^{+})^{4}$ we thus find that this
expression is formally {\it identical} to the one that we obtained for
the same four point function at the $T=0$ quasistatic depinning
threshold (Equation 5.4 in \cite{LeDoussalWiese2003a}).  This is quite
remarkable given that the method of calculation there, i.e.\ via the
non-analytic dynamical field theory, is very different. Of course the
two physical situations are different and here one must insert the
fixed point value for $\tilde R'''(0^{+})$ from the statics FRG fixed
point, while in the depinning calculation $- \tilde \Delta''(0^{+})$
takes a different value at the fixed point. In both problems the
connected four point function starts at order $O(\epsilon^4)$. However
in some cases the difference appears only to the next order in
$\epsilon$.  For instance, we can conclude that the results of
\cite{LeDoussalWiese2003a} still hold here for the {\it static random
field} to the lowest order in $\epsilon$ at which they were computed
there (of course one expects differences at next order in
$\epsilon$). On the other hand. for the static random bond case, the
result for the connected four point function will be different from
depinning even at leading order in $\epsilon$. It can easily be
obtained from the above formula following the lines of
\cite{LeDoussalWiese2003a}.

\section{Conclusion} In this article we constructed the field theory
for the statics of disordered elastic systems. It is based on the
functional renormalization group, for which we checked explicitly
renormalizability up to two loops. This continuum field theory is
novel and highly unconventional: Not only is the coupling constant a
function, but more importantly this function, and the resulting
effective action of the theory, are non-analytic at zero temperature,
which requires a non-trivial extension of the usual diagrammatic
formulation.

In a first stage, we showed that 2-loop diagrams, and {\em in some
cases even 1-loop diagrams}, are at first sight ambiguous at
$T=0$. Left unaddressed, this finding by itself puts into question all
previous studies of the problem. Indeed, nowhere in the literature the
problem was adressed that even the 1-loop corrections to the most
basic object in the theory, the 2-point function, are naively
ambiguous in the $T=0$ theory. Since the problem is controlled by a
zero-temperature fixed point there is no way to avoid this issue. An
often invoked criticism states that the problems are due to the limit
of $n\to0$ replicas.  We would like to point out that even though we
use replicas, we use them only as a tool in perturbative calculations,
which could equally well be performed using supersymmetry, or at a
much heavier cost, using disorder averaged graphs. So replicas are
certainly not at the root of any of the difficulties. Instead the
latter originate from the physics of the system, i.e.\ the occasional
occurence of quasi-degenerate minima, resulting in ambiguities
sensible to the preparation of the system.  How to deal with this
problem within a continuum field theory is an outstanding issue, and
any progress in that direction is likely to shed light on other
sectors of the theory of disordered systems and glasses.

The method we have propose to lift the apparent ambiguities is based
on two constraints: (a) that the theory be renormalizable, i.e.\ yield
universal predictions, and (b) that it respects the potentiality of
the problem, i.e.\ the fact that all forces are derivatives of a
potential. Each of these physical requirements is sufficient to obtain
the $\beta $-function at 2-loop order, and the 2-point function and
roughness exponent to second order in $\epsilon$. Next, we have
proposed several more general, more powerful and mutually consistent
methods to deal with these ambiguous graphs, which work even to higher
number of loops and allow to compute correlation functions with more
than two points.  We were then able to calculate from our theory the
roughness exponents, as well as some universal amplitudes, for several
universality classes to order $O(\epsilon^2)$.  In all cases, the
predictions improve the agreement with existing numerical and exact
results, as compared to previous 1-loop treatments. We also clarified
the situation concerning the universality (precise dependence on
boundary conditions, independence on small scale details) of various
quantities. Another remarkable finding is that the 1-loop contribution
to the 4-point function is formally identical to the one obtained via
the dynamical calculation at depinning. This hints to a general
property that {\it all} 1-loop diagrams are undistinguishable in the
statics and at depinning.  It would be extremely interesting to
perform higher precision numerical simulations of the statics, and to
determine not only exponents but universal amplitudes and scaling
functions to test the predictions of our theory. We strongly encourage
such studies.

Thus in this paper we have proposed an answer to the highly
non-trivial issue of constructing a renormalizable field theory for
disordered elastic systems. Contrarily to the closely related field
theory of depinning, which we were able to build from first
principles, we have not yet found a first principle derivation of the
theory for the statics. However, we have found that the theory is so
highly constrained, and the results so encouraging, that we strongly
believe that our construction of the field is unique. It is after all,
often the case in physics that the proper field theory is first
identified by recurrence to higher physical principles such as
renormalizability or symmetries, as is exemplified by the
Ginsburg-Landau theory of superconductivity, for which only later a
microscopic derivation was found, or gauge theories in particle
physics.

\appendix
\def\theequation{\thesection.\arabic{equation}}
\section{Symmetrization method}
\label{sec:symm}
\subsection{Continuity of the renormalized disorder and summary of the
method} The first observation is that one expects (if decomposition in
$p$-replica terms is to mean anything) that one can write the (local
disorder part of the) effective action as a sum over well defined
$p$-replica terms in the form:
\begin{eqnarray}\label{lf194}
   - \Gamma[u]& =& \sum_{p} \frac{1}{T^p} \Gamma_p[u] \nonumber
\\
& =& \sum_{p} \frac{1}{p! T^p} \sum_{a_1,...a_p}
F^{(p)}(u_{a_1},....u_{a_p)}\ ,\qquad \label{lf195}
\end{eqnarray}
where the functions $F^{(p)}$ have full permutation symmetry. The idea
of the symmetrization method is that we also expect, even at $T=0$
that these functions $F^{(p)}$ should be {\it continuous} in their
arguments when a number of them coincide.

This seems to be a rather weak and natural assumption.  Physically
these functions can be interpreted as the $p$-th connected cumulants
of a renormalized disorder, i.e.\ a random potential $V_R(u,x)$ in
each environment.  Discontinuity of the $F^{(p)}$ would mean that the
$V_R(u,x)$ would not be a continuous function. This is not what one
expects. Indeed discontinuity singularities (the shocks) are expected
to occur only in the force $F_R(u,x) = - \nabla_u V_R(u,x)$ as is
clear from the study of the Burgers equation (see e.g.\
\cite{BalentsBouchaudMezard1996} for the discussion of simple case, in
the elastic manifold formulation the shocks corresponds to rare ground
state degeneracies).  One thus expects $V_R(u,x)$ to be a continuous
function of $u$.

A further and more stringent assumption, discussed above, is the
absence of supercusp. A supercusp would mean $R'(0^+) >0$.  Thus we
assume that the non-analyticity in the effective action starts as
$|u|^3$. The usual interpretation \footnote{One should be careful in
these arguments, and consider the precise definition of $R(u)$. Indeed
one could argue that if there are many small shocks they could build a
supercusp. For instance consider the non-trivial $d=0$ limit of the
random field model, when $V(u)$ has at large $u$ the statistics of a
Brownian motion. Then, in some definitions of a coarse grained
disorder, e.g.\ such as used in \cite{LeDoussalMonthus2003} where this
model was solved exactly, $V_R(u)$ is a continuous one dimensional
Brownian motion, thus with a infinite number of small shocks and
indeed a supercusp. However, in the present paper, $V_R(u)$ is defined
from the replicated effective action, and not from the action, and
should possess - in that case - a weaker singularity.}  is that there
is a finite density of shocks and just counting how many shocks there
is in a interval between $u$ and $u'$ yields the $|u-u'|^3$
non-analyticity in $R(u)$.

Let us summarize the method before detailing actual calculations.

We thus define here the symmetrization method assuming no supercusp as
a working hypothesis. We then {\it compute} corrections to the (local
disorder part of the) effective action up to a given order in powers
of $R$, with excluded vertices for any vector such that $u_{a} \neq
u_b$ for $a \neq b$, thus with no ambiguity. This yields, as in
Section \ref{sec:unambdiag}, sums over more than two replicas with
exclusions.  These exclusions are not permutation symmetric so we
first rewrite them in an explicitly permutation symmetric way which
can be done with no ambiguity (see below).  Thus we have a sum of
terms of the form
\begin{equation}\label{lf90}
 \sum_{a_1,..a_p; 2 \neq 2} f^s(u_{a_1},..u_{a_p})\ ,
\end{equation}
where $2 \neq 2$ is a short-hand notation for $a_i \neq a_j$ for all $i
\neq j$, i.e.\ symmetrized exclusions. Each function $f$ is fully
permutation symmetric, as indicated by the $s$ superscript. Next the
non-trivial part is that we {\it explicitly verify} that these {\it
symmetrized} corrections can indeed be continued to coinciding points
unambiguously, e.g the limit $f^s(u_1,u_1,u_3,\dots ,u_{a_p})$ exist
and is independent of the direction of approach. This in itself shows
that the continuity discussed above seems to work. The existence of a
four replica term obliges us to also consider three coinciding
points. This is done by considering $f^{ss}(u_1,u_1,u_3,u_4)$, i.e.\
symmetrizing the result of two coinciding points over $u_1,u_3,u_4$
and then taking $u_3 \to u_1$. We check explicitly that this again
gives a function which can be continued unambiguously. Thus at first
sight, it would appear as the ideal method to extract the functions
$F^{(p)}$ above to order $R^3$.

\subsection{Calculations} let us reconsider the diagrams of
Fig.\ \ref{excl} We first transform them in sum with symmetrized
constraints.  We illustrate this on diagram $\beta$ where the sum can
be reorganized as:
\begin{eqnarray}\label{lf196}
\beta &\sim& \sum_{a \neq b , a \neq c}
R''_{ab}R'''_{ab}R'''_{ac}\nonumber 
 \\
&=& \sum_{ab;2\neq 2} R''_{ab}{R'''_{ab}}^{2}+ \sum_{abc;2\neq 2}
R''_{ab}R'''_{ab}R'''_{ac}\label{lf197}
\end{eqnarray}
with clearly no ambiguity. Performing similar rearrangement on all the
graphs of class A yields the sum of the graphs:
\begin{eqnarray}\label{lf198}
\delta_A R &=& 4 \sum_{ab;2\neq 2} R''_{ab}{R'''_{ab}}^{2} +
2 \sum_{abc;2\neq 2} R''_{ab} R'''_{ab} R'''_{ac} \nonumber \\
&&
- \frac{1}{2} \sum_{abc;2\neq 2} R''_{ab} R'''_{ac} R'''_{bc}
+
2 \sum_{abc;2\neq 2} R''_{ab}{R'''_{ac}}^{2} \nonumber \\
&&
+
\frac{1}{2} \sum_{abcd;2\neq 2} R''_{ab}R'''_{ac}R'''_{ad}\ . \label{lf199}
\end{eqnarray}
Now we use the property that has worked on all the examples needed
here, namely that a symmetric continuous function on $\{(x_{1},\dots
,x_{p}); i\neq j\Rightarrow x_{i}\neq x_{j} \}$ is continuous on
$R^{p}$ . Writing for any $f(x_1,..x_p)$ symmetric and continuous:
\begin{equation}\label{lf200}
\sum_{2\neq 2} f = \sum_{a_1,..a_p} \prod_{i<j} (1-\delta_{a_i a_j})
f(x_{a_1},..x_{a_p})
\end{equation}
and expanding yields, for the three and four replica sums:
\begin{eqnarray}\label{lf201}
 \sum_{abc,2\neq 2} f_{abc} &=& \sum_{abc} f_{abc} - 3 \sum_{ab} f_{aab}
+ 2 \sum_{a} f_{aaa} \nonumber \\
 \sum_{abcd,2\neq 2} f_{abcd} &=& \sum_{abcd} f_{abcd} - 6 \sum_{abc}
f_{aabc}  + 3 \sum_{ab} f_{aabb} \nonumber
\\&&  + 8 \sum_{ab} f_{aaab} - 6 \sum_{a} f_{aaaa} \label{lf202}
\end{eqnarray}
in shorthand notations such that $f_{abcd}=f(u_a,u_b,u_c,u_d)$. This
is just combinatorics.

For the three replica sums the procedure is straightforward, as
symmetrization makes manifest the continuity. One easily finds (we
drop an uninteresting single replica term):
\begin{eqnarray}
 \sum_{abc;2\neq 2}   R''_{ab} R'''_{ab} R'''_{ac} &=&
\sum_{abc} R''_{ab} R'''_{ab} R'''_{ac}  - \sum_{ab} R''_{ab} {R'''_{ab}}^{2} \nonumber \\
 \sum_{abc;2\neq 2} R''_{ab} {R'''_{ac}}^{2} &=&
\sum_{abc} R''_{ab} {R'''_{ac}}^{2} \nonumber \\
&& \hspace{-2 cm}-\sum_{ab} \left( R''_{ab} {R''' (0+)}^{2} + R''_{ab}
{R'''_{ab}}^{2}+
R'' (0) {R'''_{ab}}^{2} \right) \nonumber \\
\sum_{abc;2\neq 2} R''_{ab} R'''_{ac} R'''_{bc} &=& \sum_{abc}
R''_{ab} R'''_{ac} R'''_{bc} - \sum_{ab} R'' (0) {R'''_{ab}}^{2}
\nonumber\ ,
\end{eqnarray}
where in the first line we have applied (\ref{lf201}) to
$f_{abc}=\text{sym}_{abc} R''_{ab} R'''_{ab} R'''_{ac}$ and so on (we
define $\text{sym}_{a_1..a_p}$ as the sum over all permutations
divided by $p!$).

For the 4-replica term we find that $f_{abcd}={\rm sym}_{abcd}
R''_{ab}R'''_{ac}R'''_{ad}$ has the following limits (in a symbolic
form, omitting the free summations)
\begin{eqnarray}
f_{aabc} &=& \frac{1}{6} R'' (0) R'''_{ab}R'''_{ac} - \frac{1}{6} R''_{ab}
R'''_{ab} R'''_{bc} \nonumber \\
&& + \frac{1}{6} R''_{ac} R'''_{ac} R'''_{bc} + \frac{1}{12}
R''_{bc}{R'''_{ab}}^{2}+
\frac{1}{12}R''_{bc}{R'''_{ac}}^{2} \nonumber \\
 f_{aabb} &=& \frac{1}{3}R'' (0) {R'''_{ab}}^{2}\nonumber \\
f_{aaab} &=& \frac{1}{12}R''_{ab} {R''' (0+)}^{2}+\frac{1}{4}R''_{ab}
{R'''_{ab}}^{2}\ ,\nonumber
\end{eqnarray}
where at each step we had to symmetrize before taking coinciding point
limits (checking that this limit was unambiguous in each case).

The final result is found to be:
\begin{equation}\label{lf91}
\delta_A R_{ab} = (R''_{ab} - R''(0) ) (R'''_{ab})^2 - \frac{5}{3}
R''_{ab} (R'''(0^+))^2)\ .
\end{equation}
The same procedure applied to the repeated counter-term confirms that
it is unambiguous and give by (\ref{1-loop-rep-CC}). Thus because of
the ominous $5/3$ coefficient above, rather than the expected $1$ the
theory, using this procedure, is not renormalizable.

Diagrams of class B and C behave properly. One finds with the same
method their projections on the 2-replica part:
\begin{eqnarray}\label{lf203}
 \alpha' &=&  \frac{1}{2} R''''_{ab} R''_{ab} R''_{ab}  \\\label{lf204}
 \beta' &=&   \frac{1}{4}  ( 2 R''''(0) R''(0) R''_{ab} + R''''_{ab}
R''(0)^2 ) \\ \label{lf205}
 \gamma' &=&  \beta' \\\label{lf206}
 \delta' &=& - 2 (R''''(0) R''(0) R''_{ab} + R''''_{ab} R''_{ab}
R''(0))\qquad
\end{eqnarray}
Note the $R''''(0)$ which here is defined as
$R''''(0)=R''''(0^+)=R''''(0^-)$ since $R''''(u)$ can be continued at
zero. One has, using the expressions given in Appendix 
\ref{sec:classC}:
\begin{eqnarray}\label{lf207}
 \alpha'' &=&  R''''(0) (R''_{ab})^2 + 2  R''''_{ab} R''(0)
R''_{ab}  \\ \label{lf208}
 \beta'' + \delta'' &=&  - 2 ( R''''(0) R''(0) R''_{ab} + R''''_{ab} R''_{ab}
R''(0)) \qquad ~~ \\
 \gamma'' + \lambda'' &=&  R''''_{ab} R''(0)^2 + 2 R''''(0) R''(0) R''_{ab}  \\
 \nu'' + \eta'' &=& -  (R''''(0) (R''_{ab})^2  + R''''_{ab} (R''(0))^2)
\end{eqnarray}

These graphs (more precisely their contribution to 2-rep terms)
sum exactly to zero:

\begin{equation}\label{lf214}
\alpha'' + \beta'' + \gamma'' + \delta'' + \eta'' + \lambda''+ \nu''
=0 \ ,
\end{equation}
in agreement with the result of the ambiguous diagrammatics in the
case of an analytic function.

To conclude, although promising at first sight this method is not
satisfactory. The projection defined here seems to fail to commute
with further contractions. For instance one can check that upon
building diagrams A by contracting the subdiagram (a) in Fig.~\ref{f:sloops} onto a
third vertex does give different answers if one first projects (a) or
not. Since (a) is the divergent subdiagram this spoils
renormalizability. Since the initial assumptions of the method were
rather weak and natural, it would be interesting to see whether this
problem can be better understood in order to repair this method.

\section{Direct non-analytic perturbation theory}
\label{sec:nona}

In this Section we give some details on the method where one performs
straight perturbation theory using a non-analytic disorder correlator
$R_0(u)$ in the action. Expanding in $R_0(u)$, this involves computing
Gaussian averages of non-analytic functions, thus we start by giving a
short list of formula useful for field theory calculation of this
Section. One should keep in mind that these formula are equally useful
for computing averages of non-analytic {\it observable} in a Gaussian
(or more generally, analytic) theory.

\subsection{Gaussian averages of non-analytic functions: formulae}
\label{expabsxetc}
We start by deriving some auxiliary functions, then give a list of
expectation values for non-analytic observables of a  general
Gaussian measure.

We need
\begin{equation}\label{lf215}
\int_{0}^{\infty} \rmd q \, \left(\rme^{iqx}+\rme^{-iqx}
\right)\rme^{-\eta q} = \frac{2\eta}{\eta^{2}+x^{2}}
\end{equation}
Integrating once over $x$ starting at 0 yields
\begin{equation}\label{lf216}
\frac{1}{i}\int_{0}^{\infty} \frac{\rmd q}{q} \, \left(\rme^{iqx}-\rme^{-iqx}
\right)\rme^{-\eta q} = 2 \arctan \left(\frac{x}{\eta }\right)
\end{equation}
The r.h.s.\ reduces in the limit of $\eta \to 0$ to $\pi\, \sgn (x)$,
which gives a representation of $\sgn (x)$
\begin{eqnarray}\label{sgnx}
\sgn (x) &=& \lim_{\eta\to 0}\frac{2}{\pi} \int_{0}^{\infty} \frac{\rmd
q}{q} \sin (qx) \rme^{-\eta q}\nonumber \\
&=& \lim_{\eta\to 0}\frac{1}{\pi} \int_{-\infty }^{\infty} \frac{\rmd
q}{q} \sin (qx) \rme^{-\eta |q|}\ .
\end{eqnarray}
By integrating once more, we obtain
\begin{equation}\label{lf217}
|x| = \lim_{\eta \to 0} \frac{2}{\pi} \int_{0}^{\infty} \frac{\rmd
q}{q^{2}} \left[1-\cos (qx) \right] \rme^{-\eta q} \ .
\end{equation}
This formula is easily generalized to higher odd powers of $|x|$, by
integrating more often. The result is
\begin{equation}\label{lf218}
|x|^{2n-1} = \lim_{\eta \to 0}\frac{2}{\pi} (-1)^{n} \Gamma (2n)
\int_{0}^{\infty} \frac{\rmd q}{q^{2n}} \rme^{-\eta q}\cos
(qx)\ts_{n} \ ,
\end{equation}
where $\cos (qx)\ts_{n}$ means that one has to subtract the first
$n$ Taylor-coefficients of $\cos (qx)$, such that  $\cos (qx)\ts_{n}$
starts at order $(qx)^{2n}$:
\begin{equation}\label{lf219}
\cos  (qx)\ts_{n} := \sum_{\ell=n}^{\infty}
\frac{[-(qx)^{2}]^{\ell}}{(2\ell)! }\ .
\end{equation}

We now study expectation values. We use the measure
\begin{equation}\label{lf220}
\left(\begin{array}{cc}
\left< xx \right> & \left< xy \right>\\
\left< yx \right> & \left< yy \right>
\end{array} \right)= \left(\begin{array}{cc}
1 & t\\
t & 1
\end{array} \right)
\end{equation}
from which the general case can be obtained by 
simple rescaling $x \to x/\left< xx \right>^{1/2}$,
$y \to y/\left< yy \right>^{1/2}$.
Let us give an explicit example (we drop the convergence-generating factor
$\rme^{-\eta q}$ since it will turn out to be superfluous.)
\begin{eqnarray}\label{B4}
\left< |x| \right> &=& \frac{1}{\pi} \int_{0}^{\infty} \frac{\rmd
q}{q^{2}} \left(2- \left< \rme^{iqx} \right> -\left< \rme^{-iqx}
\right>\right) \nonumber \\
&=& \frac{2}{\pi} \int_{0}^{\infty} \frac{\rmd q}{q^{2}}
\left(1-\rme^{-q^{2}/2} \right) = \sqrt{\frac{2}{\pi }}
\end{eqnarray}
A more interesting example is
\begin{eqnarray}\label{absxabsy}
&&\!\!\!\left< \sgn (x) \sgn ( y) \right> \nonumber \\
&&= -\frac{1}{\pi^{2}}
\int_{0}^{\infty} \frac{\rmd q}{q}\int_{0}^{\infty} \frac{\rmd p}{p}
\sum_{\sigma ,\tau =\pm 1} \sigma \tau\, \left< \rme^{iq\sigma x} \rme^{ip\tau
y}  \right>\nonumber  \\
&&= -\frac{1}{\pi^{2}} \int_{0}^{\infty} \frac{\rmd
q}{q}\int_{0}^{\infty} \frac{\rmd p}{p} \sum_{\sigma ,\tau =\pm
1}\sigma \tau \,
\rme^{-\half (p^{2}+q^{2}) -\sigma \tau pqt }\nonumber \\
&&= \frac{1}{2\pi^{2}} \int_{-\infty }^{\infty} \frac{\rmd
q}{q}\int_{-\infty }^{\infty} \frac{\rmd p}{p} \,
\rme^{-\half (p^{2}+q^{2})  } \left(\rme^{pqt}-\rme^{-pqt} \right)\nonumber \\
&&= \frac{1}{2\pi^{2}} \int_{0}^{t}\rmd s\, \int_{-\infty }^{\infty}
\rmd q \int_{-\infty }^{\infty} \rmd p \,
\rme^{-\half (p^{2}+q^{2})  } \left(\rme^{pqs}+\rme^{-pqs} \right)\nonumber \\
&& = \frac{2}{\pi}\int_{0}^{t} \rmd s\, \frac{1}{\sqrt{1-s^{2}}}  =
\frac{2}{\pi} \arcsin (t)
\end{eqnarray}
Another generally valid strategy is to use a path-integral. We note
the important formula
\begin{eqnarray}\label{pathint}
\left< f (x,y) \right> &=&\frac{1}{2\pi \sqrt{1-t^{2}}}\nonumber \\
&& \times \! \int\limits_{-\infty}^{\infty}\!\rmd x
\int\limits_{-\infty}^{\infty}\!\rmd y\, f
(x,y)\,\exp\!\left[-\frac{x^{2}+y^{2}-2 t x y}{2 (1-t^{2})}
\right]\nonumber\!\! \\
&&
\end{eqnarray}
An  immediate consequence is
\begin{eqnarray}
\left< f (x)\delta (y) \right> &=& \frac{1}{2\pi
\sqrt{1-t^{2}}}\int\limits_{-\infty}^{\infty}\!\rmd x\,
 f(x)\,\exp\!\left[-\frac{x^{2}}{2 (1-t^{2})}
\right] \nonumber \\
&=&
 \frac{1}{2\pi}\int\limits_{-\infty}^{\infty}\!\rmd z\,
 f(z\sqrt{1-t^{2}})\,\exp\!\left ( -z^{2}
\right) \nonumber \\
&=& \frac{1}{\sqrt{2\pi}} \left< f (x\sqrt{1-t^{2}}) \right> \label{lf221}
\end{eqnarray}
The very existence of the path-integral representation (\ref{pathint})
also  proves that Wick's theorem remains valid. Let us give an example
which can be checked by either using (\ref{pathint}) or (\ref{B4}):
\begin{eqnarray}\label{lf95}
\left< x^{2}\,|y| \right> &=& \left< x^{2} \right> \left< |y| \right> +
2 \left< x \,y\right> \left< x\, \sgn (y) \right> \nonumber \\
&=& \left< x^{2} \right> \left< |y| \right> +2
\left< x \,y\right>^{2} \left<  \delta (y) \right> \nonumber \\
&=& \sqrt{\frac{2}{\pi}} \left(1+t^{2} \right)
\end{eqnarray}
We finish our excursion by giving a list of useful formulas, which can
be obtained by both methods:
\begin{eqnarray}\label{lf222}
 \left<|x y| \right> &=& \frac{2}{\pi} \left[ \sqrt{1-t^2} + t
\arcsin(t)  \right] \\ \label{lf223}
\left<x y|y|\right> &=& 2 \sqrt{\frac{2}{\pi}} t \\
\label{xy|xy|}
\left<x y |xy| \right> &=& \frac{2}{\pi} \left[ 3 t
\sqrt{1-t^2} + (1+ 2 t^2)   \arcsin (t)\right]\qquad 
\\
\left<|x y^3|\right> &=& \frac{2}{\pi} \left[ (2 + t^2) \sqrt{1-t^2} + 3
t \arcsin (t) \right]\label{lf226}
\end{eqnarray}

\subsection{Perturbative calculation of the 2-point
function with a non-analytic action}\label{2pointarcsine}

Let us consider the expansion of the two point function
\begin{equation}\label{}
\left<u^a_x u^b_y\right> = \frac{1}{T^2} \left<u^a_x u^b_y {\cal R}
\right> + \frac{1}{2 T^4} \left<u^a_x u^b_y {\cal R} {\cal R}\right> +
O({\cal R}^3)
\end{equation}
in powers of the disorder \footnote{These averages are connected but
this is not needed here.}, where ${\cal R} = \frac{1}{2} \int_z
\sum_{e f} R_0(u^{ef}_z)$, with $u_{z}^{ef}=u_{z}^{e}-u_{z}^{f}$. We
want to evaluate these averages at $T=0$ with a non-analytic action
$R_0(u)$. We restrict ourselves to $a \neq b$ since at $T=0$ the
result should be the same for $a=b$, and we drop the subscript $0$
from now on. As mentioned above, the Wick theorem still applies, thus
we can first contract the external legs. The term linear in $\cal R$
yields the dimensional reduction result (\ref{lf9}), thus we note
$\left<u^a_x u^b_y \right> = \left<u^a_x u^b_y \right>_{\mathrm{DR}} +
\left<u^a_x u^b_y \right>'$ and we find:
\begin{eqnarray}
\left<u^a_x u^b_y \right>' &=& 
\frac{1}{T^2} \int_{zw} \Bigg( G_{xz} G_{yw} \left< \sum_{cd} R'(u^{ac}_{z}) 
R'(u^{bd}_{w}) \right> \nonumber \\
&&- \frac{1}{2} G_{xz} G_{yz} \left< \sum_{cd} R''(u^{ab}_{z})
R(u^{cd}_{w}) \right> \Bigg)
\end{eqnarray}
up to $O({\cal R}^3)$ terms. For peace of mind one can introduce the
restrictions $c\neq a, d \neq b$ in the first sum and $c \neq d$ in
the second, but this turns out to be immaterial at the end. We need
only, in addition to (\ref{exp}):
\begin{eqnarray}
 R'(u) &=& R''(0) u + \frac{1}{2} R'''(0^+) u |u| + \frac{1}{6}
 R''''(0^+) u^3 \nonumber \\ 
 R''(u) &=& R''(0) + R'''(0^+) |u| + \frac{1}{6} R''''(0^+) u^2\ ,
\end{eqnarray}
since higher order terms in $u$ yield higher powers of $T$.  Using
(\ref{lf222}) to evaluate Gaussian averages this yields:
\begin{align}
 \left<u^a_x u^b_y \right>' =& R'''(0^+)^2 G_{zz}^2 \times \nonumber \\
&\int_{zw} \bigg( G_{xz} G_{yw}
\sum_{cd} \phi_1(t) - \frac{1}{3} G_{xz} G_{yz} \sum_{cd} \phi_2(t') \bigg) 
\end{align}
where we denote:
\begin{align}
 t &= \frac{G_{zw}}{2 G_{zz}} (\delta_{ab} + \delta_{cd} - \delta_{ad}
- \delta_{bc}) \\
 t' &= \frac{G_{zw}}{2 G_{zz}} (\delta_{ac} + \delta_{bd} - \delta_{ad}
- \delta_{bc}) \\
 \phi_1(t) &= \frac{2}{\pi} ( 3 t
\sqrt{1-t^2} + (1 + 2 t^2) \arcsin (t)  \\ 
 \phi_2(t') &= \frac{2}{\pi} \left( (2 + t'^2) 
\sqrt{1-t'^2} + 3 t' \arcsin (t') \right)
\end{align}
Note that the cross terms $R''(0) R''''(0^+)$ involve analytic
averages \footnote{$R''(0)$ can always be set formally to zero by a
trivial additive random force contribution.}  and yield zero (a
remnant of dimensional reduction). Also, to this order, no terms with
negative powers of $T$ survive for $n=0$ (see discussion below).
Performing the combinatorics in the replica sums, we find for $n=0$:
\begin{align} \label{res}
 \left<u^a_x u^b_y \right>' &= R'''(0^+)^2 G_{zz}^2 \times  \nonumber \\
& \ \int_{zw} 
\left[ G_{xz} G_{yw} \Phi_1\left(\frac{G_{zw}}{G_{zz}}\right)  +
G_{xz} G_{yz} \Phi_2\left(\frac{G_{zw}}{G_{zz}}\right) \right] \\
 \Phi_1(s) &= 2 \phi_1\left(\frac{s}{2}\right) - \phi_1(s) \\
 \Phi_2(s) &= - \frac{2}{3} \phi_2(s) + \frac{8}{3}
\phi_2\left(\frac{s}{2}\right) - 2 \phi_2(0)  
\end{align}
It is important for the following to note that cancellations
occur in the small argument behavior of these functions, namely
one has $\Phi_1(s) = -s^3/\pi + O(s^5)$ and 
$\Phi_2(s) = s^4/(4 \pi) + O(s^6)$. In $d=0$ it simplifies
(setting $G_{xy}=1/m^2$ and restoring the subscript) to:
\begin{equation}\label{sdfj}
\left<u^a u^b \right> = - \frac{R_0''(0)}{m^4} - A
\frac{R_0'''(0^+)^2}{m^8} + O(R_0^3/m^{12})
\end{equation}
with $A=(24 - 27 \sqrt{3}+8 \pi)/(3 \pi)$.  As such, this formula and
(\ref{res}) seem fine and it may even be possible to check them
numerically in $d=0$ for large $m$ using a bare disorder with the
proper non-analytic correlator $R_0(u)$.  To obtain the asymptotic $m
\to 0$ and large scale behavior in any $d$, one must resum higher
orders and use an RG procedure. The question is whether the above
formula (\ref{res}) can be used in an RG treatment.

\subsection{Discussion}\label{discussion} We found that this procedure
does not work and we now explain why.  Let us rewrite the result
(\ref{res}), including the dimensional reduction term:
\begin{eqnarray}
C_{ab}(q) &=& \frac{- R_0''(0) -
R_0'''(0^+)^2 [ A_1(q) + A_2(0) ] }{(q^2 + m^2)^2}  \nonumber \\
 A_i(q) &=& \int \rmd^d x\, \rme^{i q x} G(x)^2 \psi_i(x) \nonumber \\
\psi_i(x) &=& - \left(\frac{G(0)}{G(x)} \right)^{\!2}
\Phi_i\! \left(\frac{G(x)}{G(0)}\right)
\end{eqnarray}
with $i=1,2$. One notes that if $\psi_1(x)$ were a constant equal to
unity, one would recover the result (\ref{lf185}) obtained in Section
(\ref{sec:correlations}). However, one easily sees that while
$\psi_1(x) \approx 0.346$ approaches a constant as $x \ll a$ where $a
\sim 1/\Lambda$ is an ultraviolet cutoff, it decreases as $\psi_1(x)
\sim x^{2 - d}$ at large $x$, as a result of the above mentioned
cancellations in the small argument behavior of the functions
$\Phi_i(x)$.  Thus the infrared divergence responsible for all
interesting anomalous dimensions in the 2-point function as the
non-trivial value of $\zeta$ is killed, and the method fails. Even
more, the theory would not even be renormalizable.

We have  performed a similar calculation in the dynamical field
theory formulation of the equilibrium problem in the limit $T \to 0$,
using a non-analytic action.  There the method fails for very similar
reasons.  Only at the depinning threshold we were able to construct
the dynamical theory as explained in
\cite{ChauveLeDoussalWiese2000a,LeDoussalWieseChauve2002}.  One might
suspect that one has to start with a somehow ``normal-ordered'' theory
where self-contractions, i.e.\ terms proportional to $G_{xx}$ are
removed, since they {\em never} appear in the $T=0$ perturbation
theory.  We have not been able to find such a formulation.

Another problem with direct perturbation theory in a non-analytic
action is that there is a priori no guarantee that it has a well
defined $T=0$ limit. Let us illustrate this on a simple example in
$d=0$. The following correlation has been computed exactly by a
completely different method \cite{LeDoussalMonthus2003} for the random
field model in $d=0$ (Brownian motion plus quadratic energy landscape,
$\left< \dotsb \right>_{0}$ indicates averages over all $u$):
\begin{eqnarray}\label{lf92}
\left< u_a^2 \right>_\sigma &=& \left< u_a^2 \rme^{\sum_{cd}
\frac{\sigma}{2 T^2} |u_c - u_d| - \frac{1}{2} m^2 \sum_{c} u_c^2 }
\right>_{0} \nonumber \\
&=& C_2 \sigma^{2/3} m^{- 8/3}\ ,
\end{eqnarray}
a result which is also obtained by extrapolation from $d=4$ using the
FRG, as detailed in Section \ref{ss:Universal amplitude}.

On the other hand, the above perturbative method yields, expanding in
$\sigma$:
\begin{equation}
\left< u_a^2  \right>_\sigma = \frac{T}{m^2} + \frac{\sigma}{\sqrt{T} m^3} 
\frac{1}{\pi} \sum_{c \neq d} (1 + t^2) + O(\sigma^2)
\end{equation}
with $t=(\delta_{ac} - \delta_{ad})/\sqrt{2}$. In the zero-temperature
limit $\left< u_a^2 \right>_\sigma \approx - \frac{\sigma}{\sqrt{T}
m^3} + O(\sigma^2)$, which is ill behaved. The absence of a well
defined Taylor expansion in the zero-temperature limit is of course a
sign that the correct result (\ref{lf92}) is simply non-analytic in
$\sigma$. Although this solvable example involves a correlator
$R_0(u)$ with a {\it supercusp}, it is possible that a similar problem
occurs at higher orders (three or higher) in the expansion of the
2-point function in the case of the usual cusp non-analyticity. There
have been conflicting claims in the literature about this question
\cite{BalentsDSFisher1993,BalentsBouchaudMezard1996}, i.e.\ the
presence of fractional powers at higher orders of the expansion in a
non-analytic disorder, and it may be worth reexamining. It is however
important to note that, since the $\epsilon$-expansion proposed in the
main text is not based on such a direct expansion, it {\it does not}
yield fractional powers of $\epsilon$, contrarily to what was
conjectured in \cite{BalentsDSFisher1993}.

Finally, let us point out some properties of non-analytic {\it
observables}. Let us study e.g.\ $\left< |u^x_a| \right>$. Expansion
in powers of $\cal R$ yields a first order term $\sim
1/\sqrt{T}$. This is the sign of non-analytic behavior and indeed it
is easy to find that:
\begin{eqnarray}
\left< |u^x_a|  \right> &=& \frac{2}{\pi} \sqrt{\left< (u^x_a)^2 
 \right>_{\mathrm{DR}}} - \frac{2 \sqrt{2}}{3 \pi} R'''_0(0^+) G(0)^
 2 \times\nonumber \\
&& \int_y \left[ \sqrt{1-t^2}
(2 + t^2) + 3 t \arcsin (t) -2 \right] + O(R^2)\nonumber \\
  \label{abs}
\end{eqnarray}
where $\left< (u^x_a)^2 \right>_{\mathrm{DR}} = - \int_q
\frac{R_0(0)''}{(q^2+m^2)^2}$ and $t=\frac{G(y)}{\sqrt{2} G(0)}$.  The
first term is obtained by noting that $R''_0(0)$ acts as a Gaussian
random force which can then be separated from the non-linear force,
and the last term, evaluated using the above formula, is the only one
which survives at $T=0$ to linear order in $R_0$. The formula
(\ref{abs}) is interesting as a starting point to compute universal
ratio, such as $\left< |u^x_a| \right>^2/\left< (u^x_a)^2 \right>$ or
$\left< |u^x_a - u^y_a| \right>^2/\left< (u^x_a - u^y_b)^2 \right>$.
Indeed one notes that for $d<4$ the integral in the term proportional
to $R'''_0(0^+)$ is infrared divergent at large $y$. This is left for
future study.

\section{Diagrams of class C} \label{sec:classC} In this Appendix we
give the expression of each of the diagrams of class C represented in
Fig.\ (\ref{excl}) in the excluded (nonambiguous) diagrammatics. One
finds, including all combinatorial factors:
\begin{eqnarray}
\delta'' = \beta'' \ , \quad \lambda'' =  \gamma'' \ , \qquad  \nu'' = \eta''
\end{eqnarray}
with:
\begin{eqnarray}\label{lf230}
 \alpha'' &=&  - R''''_{ab} R''_{ac} R''_{bc} \\\label{lf231}
 \beta'' + \delta'' &=&  2 R''''_{ab} R''_{ab} R''_{bc} \\\label{lf232}
 \gamma'' + \lambda'' &=& R''''_{ab} R''_{ac} R''_{ad} \\\label{lf233}
 \eta'' + \nu'' &=& R''''_{ac} (R''_{ab})^2 
\end{eqnarray}

\section{Sloop calculation of diagrams B and C} \label{sec:sloopBC}
Let us consider the expression $\delta_B R$ for the B diagrams in the
excluded diagrammatics (\ref{deltaB}). Let us start again from a
single sloop (\ref{id1}) and (\ref{id2}) and contract this time
between $y$ and $z$ twice to produce a diagram of type B. This yields:
\begin{eqnarray} \label{lf68}
&&\hspace{-0.8cm}\diagram{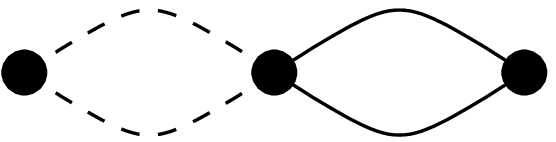}\nonumber =
\frac{1}{2} R''(0) \sum_{a \neq b} R''''_{ab} R''_{ab} \\
&& + \frac{1}{2} R''(0)
\sum_{a \neq b, b \neq c} R''''_{ab} R''_{bc}  + \frac{1}{2}
\sum_{a \neq b, b \neq c} R''_{ab} R''''_{bc} R''_{bc} \nonumber \\
&& + \frac{1}{4}
\sum_{a \neq b, a \neq c, c \neq d} R''_{ab} R''''_{ac} R''_{cd} \nonumber \\
&& + \frac{1}{4}
\sum_{a \neq c, b \neq c, c \neq d}
R''_{bc} R''''_{ac} R''_{cd} \equiv  0
\end{eqnarray}
the terms $R''(0)$ arise because the first vertex is not contracted in
the process so one must separate the (unambiguous) diagonal part to
obtain excluded sums.

If one subtracts this identity from (\ref{deltaB}) one finds that
there remain some improper three replica term (the improper four
replica term however cancels). This is because in the process of our
last contractions we have generated new sloops, but, since replica
were excluded they have to be extracted with care.

Let us rewrite the two possible ``double sloop'' from unrestricted
sums to restricted:
\begin{eqnarray}
&& \hspace{-1cm}\diagram{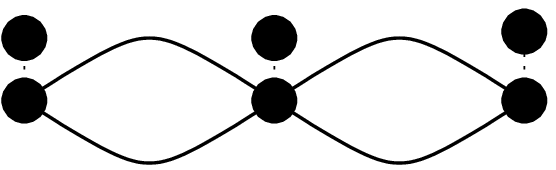} =
\frac{1}{2} \sum_{a b c d} R''_{ac} R''''_{ab} R''_{ad} \nonumber \\
&& = \frac{1}{2} \sum_{a \neq b, a \neq c, a \neq d} R''_{ac}
R''''_{ab} R''_{ad}
+ \frac{1}{2} \sum_{a \neq b} R''(0)^2 R''''_{ab} \nonumber \\
&& \qquad \qquad +
\sum_{a \neq b} R''(0) R''''_{ab} R''_{ac} \label{lf249}
\end{eqnarray}
\begin{eqnarray}\label{lf250}
&&\hspace{-1 cm} \diagram{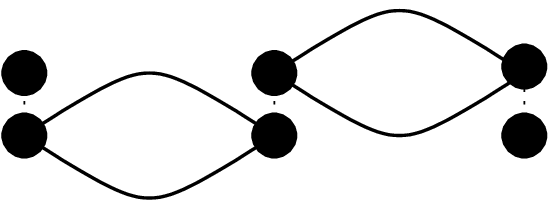} =\nonumber
\frac{1}{2} \sum_{a b c d} R''_{ac} R''''_{ab} R''_{bd} \\
&& = \frac{1}{2} \sum_{a \neq b, a \neq c, b \neq d} R''_{ac}
R''''_{ab} R''_{bd}
+ \frac{1}{2} \sum_{a \neq b} R''(0)^2 R''''_{ab} \nonumber \\
&& \qquad +
\sum_{a \neq b} R''(0) R''''_{ab} R''_{ac} \ .
\end{eqnarray}
In the process we have set to zero the terms
\begin{eqnarray}\label{lf251}
 \frac{1}{2} R''''(0) \sum_{a c d} R''_{ac} R''_{ad} &\to& 0\\\label{lf252}
 \frac{1}{2} R''''(0) \sum_{a c b d} R''_{ac} R''_{bd} &\to& 0\ ,
\end{eqnarray}
since they are proper three and four replica terms.

Defining now:
\begin{equation}\label{lf253}
\diagram{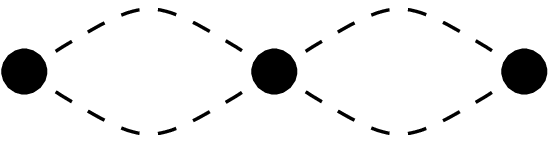}:=\frac{1}{2}\left[ \diagram{2banana2sloopsa} +
\diagram{2banana2sloopsb}\right]\ .
\end{equation}
The simplest combination which allows to extract the 2-replica part
is:
\begin{eqnarray}\label{lf254}
&& \hspace{-.5 cm}\diagram{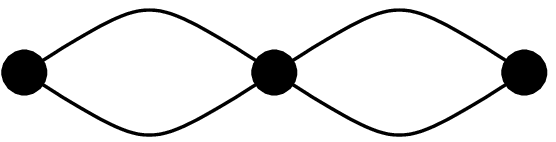}-2 \diagram{banana2sloop} +
\diagram{2banana2sloops}
\nonumber
\\
&& = \frac{1}{2} \sum _{a,b}
R''''_{ab}\left(R''_{ab}-R''_{0} \right)^{2}\ .
\end{eqnarray}

We now turn to graphs $C$. The expression for $\delta_C R$ is given as
the sum of all contributions in Appendix \ref{sec:classC}. Within the
sloop method it gives immediately zero: $\delta_C R =0$. This is
because one can start by contracting the tadpole.  Since this is a
sloop it can be set to zero:
\begin{equation}\label{tadpolesloop}
\frac{1}{8T^5} G_{xx} \sum_{abcdef} R''(u^x_{ab}) R(u^y_{cd})
R(u^z_{ef}) \equiv 0
\end{equation}
Upon further contractions, proceeding as in Section \ref{sec:sloops},
one obtains exactly that the sum of all graphs $C$ with excluded
vertices is identically zero. Graphs $C$ sum to zero since they are
all descendants of a sloop.

\section{Calculation of an integral}
\label{app:universal} We will illustrate the universality of
\begin{equation}\label{Xdef2}
X= \frac{2 \, \epsilon ( 2 I_A - I_1^2) }{ (\epsilon I_1)^2 }
\end{equation}
using a broad class of IR cutoff functions, namely
a propagator:
\begin{equation}\label{Xdefa}
\frac{1}{q^2 + m^2} \to \int \rmd x \frac{ g(x)}{q^2 + x m^2} 
\end{equation}
Here we denote $\int_x A(x) \equiv \int \rmd x g(x) A(x)$ and we
normalize $\int \rmd x g(x)=1$ (consistent with fixing the elastic
coefficient to unity).  We will show that $X=1 + O(\epsilon)$
independent of $g(x)$.

First we write:
\begin{widetext}
\begin{eqnarray}
 I_1 &=& \int_{x,x'} \int_q \frac{1}{(q^2 + x m^2)(q^2 + x' m^2)} =
\int_{x,x'} \int_{\alpha_1>0,\alpha_2>0} \rme^{- \alpha_1 (q^2 + x
m^2) + \alpha_2 (q^2 + x' m^2)}\nonumber 
\\
&& = \left(\int_q \rme^{- q^2}\right) \int_{x,x'} \int_0^{\infty}
\rmd\alpha_1 \int_0^{\infty} \rmd\alpha_2 (\alpha_1 + \alpha_2)^{-d/2}
\rme^{- m^2 (\alpha_1 x + \alpha_2 x')}
\end{eqnarray}
using the parameterization $\alpha=\alpha_1+\alpha_2$ and $\lambda
\alpha = (\alpha_1-\alpha_2)/2$ one obtains:
\begin{eqnarray}
 I_1 &=& \left(\int_q e^{- q^2}\right) \int_{x,x'}
\int_{-1/2}^{1/2}\rmd\lambda \int_0^{\infty} \rmd\alpha\,
\alpha^{1-\frac{d}{2}}
\rme^{- m^2 \alpha [\frac{x + x'}{2} + \lambda (x-x')]} \nonumber \\
&=& m^{-\epsilon} \left(\int_q e^{- q^2} \right)
\Gamma\left(\frac{\epsilon}{2} \right) \int_{x,x'} \int_{-1/2}^{1/2}
\rmd\lambda \left[\frac{x + x'}{2} + \lambda (x-x') \right]^{-\epsilon/2}
\label{xban}
\end{eqnarray}
The hat-diagram is 
\begin{eqnarray}
I_A &=& \int_{x_{i}} \int_{q_1,q_2} \frac{1}{(q_1^2 + x_{1}m^2) (q_2^2
+ x_{2}m^2) (q_2^2 + x_{2}'m^2) ((q_1+ q_2)^2+ x_{3}m^2)}\nonumber 
\\
&=& \int_{x_i} \int_{\alpha, \beta_1, \beta_2 , \gamma >0} \rme^{-
\alpha (q_1^2 + x_1 m^2) - \beta_1 (q_2^2 + x_2 m^2) - \beta_2 (q_2^2
+ x_2' m^2) - \gamma
((q_1+ q_2)^2+ x_3 m^2)} \nn \\
& =& \left(\int_{q} \rme^{-q^2}\right)^2
\int_{x_i} \int_{\alpha, \beta_1, \beta_2 , \gamma >0} 
\rme^{- m^2 ( x_1 \alpha + x_2 \beta_1 + x_2' \beta_2 + x_3 \gamma)}
\left[ \text{Det}
\left(
\begin{array} {cc} \alpha + \gamma & \gamma \\
\gamma & \beta_1 + \beta_2 +\gamma 
\end{array}
\right) \right]^{-d/2}\nn \\
& =& \left(\int_{q} \rme^{-q^2}\right)^2 \int_{x_i} \int_{\alpha,
\beta_1, \beta_2 , \gamma >0} \gamma^{3-d} \rme^{- m^2 ( x_1 \alpha +
x_2 \beta_1 + x_2' \beta_2 + x_3 )\gamma }
(\alpha + \beta_1 + \beta_2  + \alpha (\beta_1 + \beta_2))^{-d/2} \nn \\
& =& \left(\int_{q} \rme^{-q^2}\right)^2 \Gamma ( 4-d) m^{- 2 \epsilon} J
\ ,
\end{eqnarray}
where we split the divergent integral $J$ in pieces, which are either
finite or where the divergence can be calculated analytically:
\begin{eqnarray}
  J &=& \int_{x_i}  \int_0^{\infty} \rmd \alpha \int_0^{\infty} \rmd \beta \,
G(\alpha,\beta,x_i) = J_{1} + J_{2} + J_{3} \\
 G(\alpha,\beta,x_i) &=& (\alpha + \beta + \alpha
\beta)^{-2+\frac{\epsilon}{2}} \int_{-1/2}^{1/2} \rmd\lambda \left\{
x_1 \alpha + \beta \left[\frac{x_2 + x_2'}{2} + \lambda (x_2-x_2')
\right] +
x_3\right\}^{-\epsilon} \\ 
J_{1} &=& \int_{x_i} \int_0^{\infty} \rmd \alpha \int_0^{1} \rmd \beta
\, G(\alpha,\beta,x_i) = \ln 2 + O(\epsilon)
\\
 J_{2} &=& \int_{x_i} \int_0^{\infty} \rmd \alpha \int_1^{\infty} \rmd
\beta \left\{  G(\alpha,\beta,x_i) 
- \frac{1}{(1+\alpha )^{2-\frac{\epsilon}{2}} \beta^{1 + \frac{\epsilon}{2}}}
\int_{-1/2}^{1/2} \rmd\lambda
\left[\frac{x_2 + x_2'}{2} + \lambda (x_2-x_2')\right]^{-\epsilon}
\right\}\nonumber\\ 
&=& - \ln 2 + O(\epsilon)   \\
J_3 &=& \int_0^{\infty} \rmd \alpha \int_1^{\infty} \rmd \beta\,
\frac{1}{(1+\alpha )^{2-\frac{\epsilon}{2}} \beta^{1 +
\frac{\epsilon}{2}}} \int_{x_{2},x_{2}'}\int_{-1/2}^{1/2} \rmd\lambda
\left[\frac{x_2 + x_2'}{2} + \lambda (x_2-x_2') \right]^{-\epsilon}
\end{eqnarray}
This gives the final result for the hat-diagram
\begin{eqnarray} \label{IA}
I_{A} &=& \left(\int_{q} \rme^{-q^2}\right)^{\!2}\, \Gamma(4-d)\, m^{-
2 \epsilon} \left(\frac{2}{\epsilon} + 1 + O(\epsilon)\right)
\int_{x_{2},x_{2}'} \int_{-1/2}^{1/2} \rmd\lambda
\left[ \frac{x_2 + x_2'}{2} + \lambda (x_2-x_2')\right]^{-\epsilon} \nn \\
&=& \left(\frac{1}{2 \epsilon^2} + \frac{1}{4
\epsilon}+O\left(\epsilon^{0} \right)\right) \left(\epsilon I_1
\right)^2 \label{A.18} \ ,
\end{eqnarray}
where we have used that in the 1-loop integral (\ref{xban})
\begin{equation}
\int_{x,x'} \int_{-1/2}^{1/2} \rmd\lambda \left[\frac{x + x'}{2} +
\lambda (x-x') \right]^{-\epsilon/2} =
\left(1+\frac{1}{2}\alpha\epsilon +O (\epsilon^{2})\right)
\end{equation}
with $\alpha$ depending on the regulating function $g (x)$ and in the
2-loop integral (\ref{A.18})
\begin{equation}
\int_{x,x'} \int_{-1/2}^{1/2}
\rmd\lambda \left[\frac{x + x'}{2} + \lambda (x-x')
\right]^{-\epsilon}  = \left(1+\alpha\epsilon +O
(\epsilon^{2})\right) =\left(1+\frac{1}{2}\alpha\epsilon +O
(\epsilon^{2})\right)^{2} 
\end{equation}
with the {\em same} $\alpha$. 
\enlargethispage{-1cm} 
\end{widetext}

\section{Summary of all non-ambiguous diagrams, finite temperature}
\label{sec:finiteT}
\begin{figure}
\centerline{\Fig{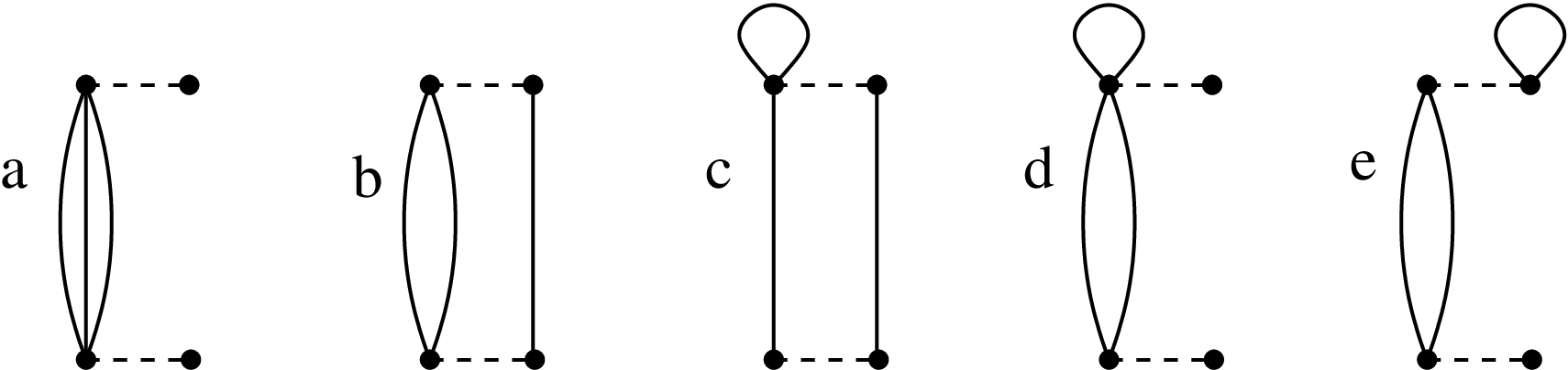}}
\caption{Diagrams to order $T R^2$  with excluded vertices.}
\end{figure}%
\begin{figure}
\centerline{\Fig{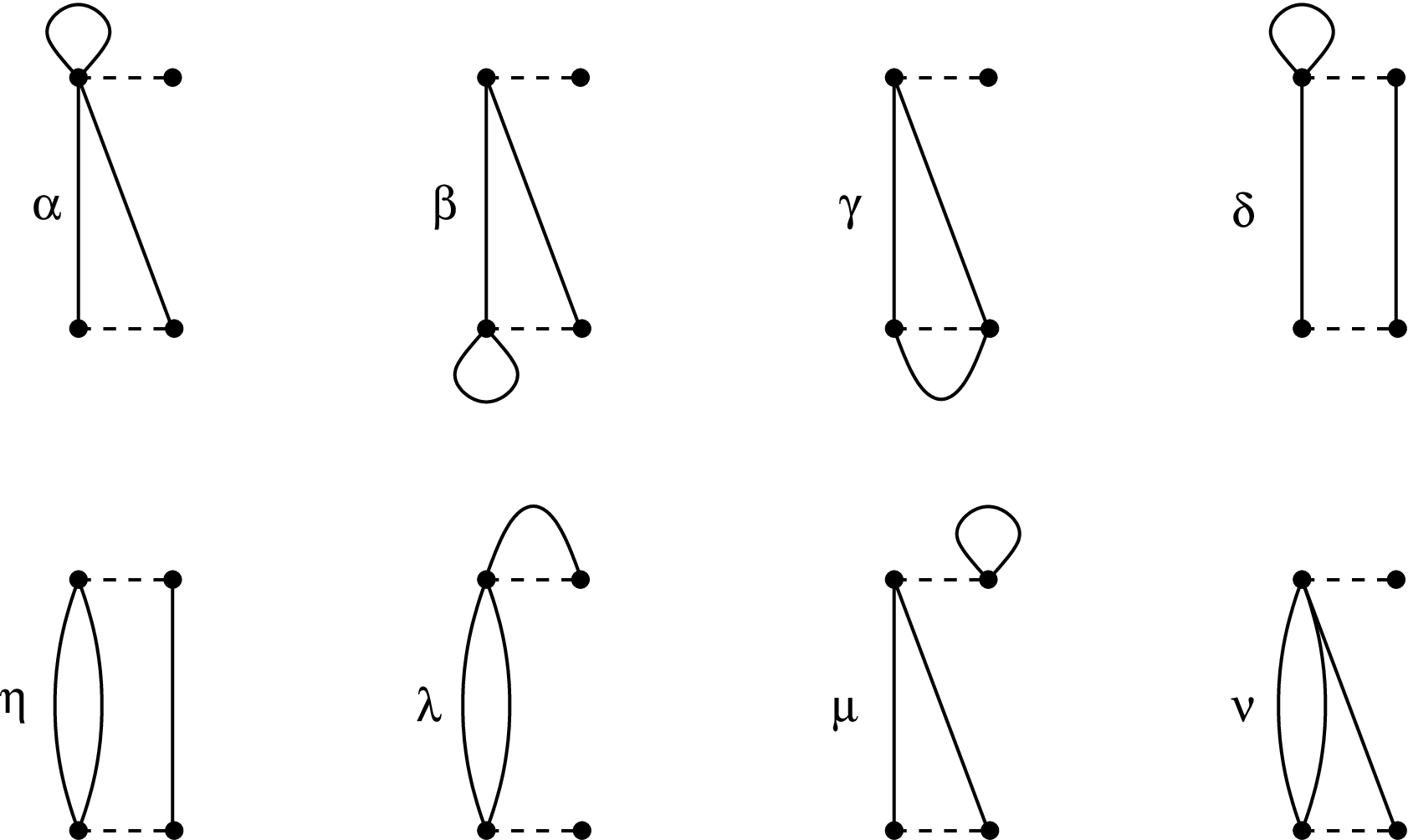}}
\caption{Diagrams to order $T R^2$  with non-excluded vertices.}
\end{figure}%
\begin{figure}
\centerline{\fig{4cm}{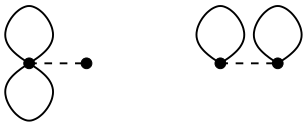}}
\caption{Diagrams to order $T^2 R$  with excluded vertices.}
\end{figure}

In this Section we give all 1-loop and 2-loop diagrams including
finite $T$, evaluated with the unambiguous diagrammatics, which have
not been given in the text. We use the unambiguous vertex $\sum_{a\neq
b}R (u_{a}-u_{b})$, denote $R_{ab}=R(u_a-u_b)$, $R'_{ab}=R'(u_a-u_b)$
etc..

\begin{figure}
\centerline{\fig{6cm}{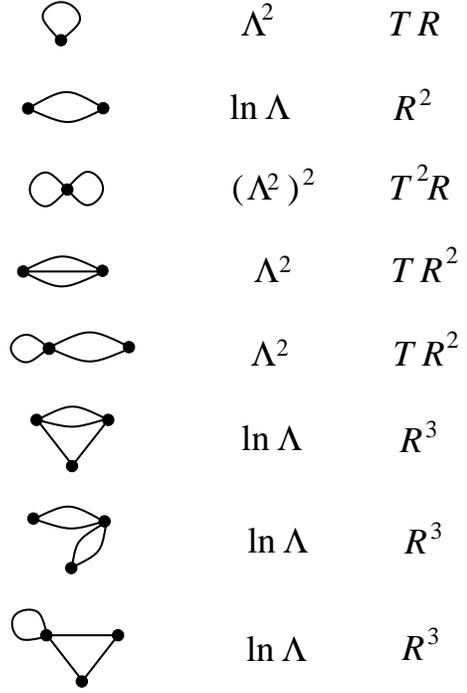}}
\caption{Divergent unsplitted diagrams to one and two loops.}
\label{f:diverg}
\end{figure}

The list of all UV-divergent diagrams up to two loops is given in
Fig.~\ref{f:diverg}. We write their contribution to the effective
action as
\begin{eqnarray}\label{lf237}
\Gamma[u]|_{u_x=u} &=& - \frac{1}{2 T^2} R \\\label{lf238}
R &=& \sum_{ab} R_{ab} + \delta^{(1)} R + \delta^{(2)} R + \dots \ .
\end{eqnarray}%
The total 1-loop contribution is
\begin{eqnarray}\label{lf239}
 \delta^{(1)} R &=& \left( \sum_{a \neq b} \frac{1}{2} (R''_{ab})^2 +
\sum_{a \neq b,a \neq c} \frac{1}{2} R''_{ab} R''_{ac} \right) I_1
\nonumber
\\
&& + \left( T \sum_{a \neq b} R''_{ab} \right) I_t\ .
\end{eqnarray}
The total 2-loop contribution is:
\begin{equation}\label{lf64}
 \delta^{(2)} R = \delta^{(2)}_A R + \delta^{(2)}_B R
+ \delta^{(2)}_C R + \delta^{(2)}_{T} R  \ ,
\end{equation}
where $\delta^{(2)}_A R$ is given in (\ref{deltaA}),
$\delta^{(2)}_B R$ is given in (\ref{deltaB})
and
\begin{eqnarray}
 \delta^{(2)}_C R &=&
\Big[ 2 \sum_{a \neq b,a \neq c} R''''_{ab} R''_{ab} R''_{ac}
- \sum_{a,b,c, 2 \neq 2} R''''_{ab} R''_{ac} R''_{bc} \nonumber
\\
&& + \!\!\!\!\sum_{a \neq b,a \neq c,a \neq d} R''''_{ab} R''_{ac} R''_{ad}
+ \!\!\! \sum_{a \neq b,a \neq c} R''''_{ab} (R''_{ac})^2 \Big] I_t
I_T \nonumber \\
&& \label{deltaC}
\end{eqnarray}
while the finite $T$ diagrams are given by
\begin{eqnarray}
 \delta^{(2)}_T R &=& \left( \frac{1}{2} T^2 \sum_{a \neq b}
R''''_{ab} \right) I_t^2
\label{deltaT} \\
&& + \left( \frac{1}{6} T \sum_{a \neq b,a \neq c} R'''_{ab} R'''_{ac}
+ \frac{1}{2} T (R'''_{ab})^2 \right) I_4 \nonumber
\\
&& + \left( T \sum_{a \neq b} R''''_{ab} R''_{ab}
+ T \sum_{a \neq b,a \neq c} R''''_{ab} R''_{ac}  \right) I_1 I_t \nonumber\\
\label{lf65}
  I_4 &=& \int_{q_1,q_2} \frac{1}{q_1^2 q_2^2 (q_1 + q_2)^2}\ .
\end{eqnarray}
For an {\em analytic} $R$ one substitutes $R_{ab} \to R_{ab} (1-
\delta_{ab})$ in the above formula and selects the 2-replica terms:
\begin{eqnarray}\label{lf240}
 \delta^{(1)} R &=& T R'' I_t + \left[\frac{1}{2} (R'')^2 - R''(0)
R''\right] I_1
\\\label{lf241}
 \delta^{(2)}_B R &=& \frac{1}{2} R'''' (R'' - R''(0))^2 I_1^2 \\\label{lf242}
 \delta^{(2)}_A R &=& (R'' - R''(0)) (R''')^2 I_A \\\label{lf243}
 \delta^{(2)}_C R &=& 0 \\\label{lf244}
 \delta^{(2)}_T R &=& \frac{1}{2} T^2 R'''' I_t^2
+ \frac{1}{2} T (R''')^2 I_4 \nonumber \\
&& +
\Big[T R'''' (R'' - R''(0)) - T R'' R''''(0)\Big] I_1 I_t\ .\nonumber \\
&&\label{lf245}
\end{eqnarray}
Let us show that if one renounces to the projection onto 2-replica
terms, one can still obtain some formal renormalizability property,
but at the cost of introducing an unmanageable series of terms with
more than two replicas.

We show how to subtract divergences by adding counter-terms of similar
form. Let us discuss only $T=0$. To cancel the 1-loop divergences we
introduce the counter-term:
\begin{equation}\label{lf66}
\delta^{(1)}_c R = \left( \sum_{a \neq b} \frac{1}{2} (R''_{ab})^2 +
\sum_{a \neq b,a \neq c} \frac{1}{2} R''_{ab} R''_{ac} \right)
I_1^{\text{div}}
\end{equation}
The repeated counter-term is
\begin{eqnarray}\label{lf246}
 \delta^{(1,1)} R &=& \Big[ R''''_{ab} (R''_{ab})^2 + R''_{ab} (R'''_{ab})^2
+ R''''_{ab} R''_{ab} R''_{ac} \nonumber \\
&& ~~ + R''_{ab} (R'''_{ac})^2 +
 R''''_{ab} R''_{ab} R''_{ac}
+ \frac{1}{2} R''''_{ab} R''_{ac} R''_{ad} \nonumber \\\label{lf247}
&& ~~ + \frac{1}{2} R''''_{ab} R''_{ac} R''_{bd}
+ R''_{ab} R'''_{ab} R'''_{ac}
+ \frac{1}{2}  R''_{ab} R'''_{ac} R'''_{ad}\nonumber \\
&&
~~ + \frac{1}{2}  R''_{ab} R'''_{ac} R'''_{ac}
- \frac{1}{2}  R''_{ab} R'''_{ac} R'''_{bc} \Big] I_1 I_1^{\text{div}}
\end{eqnarray}
omitting all (excluded) sums.
One  checks  that
\begin{equation}\label{lf67}
2( \delta^{(2)}_B R + \delta^{(2)}_A R ) \sim \delta^{(1,1)} R + O
\left(\frac{1}{\epsilon} \right)
\end{equation}
using that $2 I_A =I_1^2+O (\frac{1}{\epsilon})$. Thus
 there is some renormalizability property for $R$. One can
thus define formally a $\beta$-function:
\begin{eqnarray} \label{lf248}
- m \partial_m R &=&\epsilon R + \Big[\sum_{a \neq b} \frac{1}{2}
(R''_{ab})^2 + \sum_{a \neq b,a \neq c} \frac{1}{2} R''_{ab}
R''_{ac}\Big] (\epsilon
I_1) \nonumber \\
&& + \delta^{(2)}_A R \frac{\epsilon (I_{A}-\frac{1}{2}I_{1}^{2})}{I_{A}}
\end{eqnarray}
$R$ however includes a series of terms each with excluded sums over
$p$ replica.  Thus to be consistent one should in principle include
them from the start and pursue the method. It is not clear that it can
be closed in any way.

\begin{acknowledgments}
It is a pleasure to thank E.~Br\'ezin, W.~Krauth and A.~Rosso, for
stimulating discussions.  K.J.W.~gratefully acknowledges financial
support by the Deutsche Forschungsgemeinschaft (Heisenberg grant
Wi-1932/1-1), and additional support from NSF under grant
PHY99-07949. P.L.D.\ thanks the KITP, and K.J.W.\ thanks ENS for
hospitality during part of this work.
\end{acknowledgments}




\begin{widetext}
\tableofcontents
\bigskip 
\end{widetext}
\end{document}